\documentclass{jfm}

\usepackage{amsmath}
\usepackage{amssymb}
\usepackage{natbib}
\usepackage{verbatim}
\usepackage{graphicx}
\usepackage{psfrag}
\usepackage{xfrac}
\usepackage{bm}
\usepackage{lscape}
\usepackage{xcolor}
\usepackage{lineno}
\begin{document}
\def\Sigmap{\Sigma}
  \def\bec{}
 \def\be{\begin{equation}}
 \def\ee{\end{equation}}
 \def\bea{\begin{eqnarray}}
 \def\eea{\end{eqnarray}}
 \def\bA{\bm{A}}
 \def\bB{\bm{B}}
 \def\bE{\bm{E}}
 \def\bJ{\bm{J}}
 \def\bOmega{\bm{\Omega}}
 \def\bx{\bm{x}}
 \def\bR{\bm{R}}
 \def\bu{\bm{u}}
 \def\bcdot{\bm{\cdot}}
 \def\ra{r^\ast}
 \def\Ra{R^\ast}
 \def\Omegaa{\Omega^\dag}
 \def\Omegad{\Omega^\ast}
 \def\Omegaaa{\Omega^{\ddag}}
 \def\cF{{\cal F}}
 \def\bep{\beta_p}
 \def\bes{\beta_s}
 \def\Sigs{\Sigmap}
 \def\Sigp{\Sigmap}
 \def\calC{{\cal C}}
 \def\calP{{\cal P}}
 \def\bomega{\bm{\omega}}
 \def\bomegap{\bm{\omega}'}
 \def\tbomega{\hat{\bm{\omega}}}
 \def\tbomegap{\hat{\bm{\omega}}'}
 \def\tbomegapp{\hat{\bm{\omega}}''}
 \def\barbomega{\bar{\bm{\omega}}}
 \def\bOmega{\bm{\Omega}}
 \def\bR{{\cal \bf R}}
 \def\ck{k}
 \def\Ri{R_i}
 \def\bM{\bm{M}}
 \def\Ria{\delta}
 \def\bT{\bm{T}}
 \def\bTm{\bm{T}^{m}}
 \def\bTh{\bm{T}^{h}}
 \def\ckdag{k^{\ddag}}
 \def\bH{\bm{H}}
 \def\bhOmega{\hat{\bm{\Omega}}}
 \def\bhH{\hat{\bm{H}}}
 \def\hOmega{\hat{\Omega}}
 \def\bOmegazero{\bm{\Omega}^{(0)}}
 \def\bOmegaone{\bm{\Omega}^{(1)}}
 \def\bHzero{\bm{H}^{(0)}}
 \def\bHone{\bm{H}^{(1)}}
 \def\bomegazero{\hat{\bm{\omega}}^{(0)}}
 \def\bomegaone{\hat{\bm{\omega}}^{(1)}}
 \def\bhOmegazero{\hat{\bm{\Omega}}^{(0)}}
 \def\bhOmegaone{\hat{\bm{\Omega}}^{(1)}}
 \def\bhHzero{\hat{\bm{H}}^{(0)}}
 \def\bhHone{\hat{\bm{H}}^{(1)}}
 \def\bhomegazero{\hat{\bm{\omega}}^{(0)}}
 \def\bhomegaone{\hat{\bm{\omega}}^{(1)}}

 \def\hH{\hat{H}}
 \def\bfe{\bm{e}}
 \def\bht{\hat{\bm{\omega}}}
 \def\bt{\bm{\omega}}
 \def\hOmegaone{\hat{\Omega}^{(1)}}
 \def\hOmegatwo{\hat{\Omega}^{(2)}}
 \def\Omegadzero{\Omega^{\dag(0)}}
 \def\Omegadone{\Omega^{\dag(1)}}
 \def\Omegadtwo{\Omega^{\dag(2)}}
 \def\Mizero{M_I^{(0)}}
 \def\Mrzero{M_R^{(0)}}
 \def\Mip{M_I'}
 \def\Mrp{M_R'}
 \def\tH{\hat{\bm{\omega}} \bcdot \hat{\bm{H}}}
 \def\tO{\hat{\bm{\omega}} \bcdot \hat{\bm{\Omega}}}
 \def\tO{\hat{\Omega}_{\bm{\omega}}}
 \def\HO{\hat{\bm{H}} \bcdot \hat{\bm{\Omega}}}
 \def\HO{\hat{\Omega}_{\bm{H}}}
 \def\bOmegad{\bm{\Omega}^\ast}
 \def\bepsilon{\hat{\bm{\epsilon}}}
 \def\bhomega{\hat{\bm{\omega}}}
 \def\hOmegat{\hat{\Omega}_{\bt}}
 \def\hOmegapar{\hat{\Omega}_\parallel}
 \def\hOmegaperp{\hat{\Omega}_\perp}
 \def\bL{\bm{L}}
 \def\bsigmap{\bm{\sigma}^{(p)}}
 \def\btimes{\bm{\times}}
 \def\cG{{\cal G}}
 \def\cD{{\cal D}}
 \def\bcG{\bm{{\cal G}}}
 \def\bcD{\bm{{\cal D}}}
 \def\etaone{\eta_c^{(1)}}
 \def\etatwo{\eta_c^{(2)}}
 \def\etathree{\eta_c^{(3)}}
\def\Omegadast{\Omega^{\ast \ast}}
\def\etaoneast{\eta_1^\ast}
\def\etatwoast{\eta_2^\ast}
\def\etathreeast{\eta_3^\ast}
\def\bM{\bm{M}}
\def\bHO{\bm{H} \bm{\cdot} \hat{\bm{\Omega}}}
\def\bHOp{\bm{H} \bm{\cdot} \hat{\bm{\Omega}}'}
\def\bF{\bm{F}}
 \def\bFm{\bm{F}^{m}}
 \def\bFh{\bm{F}^{h}}
\def\bB{\bm{B}}
\def\bk{\bm{k}}
\def\Mpar{\bm{M}_{\parallel}}
\def\bMperp{\bm{M}_\perp}
\def\hbk{\hat{\bm{k}}}
\def\Fpar{\bm{F}_{\parallel}}
\def\bFperp{\bm{F}_{\perp}}
\def\bsigmam{\bm{\sigma}^M}
\def\bI{\bm{I}}
\def\bB{\bm{B}}
\def\br{\bm{r}}
\def\barB{\bar{\bm{B}}}
\def\delB{\Delta \bm{B}}
\def\barM{\bar{\bm{M}}}
\def\delM{\Delta \bm{M}}
\def\delbu{{\bf u}'}
\def\barOmega{\bar{\bm{\Omega}}}
\def\delbOmega{\Delta \bm{\Omega}}
\def\bv{\bm{v}}
\def\barphi{\bar{\phi}}
\def\delphi{\Delta \phi}
\def\bddot{\bm{:}}
\def\barbH{\bar{\bm{H}}}
\def\barbB{\bar{\bm{B}}}
\def\tbH{\hat{\bm{H}}}
\def\tbHp{\hat{\bm{H}}'}
\def\tbHpp{\hat{\bm{H}}''}
\def\barbOmega{\bar{\bm{\Omega}}}
\def\barbomega{\bar{\bm{\omega}}}
\def\barbM{\bar{\bm{M}}}
\def\hatbH{\hat{\bar{\bm{H}}}}
\def\hatbOmega{\hat{\bar{\bm{\Omega}}}}
\def\hatbomega{\hat{\bar{\bm{\omega}}}}
\def\hatbM{\hat{\bar{\bm{M}}}}
\def\delbH{\bm{H}'}
\def\delbOmega{\bm{\Omega}'}
\def\delbomega{\bm{\omega}'}
\def\delbM{\bm{M}'}
\def\barH{\bar{H}}
\def\barOmega{\bar{\Omega}}
\def\baromega{\bar{\omega}}
\def\barM{\bar{M}}
\def\delH{\bm{H}'}
\def\delbor{\bor'}
\def\delOmega{\Omega'}
\def\delbomega{\bm{\omega}'}
\def\delomega{\omega'}
\def\delM{M'}
\def\beomega{{\bf e}_{\bomega}}
\def\bepar{{\bf e}_\parallel}
\def\beperp{{\bf e}_\perp}
\def\ko{k_{\bomega}}
\def\kpar{k_{\parallel}}
\def\kperp{k_{\perp}}
\def\bfD{{\bf D}}
\def\delbF{{\bf F}'}
\def\bMzero{\bm{M}_0}
\def\bMone{\bm{M}_1}
\def\bQ{\bm{Q}}
\def\phiav{\bar{\phi}}
\def\delphi{\phi'}
\def\tH{\hat{\bm{H}}}
\def\tphi{\hat{\phi}}
\def\bk{\bm{k}}
\def\bcdot{\bm{\cdot}}
\def\tbF{\hat{\bm{F}}}
\def\barbv{\bar{\bm{v}}}
\def\barbu{\bar{\bm{u}}}
\def\tbv{\hat{\bm{v}}}
\def\tbu{\hat{\bm{u}}}
\def\bD{\bm{D}}
\def\bK{\bm{K}}
\def\bS{\bm{S}}
\def\bor{\bm{o}}
\def\barbor{\bar{\bm{o}}}
\def\borp{\bm{o}'}
\def\tbor{\hat{\bm{o}}}
\def\torp{\hat{o}^\dag}
\def\torpp{\hat{o}^\ddag}
\def\hatbor{\hat{\bar{\bm{o}}}}
\def\baru{\bar{u}}
\def\barv{\bar{v}}
\def\bDm{\bm{D}^m}
\def\bDh{\bm{D}^h}
\def\Di{D_i}
\def\Diprime{D_i^\prime}
\def\Didprime{D_i^{\prime \prime}}
\def\bDhm{\bm{D}^{h/m}}
\def\barn{\bar{n}}
\def\deln{n'}
\def\tn{\hat{n}}
\def\tphi{\hat{\phi}}
\def\barphi{\bar{\phi}}
\def\beomega{{\bf e}_{\barbomega}}
\def\beH{{\bf e}_{\barbH}}
\def\komega{k_{\barbomega}}
\def\kH{k_{\barbH}}
\def\kperp{k_{\perp}}
\def\Sigi{\Sigma_i}
\def\Da{D^\ast}
\def\Ma{M^\ast}
\def\chia{\chi^\ast}
\def\wH{\hat{\omega}_{\bH}}
\def\oH{\hat{o}_{\bH}}
\def\Dwwh{D_{\barbomega \barbomega}^h}
\def\Dwparh{D_{\barbomega \parallel}^h}
\def\Dwperph{D_{\barbomega \perp}^h}
\def\Dparparh{D_{\parallel \parallel}^h}
\def\Dparperph{D_{\parallel \perp}^h}
\def\Dperpperph{D_{\perp \perp}^h}

\def\Dwwm{D_{\barbomega \barbomega}^m}
\def\Dwparm{D_{\barbomega \parallel}^m}
\def\Dwperpm{D_{\barbomega \perp}^m}
\def\Dparparm{D_{\parallel \parallel}^m}
\def\Dparperpm{D_{\parallel \perp}^m}
\def\Dperpperpm{D_{\perp \perp}^m}

\def\Dwwhm{D_{\barbomega \barbomega}^{h/m}}
\def\Dwparhm{D_{\barbomega \parallel}^{h/m}}
\def\Dwperphm{D_{\barbomega \perp}^{h/m}}
\def\Dparparhm{D_{\parallel \parallel}^{h/m}}
\def\Dparperphm{D_{\parallel \perp}^{h/m}}
\def\Dperpperphm{D_{\perp \perp}^{h/m}}

\def\DHH{D_{\barbH \barbH}}
\def\DHperp{D_{\barbH \perp}}
\def\Dperpperp{D_{\perp \perp}}

\def\DHHh{D_{\barbH \barbH}^h}
\def\DHperph{D_{\barbH \perp}^h}
\def\Dperpperph{D_{\perp \perp}^h}

\def\DHHm{D_{\barbH \barbH}^m}
\def\DHperpm{D_{\barbH \perp}^m}
\def\Dperpperpm{D_{\perp \perp}^m}

\def\DHHhm{D_{\barbH \barbH}^{h/m}}
\def\DHperphm{D_{\barbH \perp}^{h/m}}
\def\Dperpperphm{D_{\perp \perp}^{h/m}}

\def\timesp{\cdot}

\def\etaprime{\eta^\prime}
\def\muprime{\mu^\prime}
\def\phiprime{\phi^\prime}

\def\bu{\bm{u}}
\def\bv{\bm{v}}
\def\barbv{\bar{\bm{v}}}
\def\barbu{\bar{\bm{u}}}
\def\delbv{\bm{v}^\prime}
\def\delbu{\bm{u}^\prime}
\def\bnabla{\bm{\nabla}}
\title[Particle interactions in a magnetorheological fluid]{The effect of
inter-particle hydrodynamic and magnetic interactions in a magnetorheological fluid}

\author[V. Kumaran]{V. Kumaran}

\affiliation{Department of Chemical Engineering, Indian Institute of Science, Bangalore 560 012, India.}

\date{}
\maketitle
\begin{abstract}
A magnetorheological fluid, which consists of magnetic particles suspended in a 
viscous fluid, flows freely with well-dispersed particles in a the absence of a magnetic field, but particle 
aggregation results in flow cessation when a
field is applied. The mechanism of dynamical arrest is examined by analysing
interactions between magnetic particles in a magnetic field subject to a shear flow.
An isolated spherical magnetic particle
undergoes a transition between a rotating state at low magnetic field and a static orientation at 
high magnetic field. The effect of interactions for spherical dipolar and polarisable particles with static 
orientation is examined for a  dilute viscous suspension. There are magnetic interactions due
to the magnetic field disturbance at one particle caused by the dipole moment
of another, hydrodynamic interactions due to the antisymmetric force moment
of a non-rotating particle in a shear flow, and a modification of the magnetic field due
to the particle magnetic moment density. When there is a concentration variation, 
the torque balance condition results in a disturbance to the orientation of the particle magnetic moment. 
The net force and the drift velocity due to these disturbances is calculated, and the 
collective motion generated is equivalent to an anisotropic diffusion process. 
When the magnetic field is in the flow plane, the diffusion 
coefficients in the two directions perpendicular to the field direction are negative, 
implying that concentration fluctuations are unstable in these directions. 
This instability could initiate 
field-induced dynamical arrest in a magnetorheological fluid.
\end{abstract}

\section{Introduction}
 A magnetorheological fluid is a suspension of magnetic particles of size about
 \( 1-10 \mu \) m in a viscous fluid (\cite{vincenteetal,softmatterreview}). 
 Brownian motion
 is not important in these suspensions, because the particle size exceeds
 \( 1 \mu \)m, and the viscosity of the carrier fluid could be 2-3 orders of magnitude larger than
 that of water. The volume fraction of the particles could be as low as \( 10 \% \) or lower (\cite{anupama}).
 The salient feature of these fluids is the rapid reversible transition between a low-viscosity
 state in the absence of a magnetic field, where the particles are well dispersed, and a 
 high viscosity state under a magnetic field where the particles form
 sample-spanning clusters which arrest flow in the conduit (\cite{sherman}). This transition takes place reversibly and
 rapidly within time periods of tens to hundreds of milliseconds. Due to this rapid switching, magnetorheological
 fluids are used in applications such as dampers and shock absorbers (\cite{klingenberg}).
  
 Magnetorheological fluids are characterised by measuring their `yield stress' as a function of the 
 magnetic field (\cite{sherman}). In the field of rheology, the yield stress for a Bingham plastic fluid delineates
 solid-like and fluid-like behaviour --- the material behaves as an elastic solid when the stress
 is less than the yield stress, and flows like a viscous liquid when the stress exceeds the yield
 stress (\cite{barneshuttonwalters}). In the Bingham model, the stress
 is the sum of a yield stress \( \tau_y \) and a contribution which is linear in the strain rate,
 and the slope of the stress-strain rate curve is called the plastic viscosity. 
 For characterisation, the magnetorheological fluid is placed in a rheometer, and a 
 magnetic field is applied across the sample. The strain rate is then set at progressively increasing
 values, and the stress is measured. The stress-strain rate curve is fitted to the Bingham plastic
 model to determine the yield stress and the plastic viscosity. The
 stress-strain rate curves for magnetorheological fluids typically increase continuously, and they 
 do not exhibit a discontinuous change in slope at yield. In order to determine the yield stress and
 plastic viscosity, their high strain rate behaviour of these fluids are extrapolated linearly to 
 zero strain rate. There are two dimensionless numbers that are used to characterise magnetorheological
 fluids. The first is the Mason number, the ratio of the shear stress and a reference magnetic stress,
 the latter is the ratio of the magnetic dipole-dipole interaction force between pairs of particles
 and the square of the particle diameter (\cite{sherman}). The second is the Bingham number which 
 is the ratio of the yield stress and 
 the fluid stress. The yield stress does increase as the applied magnetic field is increased,
 and it has been proposed that the Bingham number (ratio of yield stress and viscous stress)
 is inversely proportional to the Mason number (\cite{sherman}). 
%

 More sophisticated constitutive relations for magnetorheological fluids have been
 explored in simulations and experiments. Empirical fitting functions of the type
 \begin{eqnarray}
  \frac{\eta}{\eta_0} & = & 1 + \left( \frac{\mbox{Mn}^\ast}{\mbox{Mn}} \right)^{\alpha}
 \label{eq:constitutive}
 \end{eqnarray}
 have been used for the effective viscosity \( \eta \), which is the ratio of the 
 stress and strain rate (\cite{tighe}). Here, \( \eta_0 \) is the viscosity in the absence of
 a magnetic field, \( \mbox{Mn} \) is the Mason number which is proportional to
 the strain rate, and \( \mbox{Mn}^\ast \) and \( \alpha \) are fitting parameters. Equation 
 \ref{eq:constitutive} reduces to the Bingham equation for \( \alpha = 1 \). For \( \alpha < 
 1 \), there is a transition from a power law form for the constitutive
 relation at low \( \mbox{Mn} \) to a Newtonian form for high \( \mbox{Mn} \). While some
 simulation and experimental studies (\cite{sherman,marshall,bonnecaze}) on
 dipolar particles in an external field have found that \( \alpha \) is close
 to \( 1 \), others (\cite{melrose,felt,martin}) report that the exponent is less than 1. In the simulations of
 \cite{tighe}, the exponent \( \alpha \) is found to depend on the details
 of the interactions between particles. More sophisticated rheological models,
 such as the Casson model in \cite{lopez}, have also been employed for the 
 rheology of magnetorheological fluids.

 The above characterisation procedure examines
 the `unjamming' or dynamical release transition, where the particle structures formed by the magnetic field are disrupted
 due to shear. However, rapid flow cessation involves the opposite dynamical arrest, where 
 initially dispersed particles cluster and block the flow in the conduit when a magnetic field is applied. The mechanism for dynamical arrest is different from that probed in characterisation experiments. The clustering has to be initiated by interactions between well-dispersed particles
 upon application of a magnetic field in the presence of flow. Here, some insight is obtained into the initiation of
 the dynamical arrest process by considering the effect of the hydrodynamic and magnetic interactions
 between dispersed particles in the sheared state in the presence of a magnetic field.
 
 Ferrofluids (\cite{schumacher,moskowitzrosensweig,zaitsevshliomis,chavesrinaldi}) form
 another class of suspensions of magnetic particles. In this case, the particles are of
 nanometer size, Brownian diffusion is significant, and the effect of interactions
 on the fluctuating motion of the particles may be less important. Suspensions of conducting 
 particles in shear flow also experience a torque in a magnetic field (\cite{moffat,vk19,vk20a}). This is because an eddy current is induced when a particle rotates in a magnetic field,
 and this induces a magnetic dipole which interacts with the field. Suspensions of
 conducting particles are not considered here, but a similar calculation procedure can be 
 used to predict the effect of interactions.
 
 Rather than fitting the measured or simulated rheology to a specific constitutive
 relation, the approach here is to examine the effect of particle interactions in a 
 suspension with well-dispersed particles. This is similar to the effect of 
 particle interactions on the viscosity of a non-Brownian particle suspension 
 (\cite{batchelor,hinch}). The dynamics of a spheroidal particle subjected to a 
 shear flow and a magnetic field has been studied (\cite{almog,cebecki,vk20,vk21a,vk21b}).
 In the absence of a magnetic field, the particle axis rotates in closed `Jeffery orbits' 
 (\cite{jeffrey,hinchleal}) on a unit sphere. When a magnetic field is applied, the 
 magnetic torque tends to  align the particle along the field direction. When the 
 magnetic field is below a threshold, the particle rotates with frequency lower than 
 the Jeffery frequency. When the magnetic field exceeds the threshold, the particle
 has a steady orientation which progressively inclines towards the field direction
 as the field strength is increased. The transition between steady and rotating states
 depends on the particle shape factor and the magnetisation model, and there could
 also be multiple steady states. Here, the effect of interactions is studied for a
 suspension of spherical particles. Though this configuration is sufficiently simple that
 the particle orientation can be determined analytically, it does provide physical
 insight into the mechanisms that drive the collective behaviour of the particles.
 
\begin{figure}
%
  \parbox{.49\textwidth}{
 \includegraphics[width=.49\textwidth]{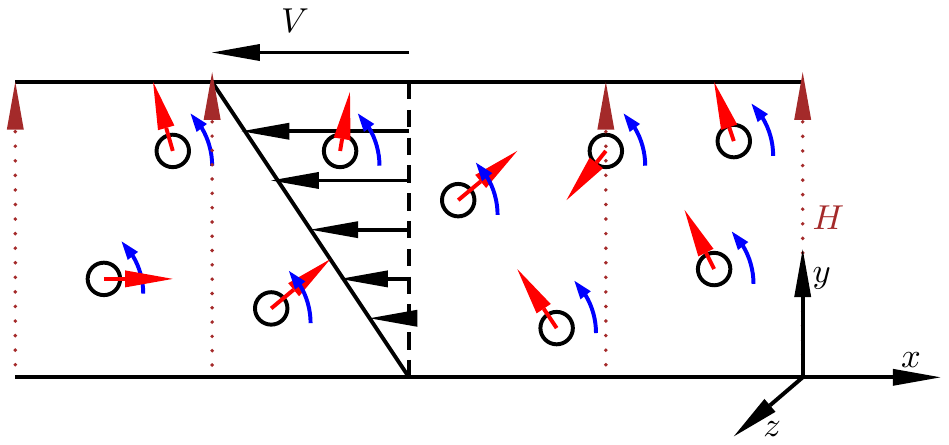}
 
 \begin{center} (a) \end{center}
 }
  \parbox{.49\textwidth}{
 \includegraphics[width=.49\textwidth]{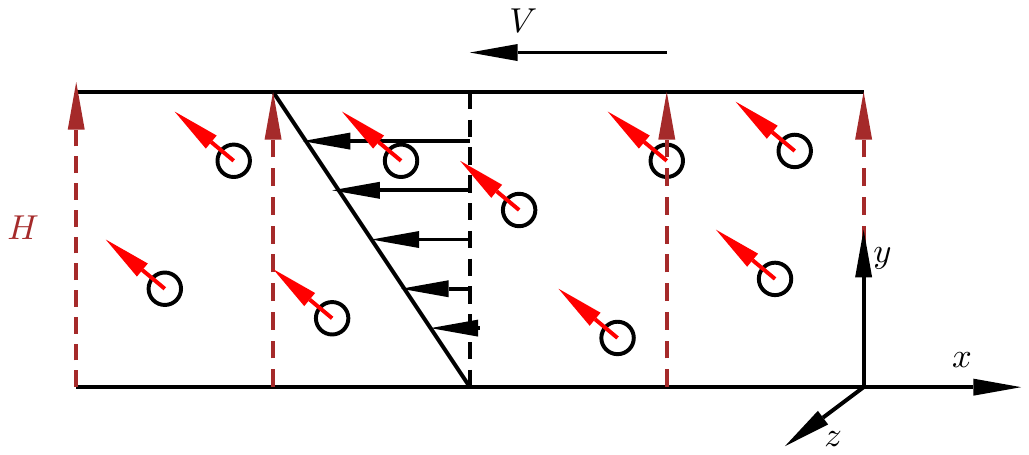}
 
 \begin{center} (b) \end{center}
 }
 
 \parbox{.49\textwidth}{
 \includegraphics[width=.5\textwidth]{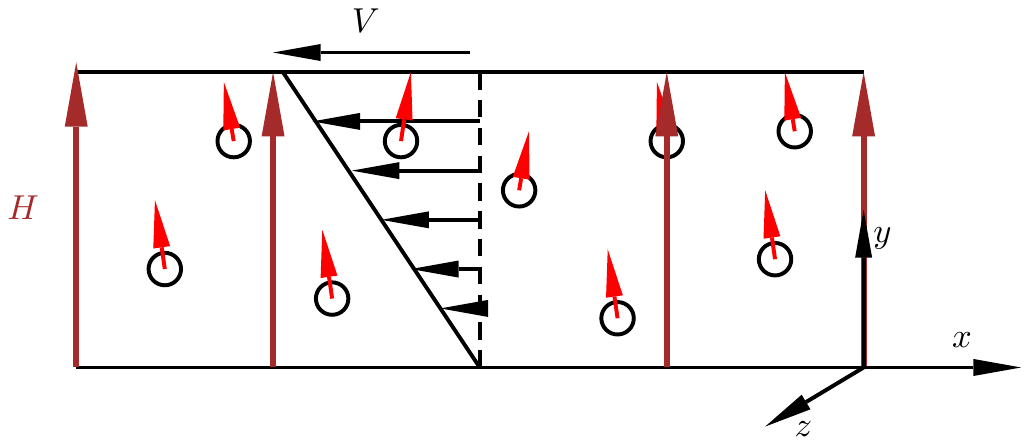}
 
 \begin{center} (c) \end{center}
 }
  \parbox{.49\textwidth}{
 \includegraphics[width=.5\textwidth]{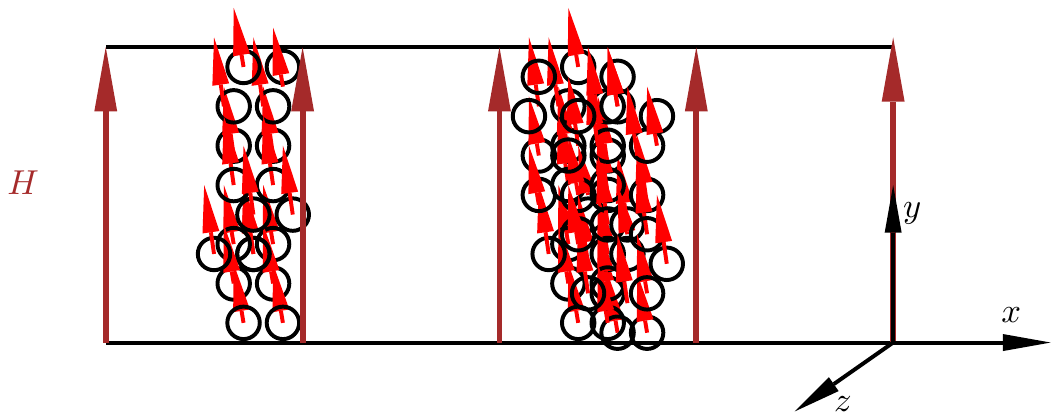}
 
 \begin{center} (d) \end{center}
 }
%
\caption{\label{fig:mechanism} A suspension of dipolar spherical particles in a magnetic field; (a) for low
field intensity, the rotation rate is smaller then the fluid rotation rate; (b) above a threshold, the particles
align in the direction determined by a balance between the hydrodynamic and magnetic torque; (c) at high
magnetic field, the particles align close to the field direction; (d) dynamical arrest occurs due to sample-spanning
aggregation of particles. 
}
\end{figure}

Consider a suspension of spherical magnetic particles sheared between two plates, as shown in figure
\ref{fig:mechanism}. The red arrows show the magnetic moment of the particles, and the blue arrows
indicate the direction of rotation of the particles due to the fluid shear. When there is no magnetic
field, there is no magnetic torque, and the hydrodynamic torque is zero in the viscous limit. The particles
rotate with angular velocity equal to the local fluid rotation rate, which is one half of the 
vorticity. When a small magnetic field is applied, there 
is a torque on the particle which depends on the particle orientation. The particles do rotate, but 
with average angular velocity smaller than the fluid rotation rate, as shown in figure 
\ref{fig:mechanism} (a). When the magnetic field is increased beyond a threshold, the particles
do not rotate; the static orientation is determined by a balance between the magnetic and
hydrodynamic torques, as shown in figure \ref{fig:mechanism} (b). There is a transition between rotating
and steady states when the dimensionless parameter \( \Sigma \), defined later in equation \ref{eq:eq221},
exceeds $\frac{1}{2}$. In the limit of large magnetic field, the particle magnetic moments align closer
to the field direction, as shown in figure \ref{fig:mechanism} (c).
 
 The formation of sample-spanning clusters for dynamical arrest in magnetorheological fluids, shown in figure \ref{fig:mechanism} (d),
 requires an additional mechanism not present in the single-particle dynamics. While alignment of particle
 dipoles is expected for high magnetic field, it is not clear how the non-Brownian particles approach and cluster 
 in the stream-wise direction in a highly viscous fluid where particle contact is prevented by lubrication.
 Formation of isotropic clusters is not sufficient to explain the dynamical arrest phenomenon; such clusters
 would be rotated and stretched by the shear flow (\cite{vargaetal}). Here, we show that anisotropic clustering 
 is initiated by interactions between dispersed particles. The interactions are of two types, magnetic and 
 hydrodynamic. It is shown that hydrodynamic interactions between particles amplify concentration 
 variations along the flow direction, while magnetic interactions dampen concentration variations along the 
 cross-stream direction. The amplification of density of waves in the stream-wise direction, combined with
 the rapid equalisation of concentration in the cross-stream direction, could result in the formation of
 sample-spanning aggregates that arrest the flow.

The effect of magnetic interactions between particles
on magnetophoresis has been calculated (\cite{morozov1,morozov2}), and this has
been included in models for suspensions of magnetic particles in the presence
of magnetic fields (\cite{elfimova,ivanov}). Magnetic interactions enhance
the diffusion coefficient, dampen concentration and
fluctuations and stabilise the suspension. This is contrary to the aggregation
required for dynamical arrest in magnetorheological fluids. The following simple
calculation illustrates the diffusion enhancement due to magnetic interactions.

The force on a particle in a magnetic field
is assumed of the form \( {\bf F} = \mu_0 \bnabla (\bM \bcdot \bH) \), 
where \( \mu_0 \) is the magnetic permeability which is considered
a constant, \( \bM \) is the particle
moment and \( \bH \) is the magnetic field. The net force is zero
when the magnetic field is uniform in the case of non-interacting particles. 
The magnetic field disturbance \( \delH(\bx) \) on a particle at the location 
$\bx$ due to the presence of another 
particle at the location $\bx'$ with magnetic moment \( \bM \) is,
\begin{eqnarray}
 \delH(\bx) & = & \frac{1}{4 \pi} \left( \frac{3 (\bx - \bx')(\bx - 
 \bx')}{|\bx - \bx'|^5} - \frac{\bI}{|\bx - \bx'|^3} \right) \bcdot \bM. \label{eq:eq21a}
\end{eqnarray}
In a uniform suspension where the number density is a constant, the integral of
the right side of \ref{eq:eq21a} over all space is zero by symmetry.
In a suspension of particles with a spatially varying number density \( n(\bx) \), the 
magnetic field at the particle location \( \bx \) due to interaction with other particles is,
\begin{eqnarray}
 \delH(\bx) & = & \frac{1}{4 \pi} \int d \bx' \left( \frac{3 (\bx - \bx')(\bx - 
 \bx')}{|\bx - \bx'|^5} - \frac{\bI}{|\bx - \bx'|^3} \right) \bcdot (\bM n(\bx')). \label{eq:eq21}
\end{eqnarray}
These disturbances are expressed in Fourier space using the transform,
\begin{eqnarray}
 \hat{\star}_{\bk} & = & \int d \bx \exp{(\imath \bk \bcdot \bx)} \star'(\bx),
 \label{eq:fouriertransform}
\end{eqnarray}
where the field variable \( \star \) could be the magnetic field, velocity, vorticity, concentration
or the orientation vector. The Fourier transform of the disturbance to the magnetic field is,
\begin{eqnarray}
 \tbH_{\bk} & = & - \frac{\bk \bk \bcdot \bM \tn_{\bk}}{k^2}, \label{eq:eq01}
\end{eqnarray}
where \( \tn_{\bk} \) is the Fourier transform of the number density variations. The Fourier transform
of the force \( \bF = \mu_0 \bnabla (\bM \bcdot \bH) \) acting on the particle is,
\begin{eqnarray}
 \tbF_{\bk} & = & - \imath \bk \mu_0 (\bM \bcdot \tbH_{\bk}) = \imath \bk 
 \left( \frac{\mu_0 (\bM \bcdot \bk)(\bM \bcdot \bk) \tn_{\bk}}{k^2} \right).
 \label{eq:eq02}
\end{eqnarray}
The particle velocity drift due to the magnetic interaction force is the ratio of the force and the friction coefficient, 
\begin{eqnarray}
 \tbu_{\bk} & = & \frac{\bF_{\bk}}{3 \pi \eta d} = \imath \bk 
 \left( \frac{\mu_0 (\bM \bcdot \bk)(\bM \bcdot \bk) \tn_{\bk}}{3 \pi \eta d k^2} \right),
\label{eq:eq03}
\end{eqnarray}
where the Stokes expression for the friction coefficient for a spherical particle, \( 3 \pi \eta d \),
has been used, \( \eta \) is the fluid viscosity and \( d \) is the particle diameter. 
The variation in number density due to the drift velocity is determined from the conservation
equation,
\begin{eqnarray}
 \frac{\partial \tn_{\bk}}{\partial t} - \imath \bk \bcdot (\barn \tbu_{\bk}) & = & 0.
 \label{eq:eq04}
\end{eqnarray}
Here \( \barn \) is the average number density of the particles, and it is assumed that the variation
in number density \( \tn_{\bk} \) is much smaller than \( \barn \). This results in a diffusion-type
equation,
\begin{eqnarray}
 \frac{\partial \tn_{\bk}}{\partial t} + \bk \bk \bm{:} ( \bD \tn_{\bk}) & = & 0, \label{eq:eq05}
\end{eqnarray}
where the diffusion tensor \( \bD \) is,
\begin{eqnarray}
 \bD & = & \frac{\mu_0 \bM \bM \barn}{3 \pi \eta d}. \label{eq:eq06}
\end{eqnarray}
The diffusion tensor depends on the magnetic moment of the particles, and not the magnetic
field, because it is caused by the interaction force between the particles, as noted in 
\cite{sherman} and \cite{klingenbergetal}.
This effect is anisotropic, and it operates only in the direction of the particle moment. Importantly, 
the effect of interactions dampens fluctuations in the direction of the magnetic moment since the 
diffusivity is positive,  and it has no effect in the direction perpendicular to the particle magnetic moment. 

The clustering of particles can be predicted if the variation of the magnetic field
due to the magnetic moments of the particles is included in an effective medium
approach. The force on the particle is \( {\bf F} = \mu_0 \bnabla [\bM \bcdot (\bH + n \bM)], 
\) where \( n \bM \) is the magnetisation
or the magnetic moment per unit volume. The force on the particles is, instead of \ref{eq:eq02},
\begin{eqnarray}
 \tbF_{\bk} & = & - \imath \bk \mu_0 (\bM \bcdot \tbH_{\bk} + \tn_{\bk} \bM \bcdot \bM) = \imath \bk 
 \left( \frac{\mu_0 (\bM \bcdot \bk)(\bM \bcdot \bk) \tn_{\bk}}{k^2} - M^2 \tn_{\bk} \right).
 \label{eq:eq07}
\end{eqnarray}
The last term in the bracket on the right arises from the modification of the variation in the 
magnetic moment per unit volume due to the variation in the particle number density. With this
inclusion, the diffusion tensor is,
\begin{eqnarray}
 \bD & = & \frac{\mu_0 (\bM \bM - \bI M^2) \barn}{3 \pi \eta d}, \label{eq:eq08}
\end{eqnarray}
where \( \bI \) is the identity tensor. The above diffusion tensor is diagonal in a 
co-ordinate system where one the co-ordinate directions is along the particle magnetic
moment. The diffusion coefficient along the magnetic moment is identically zero, while that
in the two directions perpendicular to the magnetic moment is negative. This predicts 
spontaneous amplification of concentration fluctuations in the direction perpendicular to 
the magnetic moment of the particles, and no amplification or damping along the particle
magnetic moment. This is consistent with the formation of particle chains along the 
magnetic field direction if the magnetic moment is aligned along the magnetic field. 
However, this is still an effective medium or a mean-field
approach for the initial growth of perturbations.

The contrast in the electrical permittivity or magnetic permeability on the stability
of a suspension was considered by \cite{vonpfeil} for a sheared particle suspension.
In the dilute limit, the effective magnetic permeability is expressed as
\( \mu_{eff} = \mu_0 (1 + \muprime \phi) \), where \( \phi \) is the volume fraction,
\( \muprime = 3 (\mu_R - 1)/(\mu_R + 2) \), 
where \( \mu_0 \) is the magnetic permeability of the fluid in the absence of particles and
\( \mu_R \) is the ratio of the magnetic permeabilities of the particles and fluid. 
For \( \mu_R 
> 1 \) where the magnetic permeability of the particles is higher than that for the fluid, the 
magnetic flux lines preferentially pass through regions of higher particle concentration. 
In the present analysis, the change in the magnetic field due 
to the particles is the magnetisation or the magnetic moment per unit volume, and
latter is the product of the particle moment and the number density.

When there is fluid flow, there is
one other effect, which is the variation in the effective viscosity of the suspension with the 
particle concentration. For a dilute suspension of spherical particles subject to shear flow,
the effective viscosity is \( \eta = (1 + \etaprime \phi) \), where \( \etaprime = \frac{5}{2} 
\) and \( \phi \) is the volume fraction.
The 
viscosity variation with volume fraction is included in the 
analysis in sections \ref{subsec:diffusionpermanent} and \ref{subsec:diffusioninduced}, where an
effective diffusion coefficient is derived in the equation for the concentration fluctuation.
This effective diffusion coefficient is proportional to \( \etaprime \). 

The objective of the present analysis is to analyse the stability of a flowing suspension
incorporating magnetic and hydrodynamic interactions, and the effect of particle concentration
on the viscosity and magnetic field. When there is
a difference between the particle and fluid rotation rates, there is a torque exerted by the 
particle on the fluid. This torque results in an antisymmetric dipole force moment at the particle 
center, which generates a velocity disturbance. The resulting torque at the center of a test particle 
due to interactions with other particles is zero if the number density is uniform (see Appendix D1 in 
\cite{vk19}). However, when there is a 
perturbation in the number density, there is a net torque exerted due to hydrodynamic interactions.
In the presence of spatial concentration variations, there is also a net torque exerted at the 
test particle due to magnetic interactions with other particles. There is a small spatial variation 
in the particle orientation vector in order to satisfy the zero torque condition. This results in a 
small correction to the magnetic moment of the particle \( \bM \), and consequently an additional
contribution to the force. The two additional effects of hydrodynamic interactions and the zero torque condition are incorporated in the force due to interactions, and it is shown that this force amplifies
concentration fluctuations and destabilises the suspension.

The expression for the force, \( \bF = \bnabla (\bM \bcdot \bB) \), results from the 
Ampere model for a magnetic dipole, where the dipole is considered as an infinitesimal current
loop (\cite{jackson,griffiths}). Here, \( \bB = \mu_0 (\bH + n \bM) \) is the magnetic flux
density. An alternate expression is based on the Gilbert model, 
\( \bF = \bM \bcdot \bnabla \bB \), where the dipole is considered the superposition of two 
magnetic `charges' of opposite signs separated by an infinitesimal distance. Though the two models are equivalent in most
cases, there are differences in special situations, such as the magnetic field of a proton which fits the Ampere
model. The present analysis is another situation where there are differences in the results
of the two models. For small perturbations \( \bM'(\bx) \) and \( \bB'(\bx) \)
about the base state where the particle moment \( \barbM \) and magnetic field \( \barbB \)
are spatially uniform, the force in the Gilbert model linearised in the perturbations 
is \( \barbM \bcdot \bnabla \bB' \), whereas that in the Ampere model is \( 
\bnabla (\bM'(\bx) \bcdot \barbB + \barbM \bcdot \bB'(\bx)) \). 
Variations in the magnetic permeability with position due to a contrast in the 
permeabilities of the particles and fluid are also incorporated
in the Ampere model. The force in the Gilbert model not
depend on the perturbations to the particle magnetic moment, while that in the Ampere 
model does.
Another distinction is that the force in the Ampere model can be written
as \( - \bnabla U_M \), where \( U_M = - \bM \bcdot \bB \) is the magnetic energy,
whereas there is no equivalent relation to an energy for the Gilbert model. Since the 
Ampere model is also known to fit experimental results, such as the magnetic field of a 
proton measured in hyperfine splitting (\cite{griffiths2}), the Ampere model is used here for the force on the particle.

The analysis is carried out for two magnetisation models, the permanent dipole
in section \ref{sec:permanent}, where the magnitude of the magnetic moment is 
independent of the field, and the induced dipole model in section
\ref{sec:induced}, where the magnitude of the moment is proportional to the 
component of the magnetic field along the orientation axis. These models
have been used earlier to study the single-particle dynamics of
a spheroid in a magnetic field(\cite{almog,cebecki,vk20,vk21a,vk21b}).
In the induced dipole model, it is assumed that the magnetic moment
of the particle is proportional to the component of the magnetic
field along that axis. This model is applicable to superparamagnetic particles, 
which are usually nanometer sized single-domain particles polarisable along
one axis. Superparamagnetic particles have very little hysteresis, and so the 
constant susceptibility assumption is valid for low magnetic field. 
This also applies to micron sized multi-domain ferromagnetic particles, or soft 
magnets. These are magnetised along an `easy axis' \cite{shapeanisotropy} for
several reasons. Though the domains within the particles are aligned along
the magnetic field for soft magnets, those on the surface are aligned tangential
to the surface, resulting in a magnetisation along the particle axis for 
non-spherical particles. In addition, strain anisotropies could also result
in a higher susceptibility along the easy axis. At low magnetic field,
this results in a constant anisotropic susceptibility tensor. It was 
shown \cite{vk21a} that for axisymmetric particles, if the susceptibility
is axisymmetric about the easy axis, the magnetic moment can be modeled
as a vector directed along the easy axis whose magnitude is proportional
to the component of the magnetic field along the easy axis. The induced
dipole model used here applies to this case, provided the magnetic moment
is small compared to the saturation moment.

The permanent dipole model applies to particles in ferrofluids, which are
suspensions of nanometer sized single-domain magnetic particles stabilised
using surfactants. Here, the orientations of the particles are randomised
by Brownian motion in the absence of a magnetic field, but the particles 
align when a field is applied. The analysis here also applies to superparamagnetic or 
ferromagnetic particles at high magnetic field when the 
magnetic moment reaches it saturation value. In contrast to permanent magnets,
the magnetic moments align along the component of the field along the easy axis,
and magnetic moment reverses when the direction of the field is reversed.
Therefore, the orientation dynamics for rotating states is different from
that for a permanent magnet (\cite{vk20,vk21a,vk21b}). However, for states
with a steady orientation that are studied here, the effect of interactions 
is identical to those for permanent dipolar particles provided the orientation
disturbance is small and it does not reverse the orientation of the magnetic
moment.

For the permanent and induced dipole 
models, the dynamics of an isolated spheroid subjected to a shear flow and 
a magnetic field has been studied (\cite{almog,cebecki,vk20,vk21a,vk21b}).
The results for a spherical particle from these earlier studies are used
here. For a permanent dipole, the single-particle dynamics depends on the parameter
\( \Sigma \) defined in equation \ref{eq:eq221}, which is the ratio of the 
magnetic and hydrodynamic torques. The effective diffusion coefficient
is calculated in section \ref{subsec:diffusionpermanent}, and the rotating and steady 
states in the absence of particle interactions are summarised in section 
\ref{subsec:steadypermanent}. The effect of particle interactions for a parallel
magnetic field, where the field is in the flow plane, is discussed in 
\ref{subsec:parallelpermanent}, and this is generalised to an oblique
magnetic field in section \ref{subsec:obliquepermanent}. The analysis for
an induced dipole is provided in section \ref{sec:induced}, where the single
particle dynamics depends on the parameter \( \Sigi \) defined in equation
\ref{eq:eq315}. The diffusion
coefficient is calculated in section \ref{subsec:diffusioninduced}, and 
the orientation of an isolated particle is outlined in section
\ref{subsec:steadyinduced}. The diffusion coefficients for a parallel and
oblique magnetic field are provided in sections \ref{subsec:parallelinduced}
and \ref{subsec:obliqueinduced} respectively. In section \ref{sec:conclusions},
the major conclusions are  summarised, and numerical estimates for the diffusion coefficients are provided.
\section{Permanent dipole}
\label{sec:permanent}
\subsection{Diffusion due to interactions}
\label{subsec:diffusionpermanent}
The magnetic force and torque on a particle with magnetic moment \( \bM \) in a magnetic 
field \( \bH \) in a suspension of particles with number density \( n \) are,
\begin{eqnarray}
 \bFm & = & \bnabla (\bM \bcdot \mu_0 (\bH + n \bM)), \label{eq:mforce} \\
 \bTm & = & \mu_0 \bM \btimes \bH, \label{eq:mtorque}
\end{eqnarray}
where \( \mu_0 \) is the magnetic permeability. In the expression for the force,
\ref{eq:mforce}, the magnetic flux density is expressed as \( \bm{B} = \mu_0 (\bH + n \bM) \),
where \( n \bM \) is the magnetic moment per unit volume. This term is an
effective-medium approximation for the variations in the magnetic field 
due to variations in the number density of particles. The contribution to the 
magnetic moment per volume in the expression for the torque \ref{eq:mtorque} is 
zero, since it contains the cross product of the particle moment at a location with 
itself.

The hydrodynamic force and torque are,
\begin{eqnarray}
 \bFh & = & 3 \pi \eta d (\bv - \bu), \label{eq:hforce} \\
 \bTh & = & \pi \eta d^3 (\tfrac{1}{2} \bomega - \bOmega), \label{eq:htorque}
\end{eqnarray}
where \( d \) is the particle diameter, \( \bv \) and \( \bomega \) are the fluid velocity and vorticity at the particle center,
\( \bu \) and \( \bOmega \) are the particle linear and angular velocities.
The imposed magnetic field is denoted \( \barbH \), and the vorticity due to the imposed shear flow at 
the particle center is \( \barbomega \). The particle magnetic moment is expressed as
\( \bM = M \bor \), where \( M \) is the magnitude of the magnetic moment and \( \bor \) is 
the orientation vector. Linear models are used to incorporate the effect of the variations in
the volume fraction on the viscosity and the magnetic permeability,
\begin{eqnarray}
\eta & = & \eta_0 (1 + \etaprime \phiprime), \label{eq:densitydependence}
\end{eqnarray}
where \( \phiprime=\phi-\barphi \), \( \phi \) is the local volume fraction of the particles,
\( \barphi \) is the average volume fraction in the uniform suspension, and 
\( \eta_0 \) is the 
viscosity for the uniform suspension. For a viscous suspension of spherical particles
in the limit of low volume fraction, the result \( \etaprime = \frac{5}{2} \)
is obtained for the Einstein correction to the viscosity. 

The orientation vector of the 
particle \( \bor \) is determined from the torque balance equation, \( \bTm + \bTh = 0 \),
\begin{eqnarray}
 \pi \eta d^3 (\tfrac{1}{2} \bomega - \bOmega) + \mu_0 M (\bor \btimes \bH) & = & 0.
 \label{eq:torquebalancepp}
\end{eqnarray}
The magnetic force on the particle \ref{eq:mforce} is zero in a uniform suspension.
The force balance requires that the hydrodynamic force is also zero, that is, the particle
moves with the same velocity as the fluid, \( \barbu = \barbv \). 

For a stable stationary state where the particle orientation is steady, the component
of the angular velocity \( \bOmega \) perpendicular to the orientation vector is zero. 
However, the particle does spin around the orientation vector, and the components of the 
particle angular velocity and fluid rotation rate (one half of the vorticity) along the 
orientation vector are equal. This is because the torque exerted by the 
magnetic field is perpendicular to the direction of the magnetic moment and the magnetic
field, and consequently there is no magnetic torque along the orientation vector. Therefore, 
the particle angular velocity is necessarily along the orientation vector, \( \bOmega =
\Omega \bor \). It is necessary to consider the torque balance equations parallel and
perpendicular to the orientation vector in order to determine the particle orientation. 
The torque balance equation along the orientation vector is obtained by taking the dot
product of equation \ref{eq:torquebalancepp} with \( \bor \), and this is solved to obtain,
\begin{eqnarray}
 \Omega = \tfrac{1}{2} \bomega \bcdot \bor. \label{eq:angvel}
\end{eqnarray}
The above solution \ref{eq:angvel} for the angular velocity is substituted into the
torque balance equation \ref{eq:torquebalancepp} to obtain,
\begin{eqnarray}
 \tfrac{1}{2} \pi \eta d^3 (\bI - \bor \bor) \bcdot \bomega + \mu_0 M (\bor \btimes \bH) & = & 0,
 \label{eq:torquebalance}
\end{eqnarray}
where \( \bI \) is the identity tensor, and \( (\bI - \bor \bor) \bcdot \) is 
the transverse projection operator which projects a vector on to the plane
perpendicular to \( \bor \). One relation between the vorticity, magnetic field and orientation vector is obtained by taking the dot product of equation \ref{eq:torquebalance} with \( \bH \),
\begin{eqnarray}
 \bH \bcdot \bomega - (\bor \bcdot \bH)(\bor \bcdot \bomega) & = & 0. \label{eq:eq10}
\end{eqnarray}
The torque balance for the base state in the absence of interactions is obtained by
substituting \( \barbor, \barbomega \) and \( \barbH \) for \( \bor, \bomega \) and 
\( \bH \) in equation \ref{eq:torquebalance}.

There is a disturbance to the magnetic field \( \delH \) and the fluid vorticity \( \delbomega \) 
due to a number density fluctuation \( \deln(\bx) \) imposed on a uniform suspension with
number density \( \barn \). The orientation vector of the particle is expressed as 
\( \bor = \barbor + \borp(\bx) \), where \( \borp(\bx) \) is the disturbance to 
the orientation vector due to hydrodynamic and magnetic interactions resulting from 
concentration inhomogeneities; the latter is calculated from the torque 
balance equation. In the linear approximation where terms quadratic in the primed
quantities are neglected, the mean and disturbance to the orientation vector satisfy
\begin{eqnarray}
 |\barbor| = 1, \: \: \barbor \bcdot \borp = 0. \label{eq:orientationvector}
\end{eqnarray}
The reason for the above conditions is that both \( \barbor \) and \( (\barbor + \borp) \) 
are unit vectors, that is, \( \barbor \bcdot \barbor = 1 \)
and \( (\barbor + \borp) \bcdot (\barbor + \borp) = 1. \) In the linear approximation, this implies that
\( (\barbor \bcdot \barbor + 2 \barbor \bcdot \borp) = 1 \). Therefore, \( \barbor \bcdot \borp = 0 \).

The Fourier transform \( \tbor_{\bk} \)
of the disturbance to the orientation vector is defined using equation 
\ref{eq:fouriertransform}. The Fourier transform of the force on a particle due to
interactions is determined from equation \ref{eq:mforce},
\begin{eqnarray}
 \tbF_{\bk} & = & - \imath \bk \mu_0 M [\barbor \bcdot \tbH_{\bk} + \barbH \bcdot \tbor_{\bk}
 + M \tn_{\bk}],
 \label{eq:eq11}
\end{eqnarray}
where \( M \) is the magnitude of the magnetic moment.
The Fourier transform of
expression \ref{eq:densitydependence} has been substituted
for the magnetic permeability, and linearised in the perturbation to the volume
fraction field, to derive the third term in the square brackets on the right in
equation \ref{eq:eq11}. The last term in the square brackets on the right in
equation \ref{eq:eq11} is the force due to the variation in the number
density in the expression \ref{eq:mforce}.

The conservation equation for the particle number density is,
\be
\frac{\partial n}{\partial t} + \bnabla \bcdot (\bu n) = D_B \bnabla^2 n,
 \label{eq:eq11a}
\ee
where \( \bu \) is the particle velocity and \( D_B \) is the Brownian
diffusion coefficient. The particle velocity is expressed as the sum
of the fluid velocity \( \bv \) and the particle velocity relative
to the fluid, \( (\bu - \bv) + \bv \). Since there is no net force
on the particles in a uniform suspension, \( \barbu = \barbv \).
In the linear approximation, the particle conservation equation is
expressed as,
\be
\frac{\partial \deln}{\partial t} + \bnabla \bcdot (\barbv \deln + 
\delbv \barn) + \bnabla \bcdot ((\delbu -\delbv) \barn) = D_B \bnabla^2 n,
 \label{eq:eq11b}
\ee
The Fourier transform of the above equation is,
\be
\frac{\partial \tn_{\bk}}{\partial t} - \imath \bk \bcdot (\barbv \tn_{\bk}) -
\imath \bk \bcdot (\barn \tbv_{\bk}) - \imath \bk \barn (\tbu_{\bk} - 
\tbv_{\bk})) + D_B k^2 \tn_{\bk} = 0.  \label{eq:eq11c}
\ee
The first two terms on the left of \ref{eq:eq11c} are the rate of
change and the convection due to the base flow of concentration 
fluctuations. The third term on the left is the rate of change
of concentration disturbances due to fluid velocity perturbation.
The perturbation to the fluid velocity due to the rotation of
other particles, which is calculated later in \ref{eq:eq25a}, is perpendicular
to \( \bk \). Consequently, the third term on the left in 
\ref{eq:eq11c} is zero. The fourth term on the left is the 
relative motion between the particles and the fluid due to 
the forces on the particle.

In a dilute suspension, the 
difference between the particle and fluid velocity disturbances,
\( \tbu_{\bk} \) and \( \tbv_{\bk} \), is given by
Stokes law,
\begin{eqnarray}
 \tbu_{\bk} - \tbv_{\bk} & = & \frac{\tbF_{\bk}}{3 \pi \eta_0 d}. \label{eq:eq12}
\end{eqnarray}
The drift velocity does not depend on the coefficient 
\( \etaprime \) in \ref{eq:densitydependence} in the linear approximation,
because the force on a particle is zero in a uniform suspension.

The expression \ref{eq:eq12} for the drift velocity is substituted into the 
linearised conservation \ref{eq:eq11c},
\begin{eqnarray}
 \frac{\partial \tn_{\bk}}{\partial t} - \imath \bk \bcdot (\tn_{\bk} \barbv)
 + D_B k^2 \tn_{\bk} \hspace{1.5in} \mbox{} & & \nonumber \\ \mbox{}
 - \frac{k^2 \barn \mu_0 M (\barbor \bcdot \tbH_{\bk} + \barbH \bcdot \tbor_{\bk}
 + M \tn_{\bk})}{3 \pi \eta_0 d} 
  & = & 0. \label{eq:eq13}
\end{eqnarray}
The disturbance \( \tbH_{\bk} \) due to magnetic interactions is expressed as a function of 
\( \tn_{\bk} \), and the orientation disturbance \( \tbor_{\bk} \), currently
unknown, is determined from the first correction to the torque balance. 
These are substituted into equation \ref{eq:eq13} to obtain an equation of the form,
\begin{eqnarray}
 \frac{\partial \tn_{\bk}}{\partial t} - \imath \bk \bcdot (\tn_{\bk} \barbu)
 + \bk \bcdot \bD \bcdot \bk \tn_{\bk} + D_B k^2 \tn_{\bk} & = & 0, \label{eq:eq14}
\end{eqnarray}
where \( \bD \) is the magnetophoretic diffusion tensor due to hydrodynamic and
magnetic interactions. 
In the following calculation, the disturbance to the 
magnetic field and the vorticity due to particle interactions are first 
discussed, and then the disturbance to the orientation vector
is calculated using torque balance. This is inserted into equation \ref{eq:eq13} 
to extract the diffusion tensor \( \bD \).

The change in the magnetic field at the location \( \bx \) due to interactions 
with other particles is a generalisation of equation \ref{eq:eq21} which
incorporates the disturbance to the orientation vector,
\begin{eqnarray}
 \delH(\bx) & = & \frac{1}{4 \pi} \int d \bx' \left( \frac{3 (\bx - \bx')(\bx - \bx')}{|\bx - \bx'|^5} - \frac{\bI}{|\bx - \bx'|^3} \right) \bcdot [M (\barbor \deln(\bx') +
 \barn \delbor(\bx'))], \label{eq:eq22}
\end{eqnarray}
where \( d \bx' \) is the volume element.
It is easily verified that the contribution to \( \delH \) due to the product of
the background constant particle density and magnetic field, \( M \barn \barbor 
\), is zero. 
Equation \ref{eq:fouriertransform} is used to determine the disturbance to magnetic field in reciprocal space,
\begin{eqnarray}
 \tH_{\bk} & = & \mbox{} - \frac{\bk \bk \bcdot [M (\barbor \tn_{\bk} + \barn \tbor_{\bk})]}{k^2} = \tbHp_{\bk} + \tbHpp_{\bk} \bcdot \tbor_{\bk}, \label{eq:eq24}
\end{eqnarray}
where
\begin{eqnarray}
 \tbHp_{\bk} & = & - \frac{M \tn_{\bk} \bk \bk \bcdot \barbor}{k^2}, \label{eq:eq24p}
 \: \:
  \tbHpp_{\bk} = - \frac{M \barn \bk \bk}{k^2}. \label{eq:eq24pp}
\end{eqnarray}
Note that \( \tbHp_{\bk} \) is a vector, and \( \tbHpp_{\bk} \) is a 
second order tensor.

For a particle located at $\bx'$ 
in a fluid with vorticity 
$\barbomega$ in the absence of the particle, the fluid velocity disturbance \( \bv' \) at the location $\bx$ is,
\begin{eqnarray}
 \bv'(\bx) 
 & = & \int d \bx' n(\bx') (\Omega(\bx') - \mbox{$\frac{1}{2}$} \bomega(\bx')) \btimes \frac{d^3 (\bx - \bx')}{8 |\bx - \bx'|^3} \nonumber \\ 
 & = & - \int d \bx' n(\bx') [\bI - \bor(\bx') \bor(\bx')] \bcdot \bomega(\bx') \btimes \frac{d^3 (\bx - \bx')}{16 |\bx - \bx'|^3} \nonumber \\
 & = & - \int d \bx' [ \deln(\bx') (\bI - \barbor \barbor) \bcdot \barbomega + \barn (\bI - \barbor \barbor) \bcdot \delbomega(\bx') \nonumber \\ & & \mbox{} 
 - \barn (\barbor \barbomega + \bI \barbomega \bcdot \barbor) \bcdot \delbor(\bx')] \btimes 
 \frac{d^3 (\bx - \bx')}{16 |\bx - \bx'|^3}. \label{eq:eq25}
\end{eqnarray}
The expression \ref{eq:angvel} for the particle angular velocity has been used 
to simplify in the second step in equation \ref{eq:eq25}. The linearisation
approximation has been used in the third step in equation \ref{eq:eq25}, 
resulting in an equation that is linear in the disturbances due to 
particle interactions. The Fourier transform of the velocity field is,
\begin{eqnarray}
 \tbv_{\bk} & = & 
 - [\tn_{\bk} (\bI - \barbor \barbor) \bcdot \barbomega 
 + \barn (\bI - \barbor \barbor) \bcdot 
 \tbomega_{\bk} - \barn \tbor_{\bk} \bcdot (\barbomega \barbor + {\bf I} 
 \barbomega \bcdot \barbor)] \btimes \frac{\pi d^3 \imath \bk}{4 k^2}.
 \label{eq:eq25a}
\end{eqnarray}
The Fourier transform of the disturbance to the fluid vorticity is,
\begin{eqnarray}
 \tbomega_{\bk} & = & - \imath \bk \btimes \tbv_{\bk} \nonumber \\ & = &
\frac{\pi d^3 (\bk \bk - \bI k^2) \bcdot 
[\tn_{\bk} (\bI - \barbor \barbor) \bcdot \barbomega 
 + \barn (\bI - \barbor \barbor) \bcdot  \tbomega_{\bk} - 
 \barn (\barbor \barbomega + {\bf I} 
 \barbomega \bcdot \barbor) \bcdot \tbor_{\bk}]}{4 k^2}.
 \label{eq:eq28a}
\end{eqnarray}
The above implicit equation is solved for \( \tbomega_{\bk} \),
\begin{eqnarray}
\left( \bI - \frac{\pi d^3 \barn (\bk \bk - \bI k^2) \bcdot (\bI - \barbor \barbor)}{4 k^2} 
\right) \bcdot \tbomega_{\bk} & = & \tbomegap + \tbomegapp \bcdot \tbor_{\bk},
 \label{eq:eq28}
\end{eqnarray}
where
\begin{eqnarray}
 \tbomegap & = & \frac{\pi \tn_{\bk} d^3 (\bk \bk - \bI k^2) \bcdot (\bI - \barbor \barbor)
 \bcdot \barbomega}{4 k^2}, \label{eq:eq28p} \\
  \tbomegapp & = & \mbox{} - \frac{\pi \barn d^3 (\bk \bk - \bI k^2) \bcdot (\barbor \barbomega
 + \bI \barbomega \bcdot \barbor)}{4 k^2}. \label{eq:eq28pp}
\end{eqnarray}
Note that \( \tbomegap_{\bk} \) is a vector, and \( \tbomegapp_{\bk} \) is a 
second order tensor. The latter is proportional to \( (d^3 \barn) \) or the volume 
fraction. In the limit of small volume fraction, this term is small compared to $1$. 

The velocity due to the force exerted by particles is higher order in gradients or higher power in $\bk$, 
and so it will result in a higher gradient of the concentration field. The contribution to the velocity
due to the magnetic field perturbation, in Fourier space, is $(|\tbF_{\bk}|/ 3 \pi \eta d) \sim$ $(k \mu_0 M
\tbH_{\bk}/3 \pi \eta d) \sim$ $(k \mu_0 M^2 \tn_{\bk}/3 \pi \eta d)$, where $k$ is the magnitude of the wave
number. Here, \ref{eq:eq11} is used for the force and \ref{eq:eq24}, \ref{eq:eq24p} are used for the magnetic 
field perturbation. The magnitude of the fluid velocity due to particle rotation, from \ref{eq:eq25a}, is 
$|\tbv_{\bk}| \sim \pi |\barbomega| \tn_{\bk} d^3/k$.
The ratio of the velocities due to the force and torque is $(k^2 \mu_0 M^2/3 \pi \eta d^4 |\barbomega|) 
\sim (k d)^2 \Ma \Sigma$, where the dimensionless groups \( \Sigma \) and \( \Ma \) are defined later in 
\ref{eq:eq221} and \ref{eq:eq221a}. The continuum approximation is valid only for $(k d) \ll 1$, that is,
the length scale for the gradients in the concentration field is much larger than the particle diameter.
In this limit, the velocity disturbance due to the particle force is neglected compared to that
due to the torque exerted on the fluid.

The disturbance to the magnetic field and the vorticity at the particle location causes a
disturbance to the particle orientation \( \bor \), which is determined from the correction
to the torque balance equation in Fourier space. The torque balance equation, \ref{eq:torquebalance},
is divided by \( \mu_0 M \), and linearised in the primed quantities to obtain,
\begin{eqnarray}
 \frac{\pi d^3 \eta_0}{2 \mu_0 M} [(\bI - \barbor \barbor) \bcdot \tbomega_{\bk} - \barbor (\tbor_{\bk}
 \bcdot \barbomega) - \tbor_{\bk} (\barbor \bcdot \barbomega) + \etaprime \tphi_{\bk}
 (\bI - \barbor \barbor) \bcdot \barbomega] & & \nonumber \\ 
 + (\barbor \btimes \tbH_{\bk} + \tbor_{\bk}  \btimes \barbH)  & = & 0,
 \label{eq:eq29a}
\end{eqnarray}
where \( \tphi_{\bk} = \tn_{\bk} (\pi d^3/6) \) is the Fourier transform of the 
fluctuation in the volume fraction.
The expressions \ref{eq:eq24} and \ref{eq:eq28} for \( \tbH_{\bk} \) and 
\( \tbomega_{\bk} \) are substituted into equation \ref{eq:eq29a},
\begin{eqnarray}
 \frac{\pi d^3 \eta_0}{2 \mu_0 M} \{(\bI - \barbor \barbor) \bcdot \tbomegap_{\bk} + 
 [(\bI - \barbor \barbor) \bcdot \tbomegapp_{\bk} - \barbor 
 \barbomega - (\barbor \bcdot \barbomega) \bI] \bcdot \tbor_{\bk} 
 & & \nonumber \\ + \etaprime \tphi_{\bk}
 (\bI - \barbor \barbor) \bcdot \barbomega \}
 + [\barbor \btimes \tbHp_{\bk} + \tbor_{\bk} \btimes \barbH + \barbor \times
 (\tbHpp_{\bk} \bcdot \tbor_{\bk})]  & = & 0. \label{eq:eq29}
\end{eqnarray}
The first term on the left side of equation \ref{eq:eq29} is the correction to the 
hydrodynamic torque due to interactions and due to the concentration dependence of
the viscosity and magnetic permeability. In this, the contribution proportional to 
\( \etaprime \tphi_{\bk} \) is the correction to the 
hydrodynamic torque due to the concentration dependence of the viscosity.
In the first term on the left in \ref{eq:eq29}, within the flower brackets, the prefactor 
of \( \tbor_{\bk} \) contains one 
contribution proportional to \( \tbomegapp \), and the second which is proportional to 
\( |\barbomega| \). From equation \ref{eq:eq28pp}, the former is proportional to \( \barn d^3
|\barbomega| \), which is small compared to the latter, because the volume fraction is
small. Therefore, the term \( (\bI - \barbor \barbor) \bcdot \tbomegapp_{\bk} \) is neglected 
in the first term on the left in equation \ref{eq:eq29}. The second term on the left in equation \ref{eq:eq29}
is the contribution due to the magnetic torque. There are two terms containing
\( \tbor_{\bk} \)
in the magnetic torque, the first which is proportional to \( |\barbH| \) and the 
second proportional to \( |\tbHpp_{\bk}| \). From equation \ref{eq:eq24pp}, the latter
scales as \( \barn M \), where \( M \) is the magnetic moment of a particle. From dimensional
analysis, \( \barn M \) is the disturbance to the magnetic field at a distance \( \barn^{-1/3}
\) from a particle, comparable to the inter-particle distance in the suspension. The weak
interaction limit is considered here, where the magnetic field disturbance due to particle 
interactions is small compared to the applied field, \( \barn M \ll |\barbH| \).
Therefore, the term containing \( \barbor \btimes (\tbHpp_{\bk} \bcdot \tbor_{\bk})  \) 
is neglected in comparison to  \( \tbor_{\bk} \btimes \barbH \) in equation \ref{eq:eq29}.

Equation \ref{eq:eq29} is solved to obtain the unknown orientation disturbance
\( \tbor_{\bk} \). From equation \ref{eq:orientationvector}, \( \tbor_{\bk} \) is perpendicular 
to \( \barbor \), and it is expressed as,
\begin{eqnarray}
 \tbor_{\bk} & = & \torp_{\bk} (\barbH - (\barbor \bcdot \barbH) \barbor) + \torpp_{\bk} 
 (\barbor \btimes \barbH), \label{eq:eq210}
\end{eqnarray}
where \( \torp_{\bk} \) and \( \torpp_{\bk} \) are scalar functions of the wave number.
The specific form \ref{eq:eq210} is chosen because it satisfies the requirement \( \tbor_{\bk} \bcdot \barbor = 0 \). 
The expression \ref{eq:eq210} is inserted into the torque balance equation, 
\ref{eq:eq29},
\begin{eqnarray}
 \frac{\pi d^3 \eta_0}{2 \mu_0 M} \{(\bI - \barbor \barbor) \bcdot \tbomegap_{\bk} 
 - \barbor [\torp_{\bk} \underline{(\barbH - (\barbor \bcdot \barbH) \barbor) \bcdot \barbomega} +
 \torpp_{\bk} (\barbor \btimes \barbH) \bcdot \barbomega] & & \nonumber \\
 \mbox{} - (\barbor \bcdot \barbomega)
 [\torp_{\bk} (\barbH - (\barbor \bcdot \barbH) \barbor) + \torpp_{\bk} (\barbor \btimes \barbH)]
 + \etaprime \tphi_{\bk} (\bI - \barbor \barbor) \bcdot \barbomega
 \} \nonumber & & \\
\mbox{} + [\barbor \btimes \tbHp_{\bk} - \torp_{\bk} (\barbor \bcdot \barbH)(\barbor \btimes \barbH) 
- \torpp_{\bk} (\barbor |\barbH|^2 - \barbH (\barbor \bcdot \barbH))] & = & 0. \label{eq:eq211}
\end{eqnarray}
From equation \ref{eq:eq10}, the underlined term in the above equation is zero.
The scalar functions \( \torp_{\bk} \) and \( \torpp_{\bk} \) are evaluated by taking the
dot product of equation \ref{eq:eq211} with \( \barbor \btimes \barbH \)
and \( \barbor \) respectively,
\begin{eqnarray}
 \frac{\pi d^3 \eta_0}{2 \mu_0 M} [\tbomegap_{\bk} \bcdot (\barbor \btimes \barbH) - \torpp_{\bk}
 (\barbor \bcdot \barbomega)(|\barbH|^2 - (\barbor \bcdot \barbH)^2)
 + \etaprime \tphi_{\bk} (\barbor \btimes \barbH) \bcdot \barbomega] & & \nonumber \\ \mbox{}
 + [\barbH \bcdot \tbHp_{\bk} - (\barbor \bcdot \barbH)(\tbHp_{\bk} \bcdot \barbor) 
 - \torp_{\bk} (\barbor \bcdot \barbH)(|\barbH|^2 - (\barbor \bcdot \barbH)^2] & = & 0, \label{eq:eq212} \\
 - \frac{\pi d^3 \eta_0}{2 \mu_0 M} \torpp_{\bk} (\barbor \btimes \barbH) \bcdot \barbomega
 - \torpp_{\bk} (|\barbH|^2 - (\barbor \bcdot \barbH)^2)) & = & 0. \label{eq:eq213}
\end{eqnarray}
These equations are solved to obtain,
\begin{eqnarray}
 \torp_{\bk} 
 & = & \frac{\pi \eta_0 d^3 \tbomegap_{\bk} \bcdot (\barbor \btimes \barbH)}{2
 \mu_0 M (\barbor \bcdot \barbH)(|\barbH|^2 - (\barbor \bcdot \barbH)^2)}
 + \frac{\barbH \bcdot \tbHp_{\bk} - (\barbor \bcdot \barbH)(\barbor \bcdot
 \tbHp_{\bk})}{(\barbor \bcdot \barbH) (|\barbH|^2 - (\barbor \bcdot \barbH)^2)} \nonumber \\
 & & \mbox{} + \frac{\pi d^3 \eta_0 \etaprime \tphi_{\bk} (\barbor \btimes \barbH) \bcdot 
 \barbomega}{2 \mu_0 M (\barbor \bcdot \barbH)(|\barbH|^2 - (\barbor \bcdot \barbH)^2)},
 \label{eq:eq214} \\
 \torpp_{\bk} & = & 0.  \label{eq:eq215}
\end{eqnarray}

The magnetophoretic diffusion term in equation \ref{eq:eq13} is
simplified by neglecting the term \( \tbHpp_{\bk} \bcdot \tbor_{\bk} \)
in equation \ref{eq:eq24}, since it is much smaller than \( \tbHp_{\bk} \),
\begin{eqnarray}
 \bk \bcdot \bD \bcdot \bk 
 & = & \mbox{} - \frac{\barn \mu_0 M k^2 (\barbor \bcdot \tbH_{\bk} + \tbor_{\bk} \bcdot \barbH 
 + M \tn_{\bk} 
 }{3 \pi \eta_0 d \tn_{\bk}}
 \nonumber \\
 & = & \mbox{} - \frac{\barn k^2}{3 \pi \eta_0 d \tn_{\bk}} \left[ \mu_0 M \overbrace{\barbor \bcdot \tbH_{\bk}}^{\textcircled{1}} + 
 \frac{\pi d^3 \eta_0 \tbomega_{\bk} \bcdot (\barbor \btimes \barbH)}{2 (\barbor \bcdot \barbH)} 
 + \frac{\pi d^3 \eta_0 \etaprime \tphi_{\bk} \barbomega \bcdot (\barbor \btimes \barbH)}{2 (\barbor \bcdot 
 \barbH)} \right. \nonumber \\ & & \left. \mbox{} \hspace{.7in} + \frac{\mu_0 M (\barbH \bcdot \tbH_{\bk} - \overbrace{(\barbor 
 \bcdot \barbH)(\tbH_{\bk} \bcdot \barbor)}^{\textcircled{2}})}{(\barbor \bcdot \barbH)} 
 + \mu_0 M^2 \tn_{\bk} \right], \label{eq:eq217}
 \end{eqnarray}
 Here, the right sides of equations \ref{eq:eq210} and \ref{eq:eq214}-\ref{eq:eq215} have been substituted for \( \tbor_{\bk} \) in the second step. 
 In equation \ref{eq:eq217}, the terms \( \textcircled{1} \) and \( \textcircled{2} \) cancel. The term
 \( \textcircled{1} \) is the contribution due to the interaction between the undisturbed particle magnetic moment and the 
 disturbance to the magnetic field due to particle interactions. This exactly cancels with one of the terms resulting
 from the disturbance to the particle orientation; the remaining terms are entirely due to the interaction between the 
 undisturbed magnetic field and the disturbance to the particle orientation due to interactions,
 \begin{eqnarray}
 \bk \bcdot \bD \bcdot \bk
  & = & \mbox{} - \frac{\barn k^2 d^2 \tbomega_{\bk} \bcdot (\barbor \btimes
  \barbH)}{6 
  (\barbor \bcdot \barbH) \tn_{\bk}} - \frac{\pi \barn k^2 d^5 \etaprime \barbomega \bcdot 
  (\barbor \btimes \barbH)}{36 (\barbor \bcdot \barbH)} \nonumber \\ & & \mbox{} - 
  \frac{\barn k^2 \mu_0 M \barbH \bcdot \tbH_{\bk}}{3 \pi \eta d (\barbor \bcdot \barbH) \tn_{\bk}}
  - \frac{k^2 \barn \mu_0 M^2}{3 \pi \eta_0 d}.
 \label{eq:eq218}
\end{eqnarray}
Here, the substitution \( \tphi_{\bk} = \tn_{\bk} (\pi d^3/6) \) has been used to relate the volume
fraction and number density in the third term on the right. In equation \ref{eq:eq218},
the first term on the right is due to the disturbance to the orientation vector
caused by the hydrodynamic interactions, and the third term results from the 
disturbance to the orientation vector caused by magnetic interactions. 
The expressions \ref{eq:eq24p} and \ref{eq:eq28p} are substituted for the disturbance to the magnetic field 
and vorticity due to particle interactions to obtain,
\begin{eqnarray}
 \bk \bcdot \bD \bcdot \bk 
 & = & \mbox{} - \frac{\pi \barn d^5 \{ [\bk \bcdot (\barbomega - \barbor 
 (\barbor \bcdot \barbomega))] [\bk \bcdot (\barbor \btimes \barbH)] -
 k^2 [\barbomega - \barbor (\barbor \bcdot \barbomega)] \bcdot
 [\barbor \btimes \barbH] \}}{24 (\barbor
 \bcdot \barbH)} \nonumber \\ & & \mbox{}
 + \frac{\barn \mu_0 M^2 (\bk \bcdot \barbor)(\bk \bcdot \barbH)}{3 \pi \eta_0 d
 (\barbor \bcdot \barbH)} - \frac{\pi \barn k^2 d^5 \etaprime \barbomega \bcdot 
  (\barbor \btimes \barbH)}{36 (\barbor \bcdot \barbH)} \nonumber \\ & &
    - \frac{k^2 \barn \mu_0 M^2}{3 \pi \eta_0 d}.
  \label{eq:eq219}
\end{eqnarray}
The cross product \( \barbor \btimes \barbH \) in equation \ref{eq:eq219} is expressed in
terms of the vorticity using the torque balance equation \ref{eq:torquebalance} for the base state,
\begin{eqnarray}
 \bk \bcdot \bD \bcdot \bk 
 & = & \frac{\pi \barphi d^5 \eta_0 \{ [\bk \bcdot (\barbomega - \barbor (\barbor \bcdot \barbomega))][\bk \bcdot (\barbomega - \barbor(\barbor \bcdot \barbomega))]
 - k^2 [|\barbomega|^2 - (\barbomega \bcdot \barbor)^2] \}}{
 8 \mu_0 M (\barbor \bcdot \barbH)} \nonumber \\ & & \mbox{}
 + \frac{2 \barphi \mu_0 M^2 (\bk \bcdot \barbor)(\bk \bcdot \barbH)}{\pi^2 d^4 \eta_0 (\barbor 
 \bcdot \barbH)} + \frac{\pi k^2 d^5 \barphi \eta_0 \etaprime (|\barbomega|^2 - (\barbomega \bcdot \barbor)^2)}{12 
\mu_0 (\barbor \bcdot \barbH)} \nonumber \\ & & \mbox{} 
 - \frac{2 k^2 \barphi \mu_0 M^2}{\pi^2 \eta_0 d^4}.
 \label{eq:eq220a}
\end{eqnarray}
Here, the substitution \( \barn = \barphi (\pi d^3/6)^{-1} \) has been made, where 
\( \barphi \) is the volume fraction in the base state. The diffusion coefficient \( \bD \) 
is extracted from equation \ref{eq:eq220a}. This  is written in scaled form as the sum of a 
`hydrodynamic' contribution due to the vorticity disturbance, `magnetic' contribution due to the magnetic 
field disturbance, and an isotropic contribution due to the variation in viscosity and
magnetic permeability with particle volume fraction,
\begin{eqnarray}
 \bD & = & \barphi d^2 |\bomega| (\bDh + \bDm + \Diprime \bI), \label{eq:eq220}
\end{eqnarray}
where the scaled diffusivities are,
\begin{eqnarray}
 \bDh & = & \frac{\{ (\beomega - \barbor (\barbor \bcdot \beomega)) (\beomega - \barbor(\barbor \bcdot \beomega))
 - \bI [1 - (\beomega \bcdot \barbor)^2] \}}{
 8\Sigma (\barbor \bcdot \beH)}, \label{eq:eq220b} \\
 \bDm & = & \frac{\Sigma \Ma}{\pi} \left( \frac{(\barbor \beH + \beH \barbor)}{(\barbor \bcdot \beH)} - 2 \bI \right), \label{eq:eq220c} \\
 \Diprime & = & \mbox{} \frac{\etaprime (1 - (\beomega \bcdot \barbor)^2)}{12 
\Sigma (\barbor \bcdot \beH)}.
\label{eq:eq220d}
\end{eqnarray}
Here, \( \beomega \) and \( \beH \) are the unit vectors along the vorticity
and magnetic field, and the substitutions \( \barbomega = |\barbomega| 
\beomega \) and \( \barbH = |\barbH| \beH \) have been used. There are two
dimensionless numbers in equations \ref{eq:eq220b}-\ref{eq:eq220c}, the 
scaled ratio of the hydrodynamic and magnetic torques \( \Sigma \)
and the scaled magnetic moment \( \Ma \),
\begin{eqnarray}
 \Sigma & = & \frac{\mu_0 M |\barbH|}{\pi \eta_0 d^3 |\barbomega|}, \label{eq:eq221} \\
 \Ma & = & \frac{M}{d^3 |\barbH|}. \label{eq:eq221a}
\end{eqnarray}

In the diffusion tensor \ref{eq:eq220}, the contribution \( \bDm \) due to the magnetic
torque is proportional to \( \barphi |\barbomega| d^2 \Sigma \sim (\barphi \mu_0 M |\barbH|/ d
\eta_0) \) is independent of the vorticity, but it does depend on the viscosity, the particle
magnetic moment and the magnetic field. The contribution \( \bDh \) due to the hydrodynamic
torque is proportional to \( \barphi |\barbomega| d^2 \Sigma^{-1} \sim (\barphi d^5 \eta_0
|\barbomega|^2/\mu_0 M |\barbH|) \) is proportional to the square of the angular velocity 
and inversely proportional to the magnetic field. The isotropic component \( \Di \bI \) is
due to the variation of the viscosity and the magnetic permeability with the particle
volume fraction. This is positive, and it has a damping effect on concentration
fluctuations.
\subsection{Steady solution}
\label{subsec:steadypermanent}
\begin{figure}
\parbox{.49\textwidth}{
    \psfrag{x}[][][1.0][0]{$\Sigma$}
    \psfrag{y}[][][1.0][270]{$\barbor \bcdot \beH$}

 \includegraphics[width=.49\textwidth]{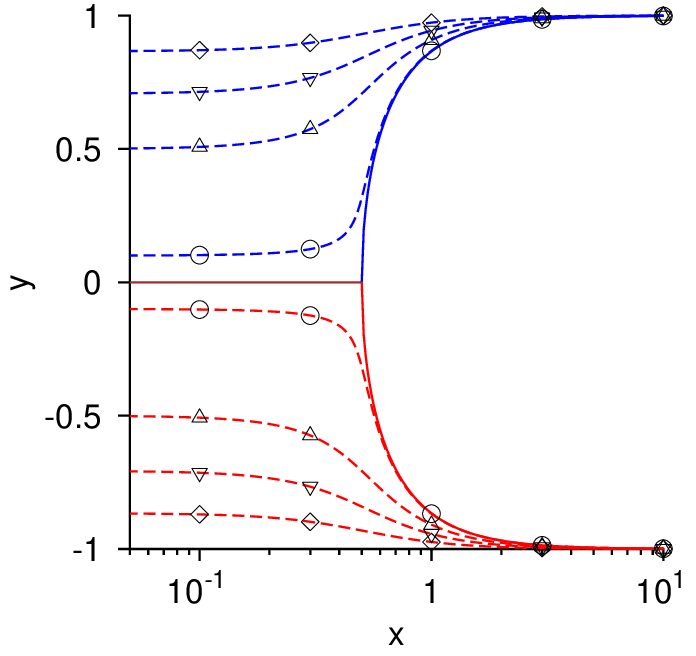}
 
 \begin{center} (a) \end{center}
 }
 \parbox{.49\textwidth}{
     \psfrag{x}[][][1.0][0]{$\Sigma$}
    \psfrag{y}[][][1.0][270]{$\barbor \bcdot \beomega$}

 \includegraphics[width=.49\textwidth]{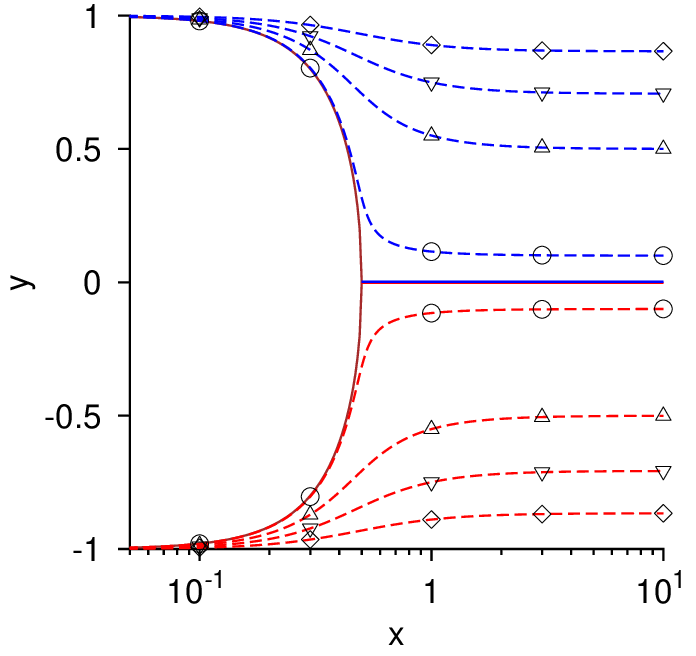}
 
 \begin{center} (b) \end{center}
 }
\caption{\label{fig:orientationp} The variation of \( \barbor \bcdot \beH \) (a) and 
\( (\barbor \bcdot \beomega) \) (b) with \( \Sigma \) for a particle with a permanent dipole.
The solid lines are the results for a parallel magnetic field \( (\beomega \bcdot \beH) = 0 \) 
and the dashed lines are the results for an oblique magnetic field with \( \beomega \bcdot
\beH = 0.1 \) ($\circ$), \( \frac{1}{2} \) ($\triangle$), \( \frac{1}{\sqrt{2}} \) ($\nabla$)
and \( \frac{\sqrt{3}}{2} \) ($\diamond$). The blue lines are the stable stationary nodes, the 
red lines are the unstable stationary nodes and the brown lines are the neutral stationary nodes around which there are periodic orbits.
}
\end{figure}

The rotating and steady solutions for the orientation vector for an isolated dipolar spheroid
in a magnetic field were derived in \cite{vk20}. Here, the results for a spherical
particle are briefly summarised.
The orientation of the particle depends on the dimensionless
parameter \( \Sigma \) in equation \ref{eq:eq221}. The projections of the
orientation vector along the directions of the magnetic field and the vorticity
are shown as a function of the parameter \( \Sigma \) in figure 
\ref{fig:orientationp}. The stable solutions are shown by the blue lines, and 
unstable solutions by the red lines and the neutral solutions by the brown lines.

In the limit \( \Sigma \rightarrow \infty \), the orientation vector is parallel to the 
magnetic field; this corresponds to \( \barbor \bcdot \beH \rightarrow 1 \) for the stable solution. 
There is also an unstable solution where the orientation vector is anti-parallel to the magnetic field,
in which case \( \barbor \bcdot \beH \rightarrow -1 \). As \( \Sigma \) decreases, there is a gradual
decrease in \( \barbor \bcdot \beH \), while \( \barbor \bcdot \beomega \) increases and tends to
\( 1 \) for the stable solution in this limit. This implies that the particle is aligned along the 
vorticity direction in the limit \( \Sigma \rightarrow 0 \).

The parallel magnetic field, \( \beomega \bcdot \beH = 0 \), is a special case. In figure 
\ref{fig:orientationp}, it is observed that there is a transition with a slope
discontinuity in this 
case for \( \Sigma = \frac{1}{2} \). The stable and unstable branches merge and bifurcate
into two center nodes, around which there are periodic closed orbits. There are no steady
solutions in this case, and the distribution of orientations in the different orbits
depends on the initial distribution. The effect of interactions for a parallel magnetic
field is examined in the following section \ref{subsec:parallelpermanent} for 
the parameter regime \( \Sigma > \frac{1}{2} \) where there are steady solutions.
The same study for an oblique magnetic field is presented in section
\ref{subsec:obliquepermanent} for \( \Sigma > 0 \), since there is a steady solution for all
values of \( \Sigma \).
\subsection{Parallel magnetic field}
\label{subsec:parallelpermanent}
\begin{figure}
\parbox{.5\textwidth}{

 \includegraphics[width=.49\textwidth]{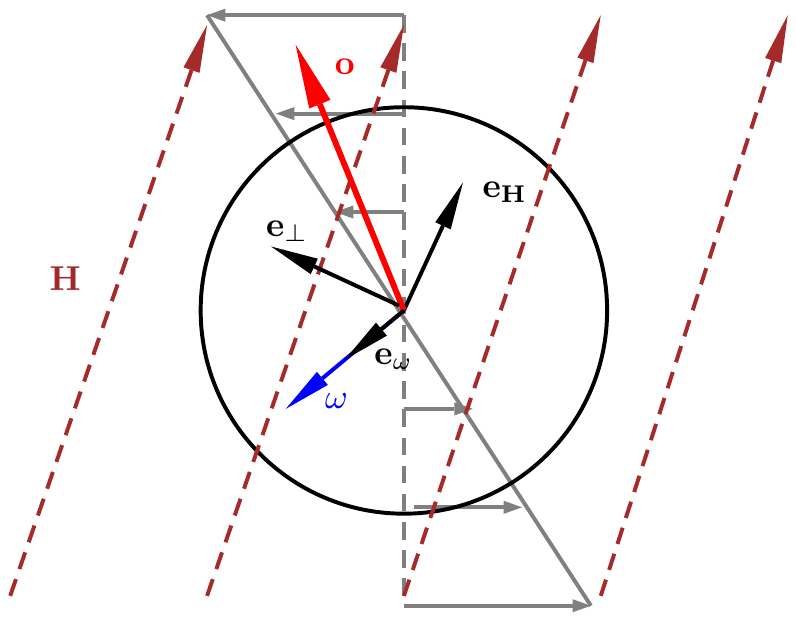}
 
 \begin{center} (a) \end{center}
 }
 \parbox{.49\textwidth}{

 \includegraphics[width=.25\textwidth]{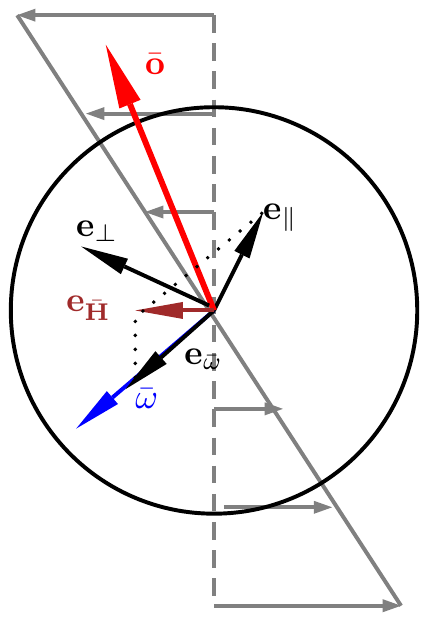}
 
 \begin{center} (b) \end{center}
 }
\caption{\label{fig:config} The configuration and co-ordinate system for analysing the effect
of interactions for a parallel magnetic field (a) and an oblique magnetic field (b). The vorticity
vector, shown in blue, is perpendicular to the plane of flow. The applied magnetic field shown
in brown and the orientation vector shown in red are in the 
plane of flow in (a) and at an angle to the plane of flow in (b). The unit vector \( \bepar \) 
is perpendicular to \( \barbomega \) in the \( \barbomega-\barbH \) plane in (b). The unit
vector \( \beperp \) is perpendicular to the \( \barbomega-\barbH \) plane.}
\end{figure}

Here, we consider the special case where the imposed magnetic field is in the flow plane 
and perpendicular to the vorticity, \( \barbH \bcdot \barbomega = 0 \). From equation \ref{eq:eq10},
\( \barbor \bcdot \barbomega = 0 \) for steady solutions, that is, the orientation vector
and the vorticity are also orthogonal. The configuration and co-ordinate system are 
shown in figure \ref{fig:config} (a). A linear shear flow, shown by grey arrows, is applied
in the fluid far from the particle. The mean vorticity is in the direction perpendicular to
the plane of shear, and the magnetic field vector is in the plane of shear.
The orthogonal unit vectors \( \beomega = (\barbomega/|\barbomega|) \) and \( \beH = 
(\barbH / |\barbH|) \) are defined along the vorticity and magnetic field directions, and
the third orthogonal vector \( \beperp = (\beomega \btimes \beH) \) is perpendicular
to the \( (\beomega-\beH) \) plane. 

The solution for the orientation vector for the 
stable stationary node is,
%
\begin{eqnarray}
 \barbor \bcdot \beH & = & \frac{\sqrt{4 \Sigma^2 - 1}}{2 \Sigma},  \: \: \label{eq:eq222a} \\
 \barbor \bcdot \beperp 
 & = & \frac{1}{2 \Sigma}. \label{eq:eq222}
\end{eqnarray}
In the equation \ref{eq:eq220} for the diffusion coefficient, the orientation vector is expressed as
\( \barbor = \oH \beH + (\barbor \bcdot \beperp) \beperp \), and equations
\ref{eq:eq222a}-\ref{eq:eq222} are substituted to obtain the diffusion tensor due to 
hydrodynamic and magnetic interactions,
\begin{eqnarray}
\bDhm & = & 
\left( \begin{array}{ccc} \beomega & \beH & \beperp \end{array} \right)
\left( \begin{array}{ccc} \Dwwhm & 0 & 0 \\
        0 & \DHHhm & \DHperphm \\
        0 & \DHperphm & \Dperpperphm \end{array} \right)
\left( \begin{array}{c} \beomega \\ \beH \\ \beperp \end{array} \right),
\label{eq:eq223}
\end{eqnarray}
where
\begin{eqnarray}
\Dwwh & = & 0, \: \: \Dwwm = \mbox{} - \frac{2 \Sigma \Ma}{\pi} \\
\DHHh & = & \mbox{} - \frac{1}{4 \sqrt{4 \Sigma^2 - 1}}, \: \: \DHHm = 0, \label{eq:eq224} \\
\DHperph & = & 0, \: \: 
\DHperpm = \frac{\Ma \Sigma}{\pi \sqrt{4 \Sigma^2 - 1}}, \label{eq:eq225} \\
\Dperpperph & = & \mbox{} - \frac{1}{4 \sqrt{4 \Sigma^2 - 1}}, \: \:
\Dperpperpm = - \frac{2 \Sigma \Ma}{\pi}. \label{eq:eq226}
\end{eqnarray}
The isotropic contributions to the diffusion tensor due to the concentration dependence
of the viscosity and the magnetic permeability are,
\begin{eqnarray}
\Diprime & = & \frac{\etaprime}{6 \sqrt{4 \Sigma^2-1}}.
\label{eq:eq226a}
\end{eqnarray}

The values of the coefficients in the limit \( \Sigma \gg 1 \) and 
\( \Sigma - \frac{1}{2} \ll 1 \) are listed in table \ref{tab:table1}. 
The eigenvalues and eigenvectors of the diffusion matrix are also
provided. Since the diffusion matrix is symmetric, the eigenvalues
are real, the eigenvectors are orthogonal, and the eigenvectors
are the principal directions of amplification (along the directions
with negative eigenvalues) or damping (along the directions with 
positive eigenvalues). The isotropic part of the diffusion tensor
\( \Di \) due to the dependence of viscosity on concentration is
not included in the calculation of the eigenvalues in table
\ref{tab:table1}. When the isotropic part is included, the 
eigenvectors remain unchanged and the eigenvalues transform
as \( \lambda \rightarrow \lambda + \Di \).

Equation \ref{eq:eq223} shows that one of the principal directions
is always the vorticity direction, and the diffusion coefficient 
in this direction is negative, resulting in concentration amplification
in this direction for all values of \( \Sigma \). The magnitude of this
diffusion coefficient
increases proportional to \( \Sigma \Ma \) for \( \Sigma \gg 1 \).
This diffusion is due to the 
modification of the magnetic field by fluctuations in the particle
number density, which is the last term in the square brackets on the 
right in \ref{eq:eq11}. 

The diffusion in the flow plane is anisotropic,
and the diffusion coefficients contain contributions due to hydrodynamic
and magnetic interactions. The principal directions for diffusion in the 
plane, which are the eigenvectors of the $2 \times 2$ square sub-matrix
in \ref{eq:eq223}, are presented in table \ref{tab:table1} for high 
magnetic field in the limit \( \Sigma \gg 1 \), and close to the transition
between static and rotating states \( (\Sigma - \frac{1}{2}) \ll 1 \).
For \( \Sigma \gg 1 \), the two principal axes are along the \( \beH \)
and \( \beperp \) directions. The eigenvalue is negative along the 
\( \beperp \) with magnitude \( (2 \Sigma \Ma/\pi) \), which is 
the same as that in the \( \beomega \) direction. The eigenvalue
in the \( \beH \) direction is positive for \( \Ma > \pi \), and 
negative otherwise, and the magnitude is proportional to \( \Sigma^{-1} \)
in this limit. This implies a weak damping of fluctuations for \( \Ma > \pi \),
and a weak amplification for \( \Ma < \pi \). Thus, there is anisotropic
clustering with a strong amplification of fluctuations in the two directions 
perpendicular to the magnetic field, and weak damping or amplification along 
the magnetic field for \( \Sigma \gg 1 \).

For \( \Sigma - \frac{1}{2} \ll 1 \), the diffusion coefficient in the 
vorticity direction tends to a finite value, \( - (\Ma / \pi) \). The
diffusion coefficients in the flow plane diverge proportional to
\( (\Sigma - \frac{1}{2})^{-1/2} \). The eigenvectors in the flow plane
are along the directions rotated by an angle \( - \pi/4 \) and \( + \pi/4 \)
respectively from the magnetic field direction. One of the eigenvalues
is always negative, while the other is positive for \( \Ma > (\pi/2) \) 
and negative otherwise. In both cases, the eigenvalues diverge 
proportional to \( (\Sigma - \frac{1}{2})^{-1/2} \). Thus, there is
strong concentration amplification along one principal direction in
the flow plane and strong damping in the perpendicular direction
for \( \Ma > (\pi/2) \), and strong amplification in both principal
directions for \( \Ma < (\pi/2) \).

It should be noted that the divergence
proportional to \( (4 \Sigma^2 - 1)^{-1/2} \) is the result of a mean-field calculation;
a more complex renormalisation group calculation is required to include the effect 
of fluctuations.
\renewcommand{\baselinestretch}{1.5}
\begin{table}
 \begin{tabular}{|c|c|c|c|c|} \hline
  & \multicolumn{2}{|c|}{Permanent dipole} & \multicolumn{2}{|c|}{Induced dipole} \\ \hline
  & $\Sigma \gg 1$ & $(\Sigma - \frac{1}{2}) \ll 1$ & $\Sigi \gg 1$ & $(\Sigi - 1) \ll 1$ \\ \hline
  $D_{\barbomega \barbomega}$ & $\mbox{} - \frac{2 \Sigma \Ma}{\pi}$ & $\mbox{} - \frac{\Ma}{\pi}$ & 
  $\mbox{} - \frac{4 \Sigi \chia}{\pi}$ & $\mbox{} - \frac{\chia}{\sqrt{2} \pi \sqrt{\Sigi-1}}$ \\ \hline
  $D_{\barbH \barbH}$ & $- \frac{1}{8 \Sigma} $ & $- \frac{1}{8 \sqrt{\Sigma - 
  \frac{1}{2}}} $ & $\mbox{} - \frac{1}{4 \Sigi}$ & $\mbox{} - \frac{1}{4 \sqrt{2} \sqrt{\Sigi - 1}}$ \\ \hline
  $D_{\barbH \perp}$ & $\frac{\Ma}{2 \pi}$ & $\frac{\Ma}{4 \pi
  \sqrt{\Sigma - \frac{1}{2}}}$ & $\frac{\chia}{\pi}$ & 
  $\frac{\chia}{2 \sqrt{2} \pi \sqrt{\Sigi - 1}}$ \\ \hline
  $D_{\perp \perp}$ & $\mbox{} - \frac{4 \Sigma \Ma}{\pi}$ & $\mbox{} - 
  \frac{1}{8 \sqrt{\Sigma - \frac{1}{2}}}$ & $- \frac{4 \Sigi \chia}{\pi}$ & 
  $\mbox{} - \frac{4 \chia + \pi}{4 \sqrt{2} \pi \sqrt{\Sigi-1}}$ \\ \hline
  $\Diprime$ & 
  $\frac{\etaprime}{12 \Sigma}$ & 
  $\frac{\etaprime}{12 \sqrt{\Sigma-\frac{1}{2}}}$
  & $\frac{\etaprime}{6 \Sigi}$
  & $\frac{\etaprime}{6 \sqrt{2} \sqrt{\Sigi-1}}$ \\ \hline
  $\lambda_1$ 
  & $\frac{(\Ma - \pi)}{8 \pi \Sigma}$ 
  & $\mbox{} \frac{(2 \Ma - \pi)}{8 \pi \sqrt{\Sigma-\frac{1}{2}}}$
  & $\frac{\chia - \pi}{4 \pi \Sigi}$
  & $\frac{2 (\sqrt{2} - 1) \chia - \pi}{4 \sqrt{2} \pi \sqrt{\Sigi-1}}$ \\ \hline
  ${\bf e}_1$ 
  & $\beH$ 
  & $\frac{\beH + \beperp}{\sqrt{2}}$ 
  & $\beH$
  & $\frac{(1+\sqrt{2}) \beH + \beperp}{2^{3/4}\sqrt{\sqrt{2}+1}}$\\ \hline
  $\lambda_2$ 
  & $\mbox{} - \frac{2 \Ma \Sigma}{\pi}$ 
  & $\mbox{} - \frac{(2 \Ma + \pi)}{8 \pi \sqrt{\Sigma-\frac{1}{2}}}$
  & $\mbox{} - \frac{4 \chia \Sigi}{\pi}$ 
  & $\mbox{} - \frac{2 (\sqrt{2} + 1) \chia + \pi}{4 \sqrt{2} \pi \sqrt{\Sigi-1}}$ \\ \hline
  ${\bf e}_2$ 
  & $\beperp$ 
  & $\frac{\beH - \beperp}{\sqrt{2}}$ 
  & $\beperp$
  & $\frac{- (\sqrt{2}-1) \beH + \beperp}{2^{3/4} \sqrt{\sqrt{2}-1}}$ \\ \hline
  $\lambda_3$ 
  & $\mbox{} - \frac{4 \Ma \Sigma}{\pi}$ 
  & $\mbox{} - \frac{\Ma}{\pi}$
  & $\mbox{} - \frac{4 \chia \Sigi}{\pi}$
  &  $\mbox{} - \frac{\chia}{\sqrt{2} \pi \sqrt{\Sigi-1}}$ \\ \hline
  ${\bf e}_3$ 
  & $\beomega$ 
  & $\beomega$ 
  & $\beomega$
  & $\beomega$ \\ \hline

 \end{tabular}
 \caption{\label{tab:table1} The asymptotic behaviour of the diffusion coefficients for 
 \( \Sigma, \Sigi \gg 1 \) and close to the transition from steady to rotating states for a
 parallel magnetic field for permanent and induced dipoles.}
\end{table}

%
\subsection{Oblique magnetic field}
\label{subsec:obliquepermanent}
Analytical solutions for the steady orientation have been derived for a spherical particle in an oblique magnetic field, where the vorticity and magnetic field are not perpendicular. The orthogonal co-ordinate system 
shown in figure \ref{fig:config} (b) is used, where the unit vector \( \beomega \) is along the 
vorticity direction, \( \bepar \) is perpendicular to \( \beomega \) in the \( \beomega-\beH \)
plane, and \( \beperp \) is perpendicular to the \( \beomega-\beH \) plane. 
The following notations are used for the dot products,
\begin{eqnarray}
 \wH & = & \beomega \bcdot \beH, \, \, \& \, \, \oH = \barbor \bcdot \beH.
 \label{eq:eq227b}
\end{eqnarray}
It should be noted that \( \wH \) is specified, since it depends
on the relative orientation of the vorticity and magnetic field, whereas
\( \oH \) is determined from the solution for the orientation vector.
From equation \ref{eq:eq10}, the dot product \( \barbor \bcdot
\beomega = (\wH/\oH) \).
The unit vectors
\( \bepar \) and \( \beperp \) are,
\begin{eqnarray}
 \bepar & = & \frac{\beH - \wH \beomega}{\sqrt{1 - \wH^2}}, 
 \label{eq:eq227a} \\
 \beperp & = & \frac{\beomega \btimes \beH}{\sqrt{1 - \wH^2}}. \label{eq:eq227}
 \end{eqnarray}
The the orientation vector for the steady solution is specified by the relation,
\begin{eqnarray}
 \barbor \bcdot \beomega & = & \sqrt{- \frac{4 \Sigma^2-1}{2} + \frac{\sqrt{(4 \Sigma^2 - 1)^2
 + 16 \Sigma^2 \wH^2}}{2}}, \label{eq:eq228a} \\
 \oH & = & \frac{1}{2 \Sigma} \sqrt{\frac{4 \Sigma^2-1}{2} +
 \frac{\sqrt{(4 \Sigma^2 - 1)^2 + 16 \Sigma^2 \wH^2}}{2}}. \label{eq:eq228}
\end{eqnarray}
The projection of the orientation vector on the unit vector \( \bepar \) is,
\begin{eqnarray}
 \barbor \bcdot \bepar & = & \frac{\wH(1 - (\barbor \bcdot \beomega)^2)}{
 (\barbor \bcdot \beomega) \sqrt{1 - \wH^2}}. \label{eq:eq229}
\end{eqnarray}
The dot product \( \barbor \bcdot \beperp \) is determined from the torque balance equation,
\begin{eqnarray}
 \barbor \bcdot \beperp & = & \frac{\barbor \bcdot (\beomega \btimes \beH)}{\sqrt{1 - (\beomega
 \bcdot \beH)^2}} = - \frac{\beomega \bcdot (\barbor \btimes \beH)}{\sqrt{1 - (\beomega
 \bcdot \beH)^2}} \nonumber \\
 & = & \frac{\beomega \bcdot (\bI - \barbor \barbor) \bcdot \beomega}{2 \Sigma \sqrt{1 - (\beomega
 \bcdot \beH)^2}} 
 = \frac{1 - (\barbor \bcdot \beomega)^2}{2 \Sigma \sqrt{1 - \wH^2}}.
 \label{eq:eq230}
\end{eqnarray}
The torque balance equation \ref{eq:torquebalance} and the definition \ref{eq:eq221} of 
\( \Sigma \) has been used in the third step in the above equation. Equations \ref{eq:eq228}-
\ref{eq:eq230} are substituted into equation \ref{eq:eq220}, to obtain equation \ref{eq:eq223}
for the diffusion coefficient,
where \( \bm{D}^h \) and \( \bm{D}^m \) are symmetric tensors with the form,
\begin{eqnarray}
\bm{D}^{h/m}& = & 
\left( \begin{array}{ccc} \beomega & \bepar & \beperp \end{array} \right)
\left( \begin{array}{ccc} \Dwwhm & \Dwparhm & \Dwperphm \\
        \Dwparhm & \Dparparhm & \Dparperphm \\
        \Dwperphm & \Dparperphm & \Dperpperphm \end{array} \right)
\left( \begin{array}{c} \beomega \\ \bepar \\ \beperp \end{array} \right).
\label{eq:eq232}
\end{eqnarray}
The solutions for the elements of the tensors \( \bm{D}^h \) and \( \bm{D}^m \),
and the asymptotic expansion of these elements for small and large \( \Sigma \),
are provided in table \ref{tab:table2}. The elements of \( \bm{D}^h \) and
\( \bm{D}^m \) do depend on the solution \ref{eq:eq228} for \( \oH \), and the 
asymptotic expansions employ the small and large \( \Sigma \) approximations for
the \( \oH \) solution,
\begin{eqnarray}
 \oH & = & \wH + 2 \Sigma^2 \wH (1 - \wH^2) \: \: \mbox{for} \: \: \Sigma \ll 1, 
 \label{eq:eq233} \\
 \oH & = & 1 - \frac{1 - \wH^2}{4 \Sigma^2} \: \: \mbox{for} \: \: \Sigma \gg 1. \label{eq:eq234}
\end{eqnarray}
It is easily verified that the elements
of \( \bm{D}^h \) and \( \bm{D}^m \) reduce to equation \ref{eq:eq220} for
a parallel magnetic field with \( \wH = 0 \).  
\renewcommand{\baselinestretch}{1}
 \begin{table}
 \begin{tabular}{|c|c|c|c|} \hline \hline
& & $\Sigma \ll 1$ & $\Sigma \gg 1$ \\ \hline
  $\Dwwh$ 
  & $\mbox{} - \frac{(\oH^2 - \wH^2) \wH^2}{8 \Sigma \oH^5}$ 
  & $\mbox{} - \frac{\Sigma (1 - \wH^2)}{2 \wH}$ 
  &$\mbox{} - \frac{\wH^2 (1 - \wH^2)}{8 \Sigma}$
  \\ \hline
  $\Dwwm$ 
  & $\mbox{} - \frac{2 \Ma \Sigma (\oH^2 - \wH^2)}{\pi \oH^2}$
  & $\mbox{} - \frac{8 \Ma \Sigma^2 (1 - \wH^2)}{\pi}$ 
  & $\mbox{} - \frac{2 \Ma \Sigma (1 - \wH^2)}{\pi}$ 
  \\ \hline \hline
    
  $\Dwparh$ 
  & $\mbox{} - \frac{\wH (\oH^2 - \wH^2)^2}{8 \Sigma \oH^5 \sqrt{1 - \wH^2}}$ 
  & $\mbox{} - 2 \Sigma^3 ( 1 - \wH^2)^{3/2}$ 
  &$\mbox{}- \frac{\wH(1 - \wH^2)^{3/2}}{8 \Sigma}$ 
  \\ \hline
  $\Dwparm$ 
  & $\frac{\Ma \Sigma \wH (1 + \oH^2 - 2 \wH^2)}{\pi \oH^2 \sqrt{1 - \wH^2}}$ 
  & $\frac{\Ma \Sigma \sqrt{1 - \wH^2}}{\pi \wH}$ 
  &$\frac{2 \Ma \Sigma \wH \sqrt{1 - \wH^2}}{\pi}$ 
  \\ \hline \hline
  
  $\Dwperph$ 
  &  $\mbox{} - \frac{\wH (\oH^2 - \wH^2)^2}{16 \Sigma^2 \oH^6 \sqrt{1 - \wH^2}}$ 
  & $\mbox{} - \frac{\Sigma^2 (1 - \wH^2)^{3/2}}{\wH}$ 
  &$\mbox{} - \frac{\wH (1 - \wH^2)^{3/2}}{16 \Sigma^2}$ 
  \\ \hline
  $\Dwperpm$ 
  & $\frac{\Ma \wH (\oH^2 - \wH^2)}{2 \pi \oH^3 \sqrt{1 - \wH^2}}$ 
  & $\frac{2 \Ma \Sigma^2 \sqrt{1 - \wH^2}}{\pi}$ 
  &$\frac{\Ma \wH \sqrt{1 - \wH^2}}{2 \pi}$ 
  \\ \hline \hline
  
  $\Dparparh$ 
  & $\frac{(\oH^2 - \wH^2) (2 \oH^2 \wH^2 - \oH^2 - \wH^4)}{8 \Sigma \oH^5 (1 - \wH^2)}$ 
  & $\mbox{} - \frac{\Sigma (1 - \wH^2)}{2 \wH}$ 
    & $\mbox{} - \frac{(1 - \wH^2)^2}{8 \Sigma}$ 
  \\ \hline
  $\Dparparm$ 
  & $\mbox{} - \frac{2 \Ma \Sigma \wH^2}{\pi \oH^2}$ 
  & $\mbox{} - \frac{2 \Ma \Sigma}{\pi}$ 
    & $\mbox{} - \frac{2 \Ma \Sigma \wH^2}{\pi}$
  \\ \hline \hline

  $\Dparperph$ 
  & $\frac{\wH^2 (\oH^2-\wH^2)^2}{16 \Sigma^2 \oH^6 (1 - \wH^2)}$
  & $\Sigma^2 (1 - \wH^2)$ 
    & $\frac{\wH^2 (1 - \wH^2)}{16 \Sigma^2}$ 
  \\ \hline
  $\Dparperpm$ 
  & $\frac{\Ma (\oH^2-\wH^2)}{2 \pi \oH^3}$
  & $\frac{2 \Ma \Sigma^2 (1 - \wH^2)}{\pi \wH}$ 
    & $\frac{\Ma (1 - \wH^2)}{2 \pi}$ 
  \\ \hline \hline

  $\Dperpperph$ 
  & $\mbox{} - \frac{\oH^2-\wH^2}{8 \Sigma \oH^3} 
  + \frac{\wH^2 (\oH^2-\wH^2)^2}{32 \Sigma^3 \oH^7 (1 - \wH^2)}$
  & $\mbox{} - \frac{2 \Sigma^3 (1 - \wH^2)}{\wH}$ 
    & $\mbox{} - \frac{1 - \wH^2}{8 \Sigma}$ 
  \\ \hline
  $\Dperpperpm$ & $\mbox{} - \frac{2 \Sigma \Ma}{\pi}$ & $\mbox{} - \frac{2 \Sigma \Ma}{\pi}$ 
  & $\mbox{} - \frac{2 \Sigma \Ma}{\pi}$ \\ \hline 
    $\Diprime$ & $\frac{\etaprime (\oH^2-\wH^2)}{12 \oH^3 \Sigma}$ 
  & $\frac{\Sigma \etaprime (1-\wH^2)}{3 \wH}$ 
  & $\frac{\etaprime (1 - \wH^2)}{6 \Sigma}$ \\ \hline

  $\lambda_{1}$ 
 &
 & $\frac{\Sigma (1 - \wH)(2 \Ma - \pi (1 + \wH))}{2 \pi \wH}$
 & $\frac{(\Ma-\pi) (1 - \wH^2)}{8 \pi \Sigma}$
 \\ \hline
 ${\bf e}_1$ 
 & 
 & $\frac{\sqrt{1 + \wH} \beomega + \sqrt{1 - \wH} \bepar}{\sqrt{2}}$
 & $\wH \beomega + \sqrt{1-\wH^2} \bepar$
 \\ \hline \hline
 $\lambda_2$ 
 & 
 & $\mbox{} - \frac{\Sigma (1 + \wH)(2 \Ma + \pi (1 - \wH))}{2 \pi \wH}$
& $\mbox{} - \frac{2 \Sigma \Ma}{\pi}$
 \\ \hline
 ${\bf e}_2$ 
 & 
  & $\frac{- \sqrt{1 - \wH} \beomega + \sqrt{1 + \wH} \bepar}{\sqrt{2}}$
   & $\mbox{} - \sqrt{1-\wH^2} \beomega + \wH \bepar$
 \\ \hline \hline
 $\lambda_3$
 &
 & $\mbox{} - \frac{2 \Sigma \Ma}{\pi}$
 & $\mbox{} - \frac{2 \Sigma \Ma}{\pi}$
 \\ \hline
 ${\bf e}_3$ 
 & 
 & $\beperp$
  & $\beperp$
 \\ \hline \hline
  \end{tabular}
 \caption{\label{tab:table2} The asymptotic behaviour of the diffusion coefficients for 
 \( \Sigma \gg 1 \) and \( \Sigma \ll 1 \) for an oblique
 magnetic field for particles with permanent dipoles.}
\end{table}
\renewcommand{\baselinestretch}{1.5}

For an oblique magnetic field, all the elements of \( \bm{D}^h \) and \( \bm{D}^m \) are non-zero.
This is in contrast to the diffusion matrix \ref{eq:eq220} for a parallel magnetic
field, where there are only four independent non-zero components. For a parallel magnetic field,
there is a transition from a
static to a rotating state for the orientation vector for a parallel magnetic field at \( \Sigma = 
\frac{1}{2} \). In contrast, there is no transition for an oblique magnetic field. Therefore,
the diffusion coefficients for a parallel magnetic field are defined only in the range \( \frac{1}{2}
< \Sigma < \infty \), whereas those for an oblique magnetic field are defined for \( 0 < \Sigma < 
\infty \) in figure \ref{fig:diffp}. 

Table \ref{tab:table2} shows that all components of the hydrodynamic contribution to the 
diffusion tensor \( \bDh \) are negative, with the exception of the component \( \Dparperph \). 
In contrast, all components of the magnetic contribution to the diffusion tensor \( \bDm \)
are positive. This implies that the magnetic interactions tend to dampen concentration 
fluctuations and stabilise the uniform state, whereas hydrodynamic interactions tend to
destabilise the concentration fluctuations. The diffusion in the direction perpendicular to 
the vorticity and magnetic field is entirely due to hydrodynamic interactions and the diffusion
coefficient \( \Dperpperp \) is negative. Therefore, concentration fluctuations are amplified
in the direction perpendicular to the vorticity and magnetic field. In other directions, the 
stability is determined by a balance between the contributions due to hydrodynamic
and magnetic interactions.

\begin{figure}
\parbox{.5\textwidth}{
    \psfrag{x}[][][1.0][0]{$\Sigma$}
    \psfrag{y}[][][1.0][0]{$\Diprime/\etaprime$}

 \includegraphics[width=.49\textwidth]{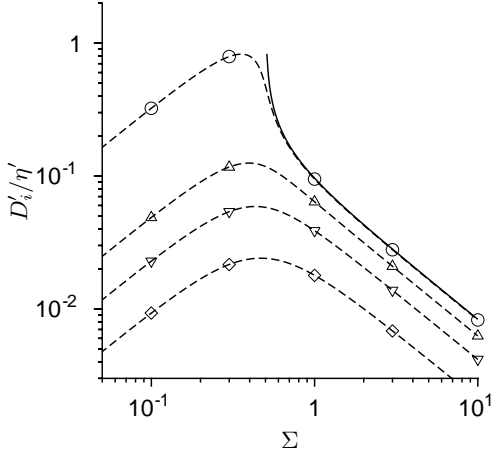}
 
 }
%
%
  
 \caption{\label{fig:diffp} The scaled isotropic part of the diffusion
 tensor, \( \Diprime/\etaprime \) 
 as a function of the parameter \( \Sigma \) for 
 particles with a permanent dipole. The orientation of the vorticity and magnetic field are 
 \( \beomega \bcdot
\beH = 0.1 \) ($\circ$), \( \frac{1}{2} \) ($\triangle$), \( \frac{1}{\sqrt{2}} \) ($\nabla$)
and \( \frac{\sqrt{3}}{2} \) ($\diamond$). The solid line is the result for
a parallel magnetic field.
 }
\end{figure}

The components of \( \bD \) are shown as a function of \( \Sigma \) for different values
of \( \wH \) in figure \ref{fig:diffp}. Also shown by solid lines in figures
\ref{fig:diffp} (a), (d), (e) and (f) are the results for a parallel magnetic field. The
components of the diffusion tensor for a parallel magnetic field do not extend to
\( \Sigma < \frac{1}{2} \), due to the transition to a rotating state; the
divergence in the diffusion coefficients \( \DHHh, \DHperpm \) and \( \Dperpperph \) 
predicted by equations \ref{eq:eq224}-\ref{eq:eq226} is apparent in figure 
\ref{fig:diffp}. The results for 
\( \wH = 0.1 \) are in close agreement with those for a
parallel magnetic field for \( \Sigma \gtrsim \frac{1}{2} \), but the 
divergence in the coefficients \( \DHHh, \DHperpm \) and \( \Dperpperph \)
is cut-off and the diffusion coefficients are finite \( \Sigma < \frac{1}{2} \).
\begin{figure}
\parbox{.49\textwidth}{
    \psfrag{x}[][][1.0][0]{$\Sigma$}
    \psfrag{y}[][][1.0][0]{\textcolor{blue}{$- \Dwwh$}, \textcolor{red}{$- \Dwwm/\Ma$}}

 \includegraphics[width=.49\textwidth]{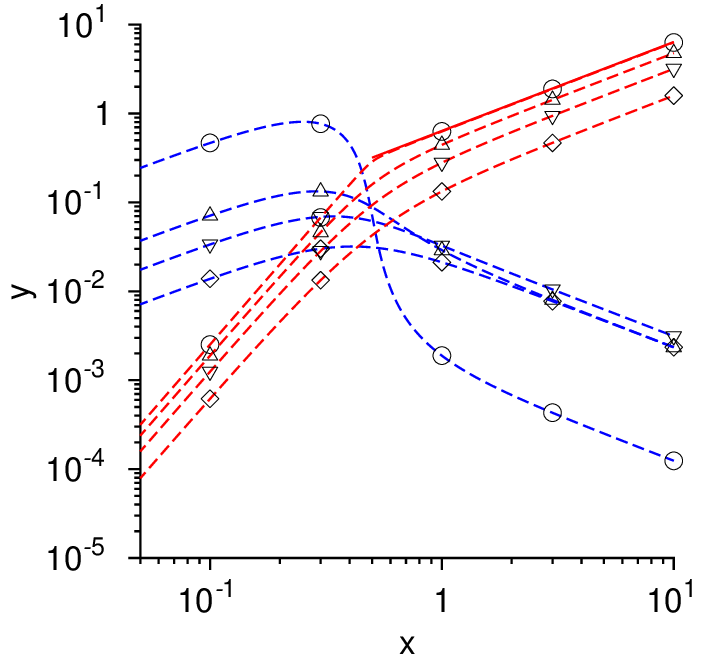}
 
 \begin{center} (a) \end{center}
 }
\parbox{.49\textwidth}{
    \psfrag{x}[][][1.0][0]{$\Sigma$}
    \psfrag{y}[][][1.0][0]{\textcolor{blue}{$- \Dwparh$}, \textcolor{red}{$\Dwparm/\Ma$}}

 \includegraphics[width=.49\textwidth]{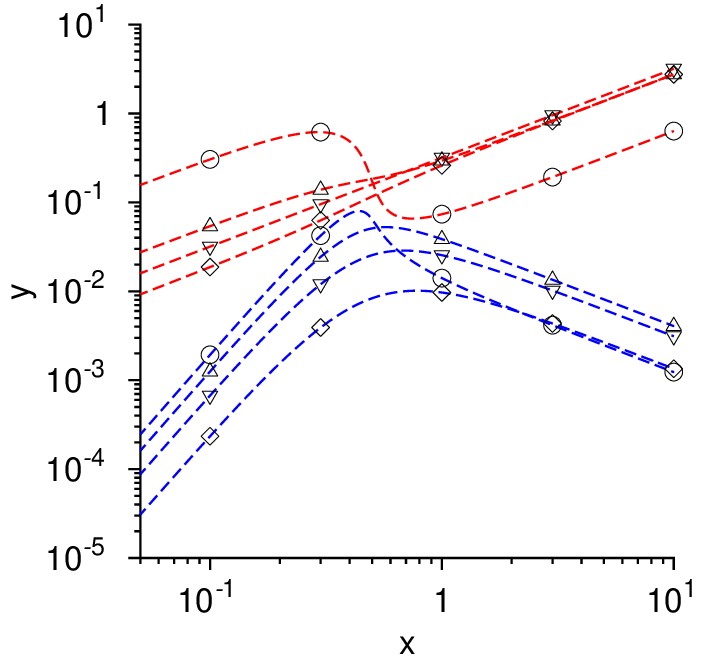}
 
 \begin{center} (b) \end{center}
 }
 
\parbox{.49\textwidth}{
    \psfrag{x}[][][1.0][0]{$\Sigma$}
    \psfrag{y}[][][1.0][0]{\textcolor{blue}{$- \Dwperph$}, \textcolor{red}{$\Dwperpm/\Ma$}}

 \includegraphics[width=.49\textwidth]{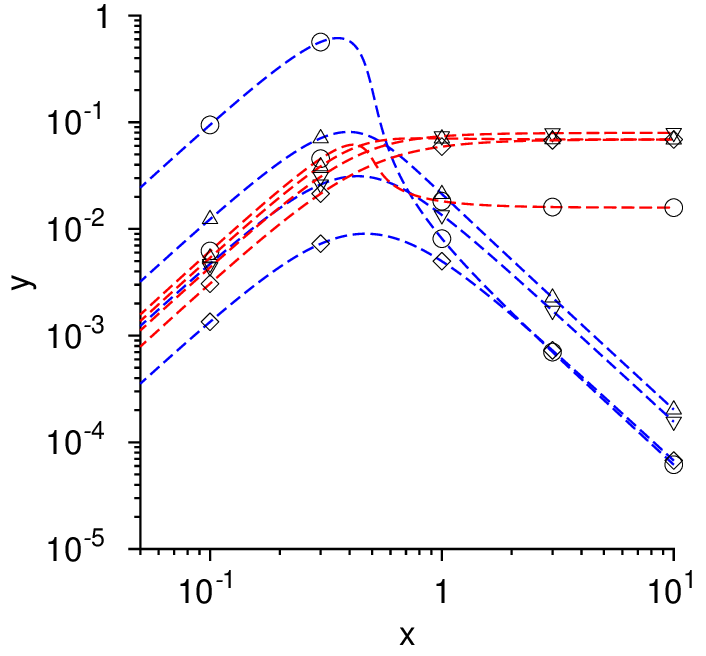}
 
 \begin{center} (c) \end{center}
 }
\parbox{.49\textwidth}{
    \psfrag{x}[][][1.0][0]{$\Sigma$}
    \psfrag{y}[][][1.0][0]{\textcolor{blue}{$- \Dparparh$}, \textcolor{red}{$- \Dparparm/\Ma$}}

 \includegraphics[width=.49\textwidth]{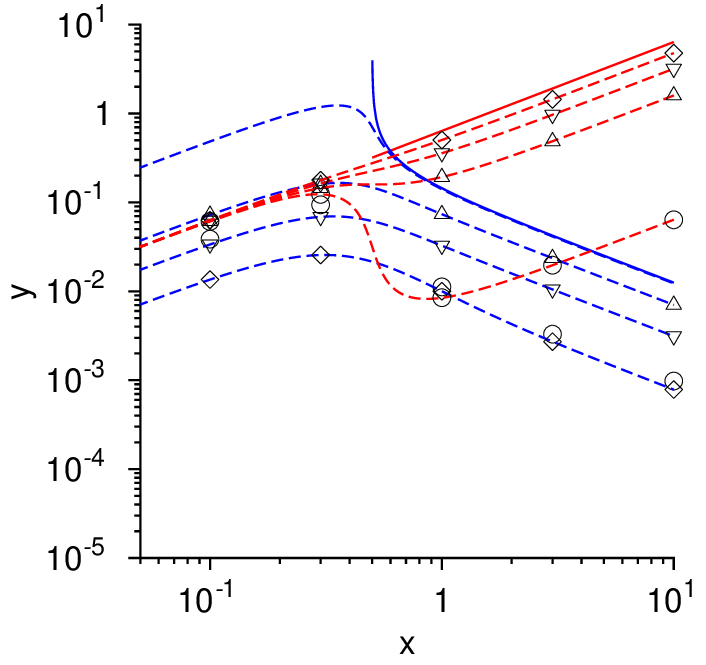}
 
 \begin{center} (d) \end{center}
 }

 \parbox{.49\textwidth}{
    \psfrag{x}[][][1.0][0]{$\Sigma$}
    \psfrag{y}[][][1.0][0]{\textcolor{blue}{$\Dparperph$}, \textcolor{red}{$\Dparperpm/\Ma$}}

 \includegraphics[width=.49\textwidth]{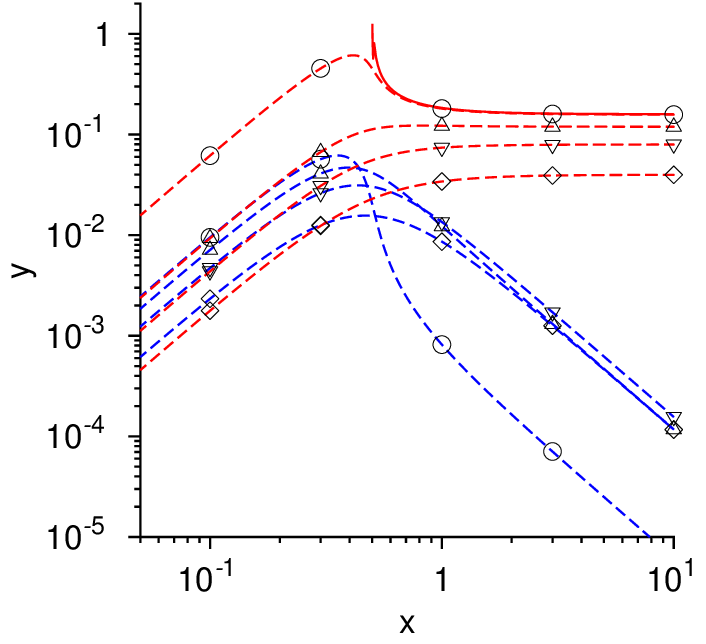}
 
 \begin{center} (e) \end{center}
 }
\parbox{.49\textwidth}{
    \psfrag{x}[][][1.0][0]{$\Sigma$}
    \psfrag{y}[][][1.0][0]{\textcolor{blue}{$- \Dperpperph$}, \textcolor{red}{$- \Dperpperpm/\Ma$}
    }

 \includegraphics[width=.49\textwidth]{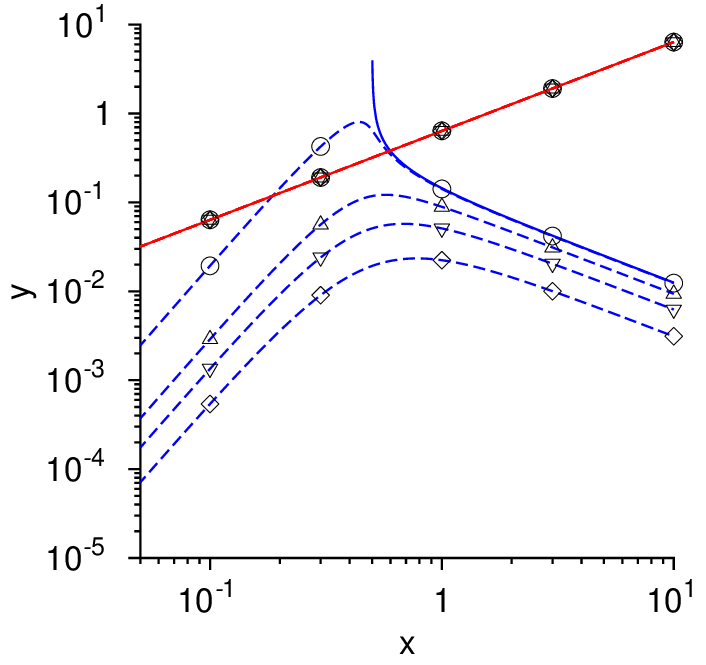}
 
 \begin{center} (f) \end{center}
 }
 
 \caption{\label{fig:diffp} The components of the diffusion tensor 
 \( - \Dwwh \& - \Dwwm/\Ma \) (a), 
 \( - \Dwparh \& \Dwparm/\Ma \) (b), 
 \( - \Dwperph \& \Dwperpm/\Ma \) (c), 
 \( - \Dparparh \& - \Dparparm/\Ma \) (d), 
 \( \Dparperph \& \Dparperpm/\Ma \) (e) and 
 \( - \Dperpperph \& - \Dperpperpm/\Ma \) (f) 
 due to hydrodynamic interactions (blue lines) and magnetic interactions (red lines) as a function of the parameter \( \Sigma \) for particles with a permanent dipole. The orientation of the vorticity and magnetic field are \( \beomega \bcdot
\beH = 0.1 \) ($\circ$), \( \frac{1}{2} \) ($\triangle$), \( \frac{1}{\sqrt{2}} \) ($\nabla$)
and \( \frac{\sqrt{3}}{2} \) ($\diamond$). The solid red and blue lines are the non-zero results for
a parallel magnetic field. 
 }
\end{figure}

The coefficients \( \Dwwh, \Dwparh, \Dwparm, \Dwperph, \Dwperpm, \Dparparm, \Dparperph \)
are all zero for a parallel magnetic field with \( \wH = 0 \). Figure
\ref{fig:diffp} shows that these coefficients do decrease as \( \wH \) 
decreases for \( \Sigma > \frac{1}{2} \), in accordance with the \( \Sigma \gg 1 \) 
expressions table \ref{tab:table2}. 
For \( \Sigma < \frac{1}{2} \), these do not as \( \wH \) decreases. 
The latter regime is not accessible for a parallel magnetic field, since there is no 
steady orientation. Therefore, the diffusion due to interactions for a nearly parallel 
magnetic field is qualitatively different from that in a parallel magnetic field, due 
to the transition to rotating states in the latter.

The magnetic contributions to
the diffusion tensor are larger than the hydrodynamic contributions for \( \Sigma \gg 1 \).
For \( \Sigma \gg 1 \), the largest contributions to the diffusion tensor are 
\( \Dwwm, \Dwparm, \Dparparm \) and \( \Dperpperpm \) which diverge proportional to \( \Sigma \).
The coefficient \( \Dperpperpm = - (2 \Ma \Sigma/\pi) \) is negative for all \( \Sigma \),
indicating that concentration fluctuations are amplified in the \( \beperp \) direction.
The diffusion in the \( \beomega-\bepar \) plane is along two principal directions.
The eigenvalue \( - (2 \Ma \Sigma/\pi) \) along \( \bm{e}_2 \) in table \ref{tab:table2} 
also increases proportional to \( \Sigma \) for \( \Sigma \gg 1 \), indicating strong amplification of
concentration fluctuations in this direction. The eigenvalue along the \( \bm{e}_1 \) 
direction decreases proportional to \( \Sigma^{-1} \) for \( \Sigma \gg 1 \), and the 
concentration fluctuations are dampened/amplified for \( \Ma \gtrless \pi \). Thus,
the behaviour of concentration fluctuations for \( \Sigma \gg 1 \) is very similar
to that for a parallel magnetic field, with strong concentration amplifications in
two directions $\beperp$ and $\bm{e}_2$, and weak amplification/damping in 
the third perpendicular direction.

In the limit \( \Sigma \ll 1 \), the diffusion coefficient along the \( \beperp \)
direction perpendicular to the vorticity and magnetic field is \( \Dperpperpm = - 
(2 \Ma \Sigma/\pi) \). The \( O(\Sigma) \) contributions to the diffusion
tensor are \( \Dwwh, \Dwwm, \Dparparh, \Dwparm \) and \( \Dperpperpm \).
For diffusion in the \( \beomega-\bepar \) plane; concentration fluctuations 
are unstable along the \( \bm{e}_2 \)
direction where the eigenvalue \( \lambda_2 \) is negative, while they are stable
in the perpendicular direction \( \bm{e}_1 \) for \( 2 \Ma > \pi (1 + \wH) \)
and unstable otherwise. 

\section{Induced dipole}
\label{sec:induced}
\subsection{Diffusion due to interactions}
\label{subsec:diffusioninduced}
For an induced dipole, the particle magnetic moment is modeled as,
\begin{eqnarray}
 \bM & = & \chi  \bor [\bor \bcdot (\bH + n \bM)], \label{eq:eq31}
\end{eqnarray}
that is, the magnetic moment is proportional to the component of the magnetic field along
the particle orientation. Here, \( \chi \) is the magnetic susceptibility of the particle.
Substituting \( \bM = M \bor \) in \ref{eq:eq31}, the magnitude of the particle
magnetic moment is,
\begin{eqnarray}
 M & = & \frac{\chi \bor \bcdot \bH}{1 - \chi n}.
\end{eqnarray}

The torque balance equation in the absence of interactions in the direction perpendicular
to the orientation vector, analogous to equation \ref{eq:torquebalance}, is
\begin{eqnarray}
 \tfrac{1}{2} \pi \eta d^3 (\bI - \bor \bor) \bcdot \bomega + \frac{\mu_0 \chi (\bor \bcdot \bH) (\bor \btimes \bH)}{
 1 - \chi n}& = & 0.
 \label{eq:torquebalancep}
\end{eqnarray}
The substitution \( \bM = \chi (\bor \bcdot \bH) \bor \) is made in equation \ref{eq:mforce} to 
obtain the expression for the force,
\begin{eqnarray}
 \bF & = & \bnabla \left[\frac{\chi (\bor \bcdot \bH)}{1 - \chi n} \bor \bcdot \left(\bH + 
 \frac{\chi n \bor (\bor \bcdot \bH)}{1 - \chi n} \right) \right] \nonumber \\
 & = & \bnabla \left[ \frac{\chi (\bor \bcdot \bH)^2}{(1 - \chi n)^2} \right].
\end{eqnarray}
When the above expression is linearised in the perturbations and transformed into Fourier space,
the equivalent of equation \ref{eq:eq11} for the interaction force is,
\begin{eqnarray}
 \tbF_{\bk} & = & \mbox{} - 2 \imath \bk \mu_0 \chi (\barbor \bcdot \barbH) \left[\frac{
 (\barbor \bcdot \tbH_{\bk} + \barbH \bcdot \tbor_{\bk})}{(1 - \chi \barn)^2}
 + \frac{\chi \tn_{\bk} (\barbor \bcdot \barbH)}{(1 - \chi \barn)^3} 
 \right]. \label{eq:eq32}
\end{eqnarray}
Since we are considering the limit where the disturbance
is small compared to the applied magnetic field, the approximation \( \barn \chi \ll 1 \)
is made in \ref{eq:eq32}. With this approximation, 
the equivalent of equation \ref{eq:eq13} for the particle concentration field is,
\begin{eqnarray}
 \frac{\partial \tn_{\bk}}{\partial t} + \imath \bk \bcdot (\barbv \tn_{\bk}) + D_B k^2 \tn_{\bk} \hfill & &
 \nonumber \\ - \left( \frac{2 k^2 \barn \mu_0 \chi (\barbor \bcdot \barbH) (\barbor \bcdot \tbH_{\bk} 
+  \barbH \bcdot \tbor_{\bk}+\chi \tn_{\bk} (\barbor \bcdot \barbH))}{3 \pi \eta_0 d}
\right)  & = & 0.
 \label{eq:eq33}
\end{eqnarray}
Comparing the expression \ref{eq:eq33} with the expression \ref{eq:eq14}, the magnetophoretic 
diffusion term in the particle number density equation is,
\begin{eqnarray}
 \bk \bcdot \bD \bcdot \bk & = & \mbox{} - \frac{2 k^2 \barn \mu_0 \chi (\barbor \bcdot \barbH) 
 (\barbor \bcdot \tH_{\bk} + \tbor_{\bk} \bcdot \barbH +
 \chi \tn_{\bk} \barbor \bcdot \barbH)}{3 \pi \eta_0 d \tn_{\bk}}.
 \label{eq:eq34}
 \end{eqnarray}

The equivalent of equation \ref{eq:eq24} for the disturbance to the magnetic 
field due to interactions is,
\begin{eqnarray}
 \tH_{\bk} & = & \mbox{} - \frac{\bk \bk}{k^2} \bcdot \left[ \frac{\tn_{\bk} \chi \barbor (\barbor \bcdot \barbH)}{
 (1 - \chi \barn)^2} + \frac{\barn \tbor_{\bk} \bcdot (\bI (\barbor \bcdot \barbH) + \barbH \barbor)}{1 - \chi \barn}
 \right]. \label{eq:eq35}
\end{eqnarray}
The second term in the square brackets on the right proportional to \( \tbor_{\bk} \), 
equivalent of the term proportional to \( \tbHpp \) in equation \ref{eq:eq24},
is neglected. 
The reason for this is discussed in the paragraph following equation \ref{eq:eq29}.
In the denominator of the first term in the square brackets on the right of \ref{eq:eq35},
\( \chi \barn \) is neglected in comparison to $1$, because the disturbance to
the magnetic field is small compared to the applied magnetic field.
The disturbance to the vorticity field is given in equation \ref{eq:eq28}. 
In this expression, the term proportional to \( \barn d^3 \) on the left
side is neglected because the volume fraction is small, and the term
proportional to \( \tbomegapp \) on the right is neglected for the 
reason discussed after equation \ref{eq:eq29}. With these approximations,
the torque balance equation, equivalent of equation \ref{eq:eq29}, is
\begin{eqnarray}
 \frac{\pi d^3 \eta_0}{2 \mu_0 \chi} [(\bI - \barbor \barbor) \bcdot \tbomega_{\bk} - \barbor (\tbor_{\bk}
 \bcdot \barbomega) - \tbor_{\bk} (\barbor \bcdot \barbomega) + \etaprime \tphi_{\bk} 
 (\bI - \barbor \barbor) \bcdot \barbomega)] 
  \nonumber & & \\ \mbox{} 
 + \{(\barbor \bcdot \barbH)(\barbor \btimes \tbH_{\bk}) 
 + (\barbor \bcdot \tbH_{\bk}) (\barbor \btimes \barbH)  
 + (\barbor \bcdot \barbH) (\tbor_{\bk} \btimes \barbH) 
 \nonumber \\ \mbox{} + (\tbor_{\bk} \bcdot \barbH)
 (\barbor \btimes \barbH) + \chi \tn_{\bk} (\barbor \bcdot \barbH)(\barbor \btimes \barbH)
 \} & = & 0, \label{eq:eq36}
\end{eqnarray}
The expression \ref{eq:eq210}, is substituted into equation \ref{eq:eq36} to obtain, 
\begin{eqnarray}
  \frac{\pi d^3 \eta_0}{2 \mu_0 \chi} \{(\bI - \barbor \barbor) \bcdot \tbomega_{\bk} 
 - \barbor \torpp_{\bk} (\barbor \btimes \barbH) \bcdot \barbomega
 - (\barbor \bcdot \barbomega)
 [\torp_{\bk} (\barbH - (\barbor \bcdot \barbH) \barbor) 
 & & \nonumber \\ \mbox{} 
 + \torpp_{\bk} (\barbor \btimes \barbH)] 
 + \etaprime \tphi_{\bk} 
 (\bI - \barbor \barbor) \bcdot \barbomega) \} 
 + \{(\barbor \bcdot \barbH)(\barbor \btimes \tbH_{\bk}) 
 & & \nonumber \\ \mbox{} + (\barbor 
 \bcdot \tbH_{\bk}) (\barbor \btimes \barbH) 
 + \torp_{\bk} [|\barbH|^2 - 2 (\barbor \bcdot \barbH)^2] (\barbor \btimes \barbH)
 & & \nonumber \\ \mbox{}
 + \torpp_{\bk} (\barbor \bcdot \barbH) [\barbH (\barbor \bcdot \barbH) - \barbor |\barbH|^2]
+ \chi \tn_{\bk} (\barbor \bcdot \barbH)(\barbor \btimes \barbH) \} & = & 0. 
\label{eq:eq37}
\end{eqnarray}
The functions \( \torp_{\bk} \) and \( \torpp_{\bk} \) are determined by taking the dot
product of equation \ref{eq:eq37} with \( \barbor \btimes \barbH \) and \( \barbor \) and respectively,
\begin{eqnarray}
 \frac{\pi d^3 \eta_0}{2 \mu_0 \chi} [\tbomega_{\bk} \bcdot (\barbor \btimes \barbH) - \torpp_{\bk}
 (\barbor \bcdot \barbomega)(|\barbH|^2 - (\barbor \bcdot \barbH)^2)
 + \etaprime \tphi_{\bk} \barbomega \bcdot (\barbor \btimes \barbH)] & & \nonumber \\
 \mbox{} + \{(\barbor \bcdot \barbH)[\barbH \bcdot \tbH_{\bk} - 
 2 (\barbor \bcdot \barbH)(\tbH_{\bk} \bcdot \barbor)]
 + (\barbor \bcdot \tbH_{\bk}) |\barbH|^2
  & & \nonumber \\ \mbox{} 
  + [|\barbH|^2 - (\barbor \bcdot \barbH)^2][\torp_{\bk} (|\barbH|^2 
 - 2 (\barbor \bcdot \barbH)^2)  +
 \chi \tn_{\bk} (\barbor \bcdot \barbH)] \} & = & 0, 
 \label{eq:eq38} 
 \end{eqnarray}
 \begin{eqnarray}
- \frac{\pi d^3 \eta_0}{2 \mu_0 \chi} [\torpp_{\bk} (\barbor \btimes \barbH) \bcdot \barbomega]
 - \{ \torpp_{\bk} (\barbor \bcdot \barbH) [|\barbH|^2 - (\barbor \bcdot \barbH)^2] 
\} & = & 0. \label{eq:eq39}
\end{eqnarray}
These are solved to obtain,
\begin{eqnarray}
\torp_{\bk} & = &  \mbox{} \frac{\pi \eta_0 d^3}{2 \mu_0 \chi} \frac{[\tbomega_{\bk} \bcdot (\barbor \btimes \barbH)
+ \etaprime \tphi_{\bk} \barbomega \bcdot (\barbor \btimes \barbH)]}{[2 
(\barbor \bcdot \barbH)^2 
- |\barbH|^2] [|\barbH|^2 - (\barbor \bcdot \barbH)^2]} + \frac{\chi \tn_{\bk} 
\barbor \bcdot 
\barbH}{2 (\barbor \bcdot \barbH)^2 - |\barbH|^2} \nonumber \\ & & \mbox{}
+ \frac{\{(\barbor \bcdot \tbH_{\bk})[|\barbH|^2 - 2 (\barbor \bcdot \barbH)^2] + 
(\barbH \bcdot \tbH_{\bk})(\barbor \bcdot \barbH)\}}{[2 (\barbor \bcdot \barbH)^2 - |\barbH|^2] 
[|\barbH|^2 - (\barbor \bcdot \barbH)^2]}. \label{eq:eq310} \\
 \torpp_{\bk} & = &  0.
 \label{eq:eq311}
\end{eqnarray}

The expression for the magnetophoretic diffusion in equation \ref{eq:eq34} is,
\begin{eqnarray}
 \bk \bcdot \bD \bcdot \bk 
  & = & \mbox{} - \frac{2 k^2 \barn \mu_0 \chi (\barbor \bcdot \barbH)}{3 \pi \eta_0 d\tn_{\bk}} \left[ \barbor \bcdot \tbH_{\bk} +
 \torp_{\bk} [|\barbH|^2 - (\barbor \bcdot \barbH)^2] + \chi \tn_{\bk} (\barbor \bcdot \barbH) 
 \right] \nonumber \\
& = & \mbox{} - \frac{2 k^2 \barn \mu_0 \chi (\barbor \bcdot \barbH)}{3 \pi \eta_0 d \tn_{\bk}} 
\left[  \frac{\pi d^3 \eta_0 [\tbomega_{\bk} \bcdot (\barbor \btimes \barbH) + 
\etaprime \tphi_{\bk} \barbomega \bcdot (\barbor \btimes \barbH)]}{2 \mu_0 \chi [2 
(\barbor \bcdot \barbH)^2 - |\barbH|^2]} \right. \nonumber \\ & & \left. \mbox{} +
 \frac{(\barbH \bcdot \tbH_{\bk}) (\barbor \bcdot \barbH)}{[2 (\barbor \bcdot \barbH)^2 - 
 |\barbH|^2]} + \frac{\chi \tn_{\bk} \barbor \bcdot \barbH [|\barbH|^2-(\barbor \bcdot 
 \barbH)]}{2 (\barbor \bcdot \barbH)^2 - |\barbH|^2} + \chi \tn_{\bk} (\barbor \bcdot 
\barbH) 
 \right]. \nonumber \\
& = & \mbox{} - \frac{2 k^2 \barn \mu_0 \chi (\barbor \bcdot \barbH)}{3 \pi \eta_0 d \tn_{\bk}} 
\left[  \frac{\pi d^3 \eta_0 [\tbomega_{\bk} \bcdot (\barbor \btimes \barbH) + 
\etaprime \tphi_{\bk} \barbomega \bcdot (\barbor \btimes \barbH)]}{2 \mu_0 \chi [2 
(\barbor \bcdot \barbH)^2 - |\barbH|^2]} \right. \nonumber \\ & & \left. \hspace{1.2in} \mbox{} +
 \frac{(\barbH \bcdot \tbH_{\bk}) (\barbor \bcdot \barbH)}{[2 (\barbor \bcdot \barbH)^2 - 
 |\barbH|^2]} + \frac{\chi \tn_{\bk} (\barbor \bcdot \barbH)^3}{[2 (\barbor \bcdot 
 \barbH)^2 - |\barbH|^2]} 
 \right].
 \label{eq:eq312}
\end{eqnarray}
Substituting equations \ref{eq:eq35} and \ref{eq:eq28} for \( \tbH_{\bk} \) and \( \tbomega_{\bk} \),
the final expression for the diffusion coefficient is,
\begin{eqnarray}
 \bk \bcdot \bD \bcdot \bk & = & \mbox{} - \frac{d^2 \barphi (\barbor \bcdot \barbH)
 [(\bk \bcdot \barbomega - (\bk \bcdot \barbor)(\barbor \bcdot \barbomega)) (\bk \bcdot (\barbor \btimes \barbH)) - k^2 (\barbor \btimes \barbH) \bcdot
 (\barbomega - (\barbomega \bcdot \barbor) \barbor)]}{2 (2 (\barbor \bcdot \barbH)^2
 - |\barbH|^2)} \nonumber \\ & & \mbox{} 
 - \frac{k^2 \barn d^2 \etaprime (\barbor \bcdot \barbH) \tphi_{\bk} 
 \barbomega \bcdot (\barbor \btimes \barbH)}{6 \tn_{\bk} (2 (\barbor \bcdot \barbH)^2 
 - |\barbH|^2)} + \frac{4 \mu_0 \chi^2 \barphi (\bk \bcdot \barbH)(\bk \bcdot \barbor) 
 (\barbor \bcdot \barbH)^3}{\pi^2 d^4 \eta_0 [2 (\barbor \bcdot \barbH)^2 - |\barbH|^2]}
 \nonumber \\ & &  \mbox{} 
 - \frac{4 k^2 \barphi \mu_0 \chi^2 (\barbor \bcdot \barbH)^4}{\pi^2 
 \eta_0 d^4 [2 (\barbor \bcdot \barbH)^2 - |\barbH|^2]}. \label{eq:eq313}
\end{eqnarray}
Here, the number density is expressed in terms of the volume fraction using the expression
\( \barn = \barphi / (\pi d^3/6) \).
Equation \ref{eq:torquebalancep} is used to express \( (\barbor \btimes \barbH) \) in the 
first term on the right in equation \ref{eq:eq313},
\begin{eqnarray}
 \bk \bcdot \bD \bcdot \bk & = & \frac{\pi \barphi \eta_0 d^5
 [(\bk \bcdot \barbomega - (\bk \bcdot \barbor)(\barbor \bcdot \barbomega)) (\bk \bcdot \barbomega
 - (\bk \bcdot \barbor) (\barbor \bcdot \barbomega)) - k^2 (|\barbomega|^2 - 
 (\barbomega \bcdot \barbor))^2]}{4 \mu_0 \chi (2 (\barbor \bcdot \barbH)^2
 - |\barbH|^2)} \nonumber \\ & & 
 \mbox{} + \frac{\pi k^2 \barphi \eta_0 d^5 \etaprime (|\barbomega|^2-(\barbomega
 \bcdot \barbor)^2)}{6 \mu_0 \chi (2 (\barbor \bcdot \barbH)^2 - |\barbH|^2)}
 + \frac{4 \mu_0 \chi^2 \phi (\bk \bcdot \barbH)(\bk \bcdot \barbor) (\barbor \bcdot \barbH)^3}{\pi^2 
 d^4 \eta_0 (2 (\barbor \bcdot \barbH)^2 - |\barbH|^2)} \nonumber \\ & & 
 \mbox{} 
  - \frac{4 k^2 \barphi \mu_0 \chi^2 (\barbor \bcdot \barbH)^4}{\pi^2 
 \eta_0 d^4 [2 (\barbor \bcdot \barbH)^2 - |\barbH|^2]}. \label{eq:eq314a}
\end{eqnarray}
The diffusion coefficient extracted from the above equation is of the form
\ref{eq:eq220}, where \( \bDh \) and \( \bDm \) are,
\begin{eqnarray}
 \bDh & = & \frac{
 [(\beomega - \barbor(\barbor \bcdot \beomega)) (\beomega
 - \barbor (\barbor \bcdot \beomega)) - \bI (1 - 
 (\beomega \bcdot \barbor))^2]}{4 \Sigi [2 (\barbor \bcdot \beH)^2
 - 1]}, \nonumber \\
 \bDm & = & \frac{2 \Sigi \chia (\barbor \bcdot \beH)^3 (\beH \barbor +
 \barbor \beH)}{\pi[2 (\barbor \bcdot \beH)^2 - 1]} - 
 \frac{4 \Sigi \chia (\barbor \bcdot \beH)^4 \bI}{\pi [2 (\barbor \bcdot \beH)^2-
 1]}, \nonumber \\
 \Diprime & = & \frac{\etaprime (1 - (\beomega \bcdot \barbor)^2)}{6
 \Sigi [2 (\barbor \bcdot \beH)^2-1]}.
 \label{eq:eq314}
\end{eqnarray}
Here, the dimensionless ratio of the magnetic and hydrodynamic torques is,
\begin{eqnarray}
 \Sigi & = & \frac{\mu_0 \chi |\barbH|^2}{\pi \eta_0 d^3 |\barbomega|}, \label{eq:eq315}
\end{eqnarray}
and the scaled susceptibility per unit volume is,
\begin{eqnarray}
 \chia & = & \frac{\chi}{d^3}. \label{eq:eq315a}
\end{eqnarray}
In going from \ref{eq:eq32} to \ref{eq:eq33}, we had made the approximation \( \barn \chi
\ll 1 \). If this approximation is not made, there is only one change in the diffusion
coefficients, \( \chia (1 + \barn \chi) \) should be substituted for \( \chia \) in the 
first term on the right in the expression for \( \bDm \).
\subsection{Steady state}
\label{subsec:steadyinduced}
The orthogonal basis vectors \( (\beomega, \bepar, \beperp) \) in figure
\ref{fig:config} (b) are used for an oblique magnetic
field, where \( \beomega \) is along the vorticity, \( \bepar \) is perpendicular to the 
vorticity in the \( \barbomega-\barbH \) plane, and \( \beperp \) is perpendicular to 
the \( \barbomega-\barbH \) plane. The unit vector \( \bepar \) and \( \beperp \) are
defined in equations \ref{eq:eq227a}-\ref{eq:eq227}. The solution \( \oH \)
(equation \ref{eq:eq227b}) 
has to be determined numerically in this case; this is in contrast to the permanent dipole
in an oblique magnetic field where it is possible to obtain an analytical solution, 
\ref{eq:eq228}.  The solutions have been derived in \cite{vk21a,vk21b}
 for a spheroid. The solution \( \oH^2 \) for a spherical particle satisfies 
 the cubic equation,
\begin{eqnarray}
 4 \Sigi^2 \oH^4 [1 - \oH^2] + [\wH^2 - \oH^2] & = & 0, \label{eq:eq320}
\end{eqnarray}
where \( \wH \) and \( \oH \) are defined in equation \ref{eq:eq227b}.

\begin{figure}
    \psfrag{x}[][][1.0][0]{$\wH$}
    \psfrag{y}[][][1.0][0]{$\Sigi$}

 \includegraphics[width=.5\textwidth]{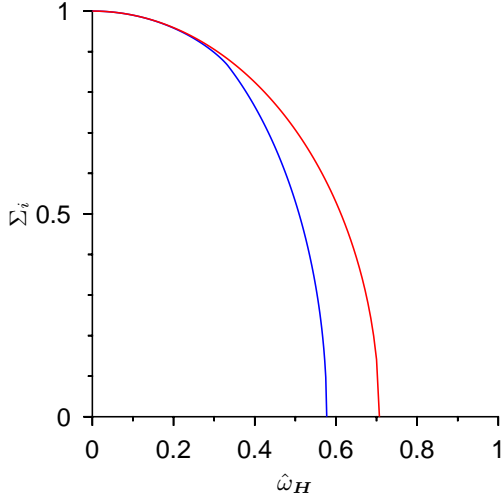}
 \caption{\label{fig:transobli} The boundary between the stable stationary
 solutions (above) and rotating states (below) for the single-particle
 dynamics, equations \ref{eq:eq328} is shown by the blue line, and the boundary
 for the dynamical transition at \( \Sigi = \sqrt{1 - 2 \wH^2} \) is shown
 by the red line in the \( \wH-\Sigi \) plane.
 }
\end{figure}

In the limit \( \Sigi \gg 1 \), the particle
aligns along the magnetic field and \( \oH \rightarrow 1 \). 
As \( \Sigi \) is decreased, there are steady stable solutions for the 
orientation vector for the parameter regimes
\begin{eqnarray}
 \Sigi^2 & > & \frac{1 + 18 \wH^2 - 27 \wH^4 -
 \sqrt{(1 - \wH^2)(1 - 9 \wH^2)^3}}{32 \wH^2} 
 \: \: \mbox{for} \: \: 0 \leq \wH^2 \leq \tfrac{1}{9}, \label{eq:eq328a} \\
 & > & \frac{9 (1 - 3 \wH^2)}{8} \: \: \mbox{for} \: \: \tfrac{1}{9}
 \leq \wH^2 \leq \tfrac{1}{3}, \label{eq:eq328b} \\
 & > & 0 \: \: \mbox{for} \: \: \wH^2 \geq \tfrac{1}{3}. \label{eq:eq328}
\end{eqnarray}
When the conditions \ref{eq:eq328a}-\ref{eq:eq328} are not satisfied, there are stable limit 
cycles and possibly an unstable steady solution. The boundary between the steady
and rotating solutions in the \( \Sigi - \wH \) parameter space is shown by
the blue line in figure \ref{fig:transobli}.

\begin{figure}
\parbox{.49\textwidth}{
    \psfrag{x}[][][1.0][0]{$\Sigi$}
    \psfrag{y}[][][1.0][270]{$\barbor \bcdot \beH$}

 \includegraphics[width=.49\textwidth]{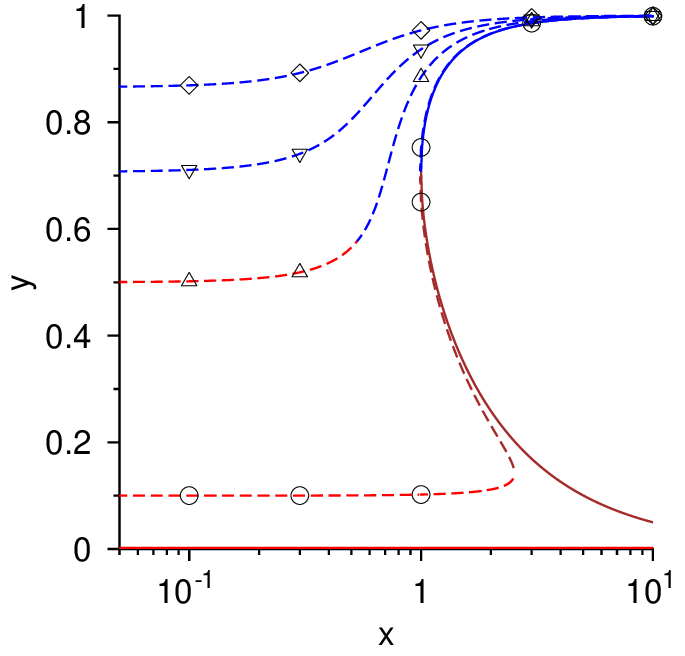}
 
 \begin{center} (a) \end{center}
 }
 \parbox{.49\textwidth}{
     \psfrag{x}[][][1.0][0]{$\Sigi$}
    \psfrag{y}[][][1.0][270]{$\barbor \bcdot \beomega$}

 \includegraphics[width=.49\textwidth]{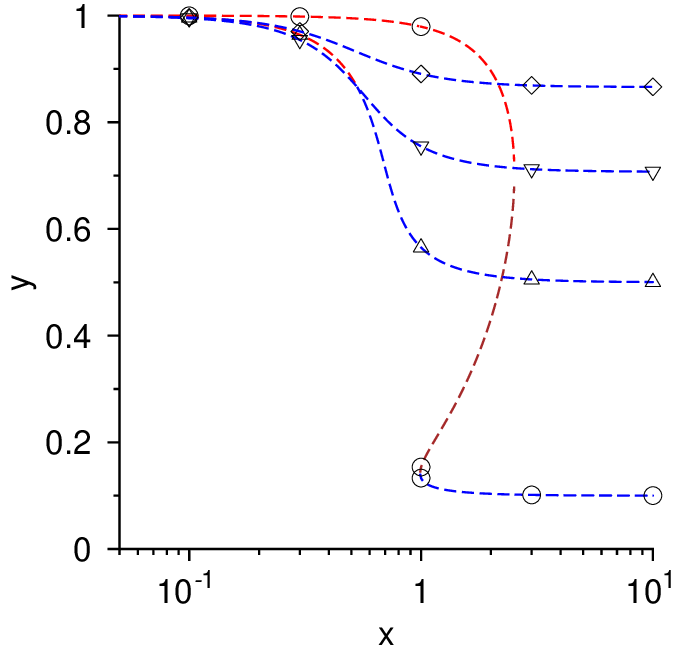}
 
 \begin{center} (b) \end{center}
 }
\caption{\label{fig:orientationi} The variation of \( \barbor \bcdot \beH \) (a) and 
\( (\barbor \bcdot \beomega) \) (b) with \( \Sigi \) for a particle with an induced dipole.
The solid lines are the results for a parallel magnetic field \( (\beomega \bcdot \beH) = 0 \) 
and the dashed lines are the results for an oblique magnetic field with \( \beomega \bcdot
\beH = 0.1 \) ($\circ$), \( \frac{1}{2} \) ($\triangle$), \( \frac{1}{\sqrt{2}} \) ($\nabla$)
and \( \frac{\sqrt{3}}{2} \) ($\diamond$). The blue lines are the stable stationary nodes, the 
red lines are the unstable stationary nodes and the brown lines are the saddle nodes. The value
of \( (\beomega \bcdot \beH) \) for parallel a magnetic field is not shown in sub-figure (b) because 
it is zero.
}
\end{figure}

The solutions of equation \ref{eq:eq320} for \( \barbor \bcdot \beH \) and 
the corresponding solutions of \( \barbor \bcdot \beomega \) are shown
as a function of $\Sigma$ in figure \ref{fig:orientationi}. These are qualitatively
different from those for particles with a permanent dipole shown in
figure \ref{fig:orientationp} because
\( \barbor \bcdot \beH \) is necessarily positive for an induced dipole;
the parameter space \( \barbor \bcdot \beH < 0 \) does not exist in this case.
The solutions for the orientation of a particle acted upon by a shear
flow and a magnetic field are derived in orientation space consisting
of the azimuthal and meridional angles of the particle orientation
\cite{vk21a,vk21b}. In this orientation space,
the red lines in figure \ref{fig:orientationi} are the unstable steady
solutions, the blue lines are the stable steady solutions and the brown
lines are saddle points. The evolution of the fixed points for
\( 0 < \wH^2 < \tfrac{1}{9} \) is illustrated by the curve for
\( \wH = 0.1 \) in figure \ref{fig:orientationi}. There is one stable
fixed point for \( \Sigi \gg 1 \). As \( \Sigi \) is decreased, there
appears one unstable fixed point and one saddle point in the phase
diagram. When there is a further decrease in \( \Sigi \), the saddle
and stable fixed point merge, and there remains one unstable fixed
point and one stable limit cycle. For \( \frac{1}{9} < \wH^2 < 
\frac{1}{3} \), the phase plot contains an stable fixed
point and an unstable limit cycle for \( \Sigi \gg 1 \), and there
is an exchange of stability to an unstable fixed point and a 
stable limit cycle for \( \Sigi \ll 1 \), as shown by the curve
for \( \wH = 0.5 \) in figure \ref{fig:orientationi}. For
\( \frac{1}{3} < \wH^2 < 1 \), there is a stable fixed point
which is alined with the magnetic field for \( \Sigi \gg 1 \),
and with the vorticity for \( \Sigi \ll 1 \).

\subsection{Parallel magnetic field}
\label{subsec:parallelinduced}
Equation \ref{eq:eq320} has analytical solutions for when the magnetic field is parallel to the flow plane,
\( \wH = 0 \). The stable solution for the orientation vector exists only for \( \Sigi > 1 \), 
and the orientation vector is given by,
\begin{eqnarray}
\oH & = & \sqrt{\frac{1}{2} + \frac{\sqrt{\Sigi^2 - 1}}{2 \Sigi}},
\: \:  \barbor \bcdot \beperp 
 = \sqrt{\frac{1}{2} - \frac{\sqrt{\Sigi^2 - 1}}{2 \Sigi}}.
 \label{eq:eq316} \\
 2 \oH^2 - 1 & = & \frac{\sqrt{\Sigi^2-1}}{\Sigi}, \: \:
 \oH(\barbor \bcdot \beperp) = \frac{1}{2 \Sigi}. \label{eq:eq316a}
\end{eqnarray}

The orthogonal basis unit vectors \( (\beomega, \beH, \beperp) \) shown in figure
\ref{fig:config} (a) are used, where \( \beomega
\) is along the vorticity direction, \( \beH \) is the along the direction of the magnetic
field and \( \beperp = \beomega \btimes \beH \). 
The diffusion coefficient due to hydrodynamic and magnetic interactions, equation 
\ref{eq:eq314}, is reduced to the form in equation \ref{eq:eq223},
where
\begin{eqnarray}
\Dwwh & = & 0, \: \: \Dwwm = \mbox{} - \frac{\chia (\Sigi + \sqrt{\Sigi^2-1})^2}{\pi
\sqrt{\Sigi^2-1}}, 
 \label{eq:eq317a} \\
\DHHh & = & \mbox{} - \frac{1}{4 \sqrt{\Sigi^2-1}}, \: \:
\DHHm = 0,
 \label{eq:eq317} \\
 \DHperph & = & 0, \: \: 
\DHperpm 
= \frac{\chia (\Sigi + \sqrt{\Sigi^2-1})}{2 \pi \sqrt{\Sigi^2-1}}, \label{eq:eq318} \\
\Dperpperph & = & \mbox{} - \frac{1}{4 \sqrt{\Sigi^2-1}}, \: \: 
\Dperpperpm = \mbox{} - \frac{\chia (\Sigi + \sqrt{\Sigi^2-1})^2}{\pi
\sqrt{\Sigi^2-1}}. \label{eq:eq319}
\end{eqnarray}
The isotropic part of the diffusion tensor due to the concentration dependence of
the viscosity and magnetic permeability is,
\begin{eqnarray}
\Diprime & = & \frac{\etaprime}{6 \sqrt{\Sigi^2-1}}.
\label{eq:eq319a}
\end{eqnarray}

The asymptotic values of the diffusion matrix for \( \Sigi \gg 1 \) and \( \Sigi - 1 \ll 1 \)
are reported in table \ref{tab:table1}. 
The qualitative characteristics of the diffusion matrix are very similar to that for permanent dipoles in a parallel magnetic
field discussed at the end of section \ref{subsec:parallelpermanent}. An important
difference is that \( \Dwwm \) and \( \Dperpperpm \) diverge for \( \Sigi \rightarrow 1 \)
where a transition occurs between static and rotating states.

The eigenvalues and eigenvectors of the diffusion matrix are provided in table 
\ref{tab:table1}. For the symmetric diffusion tensor, the eigenvalues are real,
and the orthogonal eigenvectors are the principal axes of extension/compression.
There is extension along directions with positive eigenvalues, indicating that
concentration fluctuations are damped, and compression along directions with 
negative eigenvalues resulting in amplification of concentration fluctuations.
The eigenvalues in \ref{tab:table1} are calculated without including the 
isotropic component \( \Di \bI \) in the diffusion tensor; these do not include
the effect of variations of viscosity with concentration.
The eigenvectors are unchanged when the isotropic diffusion tensor with 
the viscosity correction is included, and the eigenvalues are 
transformed as \( \lambda \rightarrow \lambda + \Di \).

The component \( \Dwwm \) of the diffusion tensor is always negative, and it
diverges as \( (- 4 \chia \Sigi/\pi) \) for \( \Sigi \gg 1 \). Therefore, perturbations
are always unstable in the vorticity direction perpendicular to the flow, and there
is clustering in this direction for a parallel magnetic field. For \( \Sigi \gg 1 \),
the principal eigenvalues are aligned along the \( \beomega, \bepar, \beperp \) directions.
The principal eigenvalue along the \( \beperp \) direction diverges
as \( (- 4 \chia \Sigi / \pi) \). The principal eigenvalue along the \( \beH \) direction 
is positive/negative for \( \chia \gtrless \pi \), and its magnitude decreases 
proportional \( \Sigi^{-1} \) in this limit. Thus, depending on the value of 
\( \chia \), there is weak amplification/damping
of fluctuations in the direction of the magnetic field, and strong amplification
of fluctuations in the other two directions. When the magnetic field is perpendicular
to the fluid velocity, this would result in the formation of long and narrow
clusters aligned along the magnetic field.

For \( \Sigi - 1 \ll 1 \) near the transition between rotating and steady states,
all three eigenvalues diverge proportional to \( (\Sigi - 1)^{-1/2} \). Two of
the eigenvalues, \( \lambda_3 \) in the \( \beperp \) direction and \( \lambda_2 \)
in the \( \beomega-\beH \) plane, are negative, and therefore strong amplification
of perturbations is predicted in these two directions. The other eigenvalue
\( \lambda_1 \) in the \( \bm{e}_1 \) direction is positive/negative for \( 2 (\sqrt{2} - 1)
\chia \gtrless \pi \), indicating damping of fluctuations if \( \chia \) exceeds
a threshold. It should be noted that the divergence exponent $-\frac{1}{2}$ is a
mean field exponent, which could be altered if fluctuations are included.
%
\subsection{Oblique magnetic field}
\label{subsec:obliqueinduced}
The orthogonal co-ordinate system shown in figure \ref{fig:config} (b) is used for an oblique
magnetic field, where \( \beomega \) is along the direction of the vorticity perpendicular to
the flow plane, the unit vector \( \bepar \) is perpendicular to \( \beomega \) in the \(
\barbomega-\barbH \) plane, and \( \beperp \) is perpendicular to the \( \barbomega-\barbH \) plane.
After the solution of equation \ref{eq:eq320} for \( \oH \) is determined, \( (\barbor \bcdot \beomega) \) is calculated using equation \ref{eq:eq10}. This completely specifies the orientation vector \( \barbor \). It is easily verified that for a parallel magnetic
field with \( \beomega \bcdot \beH = 0 \), the non-trivial solution reduces to
equation \ref{eq:eq316}.
The product \( \barbor \bcdot \bepar \) is given by equation \ref{eq:eq229},
and the product \( \barbor \bcdot \beperp \) is,
\begin{eqnarray}
 \barbor \bcdot \beperp & = & \frac{\barbor \bcdot (\beomega \btimes \beH)}{\sqrt{1 - (\beomega
 \bcdot \beH)^2}} = - \frac{\beomega \bcdot (\barbor \btimes \beH)}{\sqrt{1 - (\beomega
 \bcdot \beH)^2}} \nonumber \\
 & = & \frac{\beomega \bcdot (\bI - \barbor \barbor) \bcdot \beomega}{2 \Sigi (\barbor \bcdot
 \beH) \sqrt{1 - \wH^2}} 
 = \frac{[\oH^2 - \wH^2]}{2 \Sigi \oH^3 \sqrt{1 - \wH^2}}.
 \label{eq:eq329}
\end{eqnarray}
Here, the torque balance equation \ref{eq:torquebalancep} has been used to substitute for \( 
(\barbor \btimes \beH) \), and equation \ref{eq:eq10} is used to substitute \( \oH = \wH/(\barbor \bcdot \beomega) \).
Equations \ref{eq:eq229} and \ref{eq:eq329} are used to substitute for \( \barbor \) in 
equation \ref{eq:eq314}, and the diffusion tensor is of the form \ref{eq:eq232}, and the 
elements of the matrix are listed in table \ref{tab:table3}.

\begin{table}
 \begin{tabular}{|c|p{1.6in}|c|c|c|} \hline
  & & $\Sigi \ll 1, \wH \neq \frac{1}{\sqrt{2}}$ 
  & $\Sigi \ll 1, \wH = \frac{1}{\sqrt{2}}$ 
  & $\Sigi \gg 1$ \\ \hline \hline
  $\Dwwh$
  & $\mbox{} - \frac{(\oH^2 - \wH^2) \wH^2}{4 \Sigi \oH^4 [2 \oH^2 - 1]}$
  &  $\mbox{} - \frac{\Sigi \wH^2 (1 - \wH^2)}{(2 \wH^2-1)}$ 
  & $\mbox{} - \frac{1}{4 \Sigi}$
  & $\mbox{} - \frac{\wH^2 (1 - \wH^2)}{4 \Sigi}$
  \\ \hline
  $\Dwwm$ 
  & 
  $\mbox{} - \frac{4 \chia \Sigi \oH^2 (\oH^2-\wH^2)}{\pi (2 \oH^2-1)}$
  &  $\mbox{} - \frac{16 \chia \Sigi^3 \wH^6 (1 - \wH^2)}{\pi (2 \wH^2-1)}$ 
  & $\mbox{} - \frac{\chia \Sigi}{\pi}$
    & $\mbox{} - \frac{4 \chia \Sigi (1 - \wH^2)}{\pi}$
  \\ \hline \hline
  
  $\Dwparh$
  & $ \mbox{} - \frac{\wH (\oH^2 - \wH^2
 )^2}{4 \Sigi \oH^4 [2 \oH^2 - 1]
 \sqrt{1 - \wH^2}}$
  &  $\mbox{} - \frac{4 \Sigi^3 \wH^5 (1 - \wH^2)^{3/2}}{(2 \wH^2-1)}$ 
  & $\mbox{} - \frac{\Sigi}{4}$
  & $\mbox{} - \frac{\wH (1 - \wH^2)^{3/2}}{4 \Sigi}$
  \\ \hline
  $\Dwparm$
  & $\frac{2 \chia \Sigi \oH^2 \wH (1 + \oH^2 - 2 \wH^2)}{\pi [2 \oH^2 - 1] \sqrt{1 - \wH^2}}$
  &  $\frac{2 \chia \Sigi \wH^3 \sqrt{1-\wH^2}}{\pi (2 \wH^2 - 1)}$ 
  & $\frac{\chia}{4 \Sigi}$
  & $\frac{4 \chia \Sigi \wH \sqrt{1 - \wH^2}}{\pi}$
  \\ \hline \hline

  $\Dwperph$ 
  & $\mbox{} - \frac{\wH (\oH^2-\wH^2)^2}{8
 \Sigi^2 \oH^6 [2 \oH^2 - 1] \sqrt{1 - \wH^2}}$
  &  $\mbox{} - \frac{2 \Sigi^2 \wH^3 (1 - \wH^2)^{3/2}}{(2 \wH^2 - 1)}$
  & $\mbox{} - \frac{1}{4}$
  & $\mbox{} - \frac{\wH (1 - \wH^2)^{3/2}}{8 \Sigi^2}$
  \\ \hline
  $\Dwperpm$ 
  & $\frac{\chia \wH (\oH^2-\wH^2)}{\pi [2 \oH^2 - 1] \sqrt{1 - \wH^2}}$
  &  $\frac{4 \chia \Sigi^2 \wH^5 \sqrt{1 - \wH^2}}{\pi (2 \wH^2-1)}$
  & $\frac{\chia}{2 \pi}$
  & $\frac{\chia \wH \sqrt{1 - \wH^2}}{\pi}$
  \\ \hline \hline
  
  $\Dparparh$
  & $\frac{(\oH^2 - \wH^2) (2 \oH^2 \wH^2 - \oH^2 - \wH^4)}{4 \Sigi \oH^4 [2 \oH^2 - 1] (1 - \wH^2)}$
  &  $\mbox{} - \frac{\Sigi \wH^2 (1 - \wH^2)}{(2 \wH^2 - 1)}$ 
  & $\mbox{} - \frac{1}{4 \Sigi}$
  & $\mbox{} - \frac{(1 - \wH^2)^2}{4 \Sigi}$
  \\ \hline
  $\Dparparm$
  & 
  $\mbox{} - \frac{4 \chia \Sigi \oH^2 \wH^2}{\pi (2 \oH^2-1)}$
  &  $\mbox{} - \frac{4 \chia \Sigi \wH^4}{\pi(2 \wH^2-1)}$
  & $\mbox{} - \frac{\chia}{\pi \Sigi}$
  & $\mbox{} - \frac{4 \chia \Sigi \wH^2}{\pi}$
  \\ \hline \hline
  
  $\Dparperph$ 
  & $\frac{\wH^2 (\oH^2 - \wH^2)^2}{8 \Sigi^2 \oH^6
 [2 \oH^2 - 1] (1 - \wH^2)}$
  &  $\frac{2 \Sigi^2 \wH^4 (1 - \wH^2)}{(2 \wH^2-1)}$
  & $\frac{1}{4}$
  & $\frac{\wH^2 (1 - \wH^2)}{8 \Sigi^2}$
  \\ \hline
  $\Dparperpm$ 
  & $- \mbox{} \frac{4 \chia \Sigi \oH^4}{\pi (2 \oH^2 - 1)}$
  & $\frac{4 \chia \Sigi^2 \wH^4 (1 - \wH^2)}{\pi (2 \wH^2 - 1)}$
  & $\frac{\chia}{2 \pi}$
  & $\frac{\chia (1 - \wH^2)}{\pi}$
  \\ \hline \hline
  
  $\Dperpperph$
  & $\mbox{} - \frac{(\oH^2-\wH^2)}{4 \Sigi \oH^2 [2 \oH^2-1]} +
  \frac{\wH^2 (\oH^2 - \wH^2)^2}{16 \Sigi^3 \oH^8
 [2 \oH^2 - 1] (1 - \wH^2)}$
  &  $\mbox{} - \frac{4 \Sigi^3 \wH^4 (1 - \wH^2)}{(2 \wH^2 - 1)}$
  & $\mbox{} - \frac{\Sigi}{2}$
  & $\mbox{} - \frac{(1 - \wH^2)}{4 \Sigi}$
  \\ \hline
  $\Dperpperpm$
  & $\mbox{} - \frac{4 \chia \Sigi \oH^4}{\pi (2 \oH^2-1)}$
  & $\mbox{} - \frac{4 \chia \Sigi \wH^4}{\pi(2 \wH^2-1)}$  
  &  $\mbox{} - \frac{\chia}{\pi \Sigi}$
  & $\mbox{} - \frac{4 \chia \Sigi}{\pi}$
  \\ \hline \hline
  
  $\Diprime$ & $\frac{\etaprime (\oH^2-\wH^2)}{6 \Sigi \oH^2 (2 \oH^2-1)}$ & 
  $\frac{2 \etaprime \Sigi \wH^2 (1-\wH^2)}{3 (2 \wH^2-1)}$ & 
  $\frac{\etaprime}{6 \Sigi}$ & 
  $\frac{\etaprime (1-\wH^2)}{6 \Sigi}$
 \\ \hline

 $\lambda_{1}$ 
 &
 & $\frac{\Sigi \wH^2 (1-\wH) (2 \chia \wH - \pi (1 + \wH))}{\pi (2 \wH^2-1)}$
 & $\frac{2 \chia (\sqrt{2}-1) - \pi}{4 \pi \Sigi}$
 & $\frac{(\chia-\pi) (1 - \wH^2)}{4 \pi \Sigi}$
 \\ \hline
 ${\bf e}_1$ 
 & 
 & $\frac{\sqrt{1 + \wH} \beomega + \sqrt{1 - \wH} \bepar}{\sqrt{2}}$
 & $\frac{\beomega + (\sqrt{2}-1) \bepar}{2^{3/4} \sqrt{\sqrt{2}-1}}$
 & $\wH \beomega + \sqrt{1-\wH^2} \bepar$
 \\ \hline \hline
 $\lambda_2$ 
 & 
 & $\mbox{} - \frac{\Sigi \wH^2 (1 + \wH) (2 \chia \wH + \pi (1 - \wH))}{\pi (2 \wH^2-1)}$
 & $\mbox{} - \frac{2 \chia (\sqrt{2} + 1) + \pi}{4 \pi \Sigi}$
  & $\mbox{} - \frac{4 \chia \Sigi}{\pi}$
 \\ \hline
 ${\bf e}_2$ 
 & 
  & $\frac{- \sqrt{1 - \wH} \beomega + \sqrt{1 + \wH} \bepar}{\sqrt{2}}$
  & $\frac{- \beomega + (\sqrt{2}+1) \bepar}{2^{3/4} \sqrt{\sqrt{2}+1}}$
 & $\mbox{} - \sqrt{1-\wH^2} \beomega + \wH \bepar$
 \\ \hline \hline
 $\lambda_3$
 &
 & 
 $\mbox{} - \frac{4 \chia \Sigi \wH^4}{\pi (2 \wH^2-1)}$
 & $\mbox{} - \frac{\chia}{\pi \Sigi}$
  & $\mbox{} - \frac{4 \chia \Sigi}{\pi}$
\\ \hline
 ${\bf e}_3$ 
 & 
 & $\beperp$
 & $\beperp$
  & $\beperp$
 \\ \hline \hline
 \end{tabular}
 \caption{\label{tab:table3} The asymptotic behaviour of the diffusion coefficients for 
 \( \Sigi \gg 1 \) and \( \Sigi \ll 1 \) for an oblique
 magnetic field for particles with induced dipoles.}
\end{table}

\begin{figure}
\parbox{.5\textwidth}{
    \psfrag{x}[][][1.0][0]{$\Sigi$}
    \psfrag{y}[][][1.0][0]{$\Diprime/\etaprime$}

 \includegraphics[width=.49\textwidth]{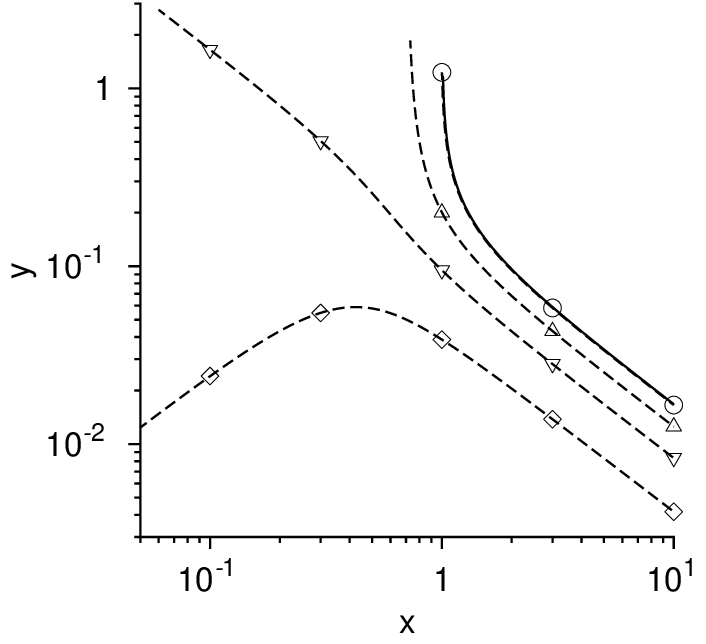}
 
 }
%
%
  
 \caption{\label{fig:diffp} The scaled isotropic part of the diffusion
 tensor, \( \Diprime/\etaprime \) 
 as a function of the parameter \( \Sigi \) for 
 particles with a permanent dipole. The orientation of the vorticity and magnetic field are 
 \( \beomega \bcdot
\beH = 0.1 \) ($\circ$), \( \frac{1}{2} \) ($\triangle$), \( \frac{1}{\sqrt{2}} \) ($\nabla$)
and \( \frac{\sqrt{3}}{2} \) ($\diamond$). The solid line is the result for
a parallel magnetic field.
 }
\end{figure}

Also presented in table \ref{tab:table3} are the expansions for the components
of \( \bm{D}^h \) and \( \bm{D}^m \) for the limits of small and large \( \Sigi \)
respectively. These are determined using the expansion for the solutions of
equation \ref{eq:eq320},
\begin{eqnarray}
 \oH & = & \wH + 2 \Sigi^2 \wH^3 (1 - \wH^2) \: \: \mbox{for} \: \: \Sigi \ll 1, \label{eq:eq336} \\
 \oH & = & 1 - \frac{1 - \wH^2}{8 \Sigi^2} \: \: \mbox{for} \: \: \Sigi \gg 1. \label{eq:eq337}
\end{eqnarray}
The denominators of the diffusion coefficients in the third column in table \ref{tab:table3} decrease 
to zero for \( \Sigi \ll 1 \) and \( \wH = \frac{1}{\sqrt{2}} \). In this case,
it is necessary to first substitute the expression \ref{eq:eq336} into the expressions
in the second column in table \ref{tab:table3}, and then substitute \( \wH = 
\frac{1}{\sqrt{2}} \) and take the limit \( \Sigi \ll 1 \). The results obtained in this
manner are presented in the fourth column in table \ref{tab:table3}.

\begin{figure}
\parbox{.49\textwidth}{
    \psfrag{x}[][][1.0][0]{$\Sigi$}
    \psfrag{y}[][][1.0][0]{\textcolor{blue}{$-\Dwwh$}, \textcolor{red}{$- \Dwwm/\chia$}}

 \includegraphics[width=.49\textwidth]{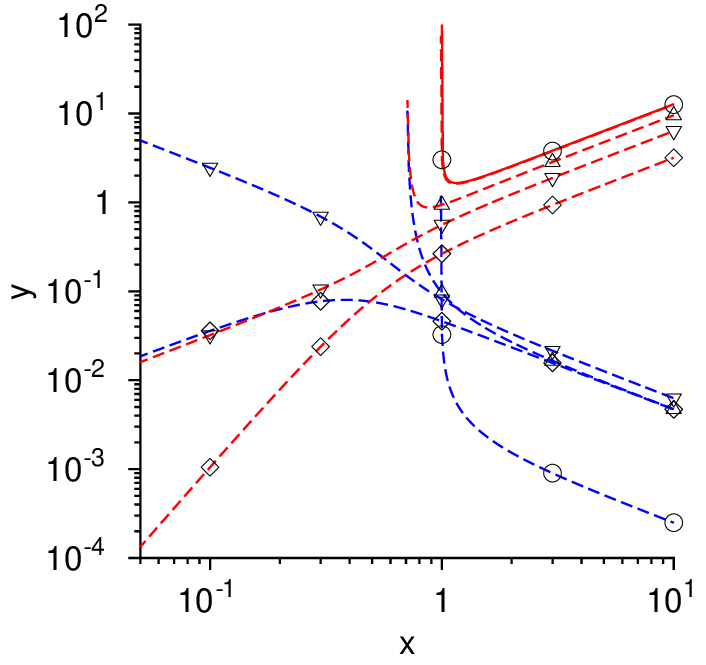}
 
 \begin{center} (a) \end{center}
 }
\parbox{.49\textwidth}{
    \psfrag{x}[][][1.0][0]{$\Sigi$}
    \psfrag{y}[][][1.0][0]{\textcolor{blue}{$-\Dwparh$}, \textcolor{red}{$\Dwparm/\chia$}}

 \includegraphics[width=.49\textwidth]{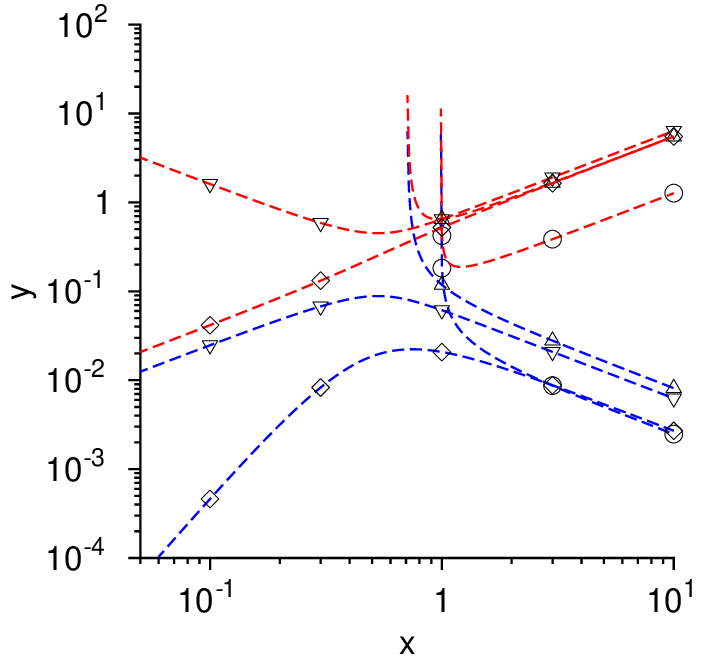}
 
 \begin{center} (b) \end{center}
 }
 
\parbox{.49\textwidth}{
    \psfrag{x}[][][1.0][0]{$\Sigi$}
    \psfrag{y}[][][1.0][0]{\textcolor{blue}{$-\Dwperph$}, \textcolor{red}{$\Dwperpm/\chia$}}

 \includegraphics[width=.49\textwidth]{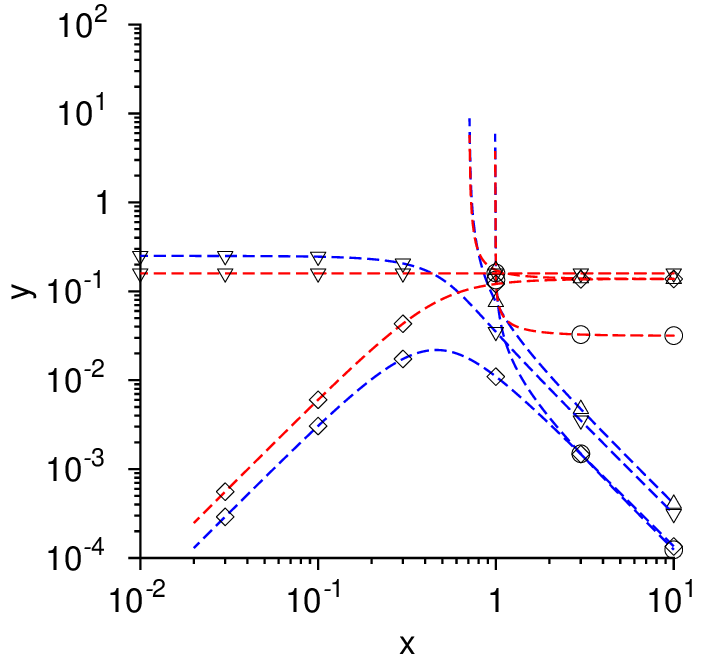}
 
 \begin{center} (c) \end{center}
 }
\parbox{.49\textwidth}{
    \psfrag{x}[][][1.0][0]{$\Sigi$}
    \psfrag{y}[][][1.0][0]{\textcolor{blue}{$-\Dparparh$}, \textcolor{red}{$- \Dparparm/\chia$}}

 \includegraphics[width=.49\textwidth]{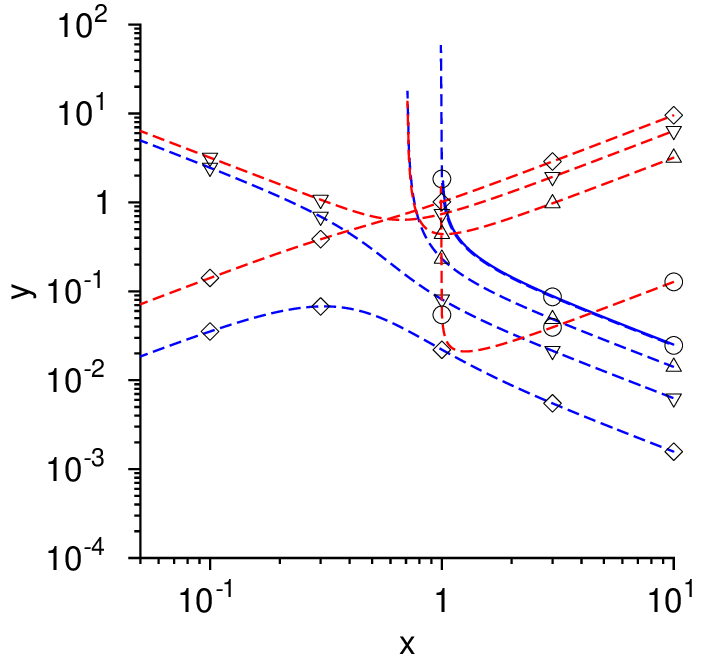}
 
 \begin{center} (d) \end{center}
 }

 \parbox{.49\textwidth}{
    \psfrag{x}[][][1.0][0]{$\Sigi$}
    \psfrag{y}[][][1.0][0]{\textcolor{blue}{$\Dparperph$}, \textcolor{red}{$\Dparperpm/\chia$}}

 \includegraphics[width=.49\textwidth]{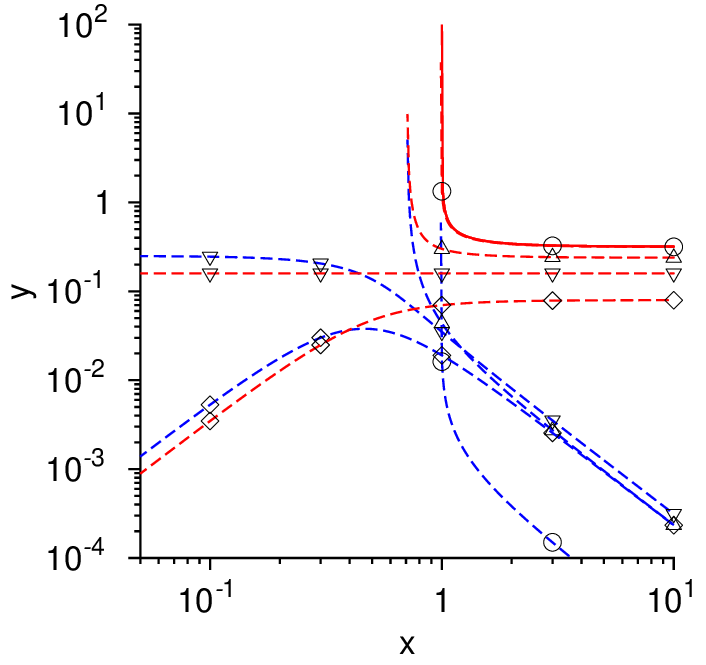}
 
 \begin{center} (e) \end{center}
 }
\parbox{.49\textwidth}{
    \psfrag{x}[][][1.0][0]{$\Sigi$}
    \psfrag{y}[][][1.0][0]{\textcolor{blue}{$-\Dperpperph$}, \textcolor{red}{$- \Dperpperpm/\chia$}}

 \includegraphics[width=.49\textwidth]{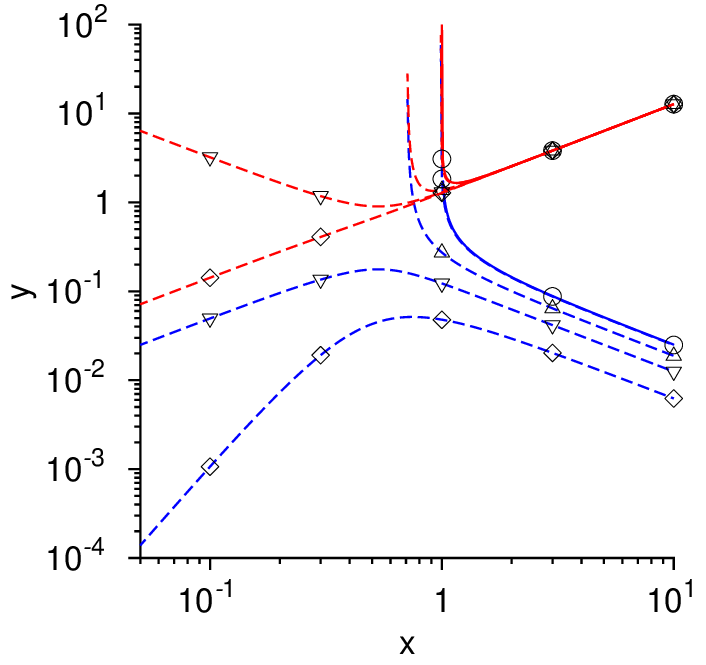}
 
 \begin{center} (f) \end{center}
 }
 
 \caption{\label{fig:diffi} The components of the diffusion tensor 
 \( - \Dwwh \& \pm \Dwwm/\chia \) (a), 
 \( - \Dwparh \& \Dwparm/\chia \) (b), 
 \( - \Dwperph \& \Dwperpm/\chia \) (c), 
 \( - \Dparparh \& \pm \Dparparm/\chia \) (d), 
 \( \Dparperph \& \Dparperpm/\chia \) (e) and 
 \( - \Dperpperph \& - \Dperpperpm/\chia \) (f) 
 due to hydrodynamic interactions (blue lines) and magnetic interactions (red lines) as a function of the parameter \( \Sigi \) for particles with an induced dipole moment. The orientation of the vorticity and magnetic field are \( \beomega \bcdot
\beH = 0.1 \) ($\circ$), \( \frac{1}{2} \) ($\triangle$), \( \frac{1}{\sqrt{2}} \) ($\nabla$)
and \( \frac{\sqrt{3}}{2} \) ($\diamond$). The solid blue and red lines are the non-zero results for a
parallel magnetic field. 
 }
\end{figure}

The striking feature of the diffusion tensor elements in table \ref{tab:table3}, is the singularity 
at \( \oH^2 = \frac{1}{2} \). From equation \ref{eq:eq320}, this corresponds
to \( \Sigi = \sqrt{1 - 2 \wH^2} \), shown by the red line in figure \ref{fig:transobli},
and this singularity exists only for \( \wH < \frac{1}{\sqrt{2}} \).
This boundary is different from the blue boundary between stationary and 
rotating states for the single-particle dynamics, and this represents a dynamical
transition due to inter-particle interactions. At this boundary, the `susceptibility'
multiplying $\torp_{\bk}$ in equation \ref{eq:eq38} decreases to zero, and therefore
the perturbation to the orientation vector \( \torp_{\bk} \) in equation
\ref{eq:eq310} diverges. The divergence at \( \oH^2 = \frac{1}{2} \) is observed 
in all components of the diffusion tensor in figure \ref{fig:diffi} with the exception
of $\Dperpperpm$. Of course, the analysis is not accurate at this point because 
it is carried out assuming \( \torp_{\bk} \ll 1 \), and it is necessary to 
include non-linear fluctuation effects in the vicinity of this point to extract
the nature of the divergence. This is a subject for future study.


For \( \Sigi \gg 1 \), the characteristics of the diffusion tensor are
similar to those for dipolar particles in figure \ref{fig:diffp}.
The largest contributions to the diffusion tensor are \( \Dwwm, \Dwparm, \Dparparm \) 
and \( \Dparperpm \); all of these diverge proportional to \( \Sigi \).
The eigenvalue \( \lambda_3 \) corresponding to the \( \beperp \)
direction is negative, and diverges proportional to \( \Sigi \), and
therefore concentration fluctuations are amplified in the direction 
perpendicular to the plane containing the vorticity and magnetic
field. In the \( \beomega-\bepar \) plane, one of the eigenvalues 
\( \lambda_2 \) diverges proportional to \( - (4 \chi \Sigi/\pi) \),
indicating strong amplification of concentration fluctuations
in the \( \bm{e}_2 \) direction. The third eigenvalue \( \lambda_1 \)
decreases proportional to \( \Sigi^{-1} \), and this is 
positive/negative for \( \chia \gtrless \pi \), indicating
weak amplification/damping of fluctuations. The two directions
\( \bm{e}_1 \) and \( \bm{e}_2 \) align with \( \beH \) and
\( \beomega \) for a parallel magnetic field \( \wH = 0 \).

The characteristics of the diffusion tensor are similar to those for dipolar particles
for \( \Sigi \ll 1 \) and \( \wH > \frac{1}{\sqrt{2}} \). The diffusion coefficients
\( \Dwwh, \Dwparm, \Dparparh, \Dparparm \) and \( \Dperpperpm \), and all three
eigenvalues, decrease proportional to \( \Sigi \). The eigenvalue of the diffusion
tensor in the \( \beperp \) direction is negative for \( \wH > \frac{1}{\sqrt{2}} \), 
\( \lambda_3 = - (4 \chia \Sigi \wH^2/\pi (2 \wH^2-1)) \) and this is an 
unstable direction. One of the eigenvalues in the \( \bm{e}_2 \) direction is also
always negative, while the third eigenvalue is positive/negative for
\( 2 \chia \wH \gtrless \pi (1 + \wH) \). Thus, concentration fluctuations
are unstable in the \( \beperp \) and \( \bm{e}_2 \) directions, and they
could be stable/unstable in the \( \bm{e}_1 \) direction depending on 
the value of \( \chia \).

For \( \wH > \frac{1}{\sqrt{2}} \), there is a dynamical transition at \( \oH = \frac{1}{2} \)
along the red line in figure \ref{fig:transobli}. This is a transition in the collective
dynamics of the particles, and is distinct from the single-particle transition between
stationary and rotating states along the blue line in figure \ref{fig:transobli}. All
the coefficients of the diffusion matrix diverge proportional to \( (\oH - \frac{1}{2})^{-1/2}
\) at this transition. For \( \wH = \frac{1}{\sqrt{2}} \), the elements of the diffusion tensor 
have a different scaling than that for \( \wH > \frac{1}{\sqrt{2}} \). The elements
\( \Dwwh, \Dwparm, \Dparparh, \Dparparm \) and \( \Dperpperpm \) actually increase proportional 
to \( \Sigi^{-1}
\) in this limit. The eigenvalues of the diffusion matrix increase proportional to \( \Sigi^{-1} 
\). Two of these
is negative, and therefore concentration fluctuations are amplified in one direction in the 
\( \beomega - \bepar \) plane and in the \( \beperp \) direction. The third is positive for 
\( 2 \chia (\sqrt{2} - 1) > \pi \),
and negative otherwise. 
\section{Conclusions}
\label{sec:conclusions}
The principal result of the present calculation is that the hydrodynamic and magnetic 
inter-particle interactions manifest as an anisotropic diffusion tensor in the equation 
for the particle concentration field for spherical particles with a permanent or
induced dipole moment. The 
magnetic dipole due to neighbouring particles results in the disturbance to the magnetic field
at a test particle. When a neighbouring particle is stationary in a shear flow, there is a hydrodynamic
disturbance in the form of an antisymmetric force moment, which causes a disturbance to the 
vorticity at the location of a test particle. In addition, there is the modification of the applied
magnetic field due to the particle magnetisation, which depends on the particle concentration.
The net effect of these disturbances is zero in 
a spatially uniform suspension. When there are variations in the particle concentration field,
the net effect of these interactions in the torque balance equation results in a disturbance to the 
orientation vector. There is a force due to the gradient in the dot product of the magnetic moment
and the magnetic field, which causes a drift velocity of the particles relative to the fluid. 
There is a contribution to the drift velocity due to the variation in the magnetic moment
per unit volume when there is a concentration variation. The 
divergence of the drift velocity in the concentration equation has the form of an anisotropic
diffusion term, and the elements of the diffusion tensor have been calculated for both permanent and 
induced dipoles. 

The dimensionless parameters are the ratio of the magnetic and hydrodynamic torque on a particle,
\( \Sigma \) for permanent dipoles (equation \ref{eq:eq221}) and \( \Sigi \) for induced dipoles
(equation \ref{eq:eq315}), and the ratio of the magnetic moment per unit volume and the magnetic field,
\( \Ma \) (equation \ref{eq:eq221a}) for permanent dipoles and \( \chia \) (equation \ref{eq:eq315a}) 
for induced dipoles. The diffusion tensor consists of two distinct contributions, one due to 
magnetic interactions which is proportional to \( \Ma \) \& \( \chia \), and the second due
to hydrodynamic interactions which does not depend on \( \Ma \) \& \( \chia \). The product
\( \Sigma \Ma = (\mu_0 M^2/\pi \eta_0 d^6 |\barbomega|) \), which is independent of the magnetic field,
represents the effects of the interaction between particles and the modification of the magnetic
field due to the particle magnetic moment. This is the inverse of the Mason number (\cite{sherman}),
for simple shear flows where the magnitudes of the strain rate and vorticity are equal. For particles
with an induced dipole moment, the product \( \Sigi \chia = (\mu_0 \chi^2 |\barbH|^2/\pi \eta_0 d^6 
|\barbomega|) \) is the inverse of the Mason number for a simple shear flows.

The components of the diffusion tensor are listed in table \ref{tab:table1} for the particular case
\( \wH = 0 \) where the magnetic field is in the flow plane. For high magnetic field, \(\Sigma, \Sigi \gg 1 \),
two principal directions of the diffusion tensor are along the magnetic field and along the vorticity
direction (perpendicular to the flow plane), while the third component is orthogonal to the first
two. For particles with a permanent and induced dipoles, the eigenvalues of the diffusion 
tensor in the two directions perpendicular to the magnetic field are negative and diverge 
proportional to \( \Sigma, \Sigi 
\); this would lead to strong amplification of concentration fluctuations in these two directions. 
The eigenvalue of the diffusion tensor along the magnetic field is positive when \( \Ma, \chia \)
exceed a threshold and negative otherwise, and it decreases proportional
to \( \Sigma^{-1} \); this would lead to weak amplification/damping of fluctuations along the direction of the 
magnetic field. When the magnetic field is perpendicular to the velocity, this explains
the experimental observation of the formation of particle chains in the direction of the field
direction when a magnetic field is applied. 

The diffusion matrix exhibits interesting behaviour close to the transition between rotating
and stationary solutions of the orientation vector, \( \Sigma - \frac{1}{2} \ll 1 \) and
\( \Sigi - 1 \ll 1 \). The component of the diffusion coefficient perpendicular to the plane of
flow is negative, indicating amplification of
concentration fluctuations perpendicular to the flow plane. The components of the diffusion
tensor in the flow plane diverge proportional to \( (\Sigma - \frac{1}{2})^{-1/2} \) and \( 
(\Sigi - 1)^{-1/2} \) for permanent and induced dipoles. The eigenvalue of the diffusion 
matrix in one of the principal directions in the flow plane is negative, indicating
strong clustering, while that in the other principal direction is positive when
\( \Ma \) or \( \chia \) exceed a threshold. This implies a strong anisotropic clustering 
tendency in the flow plane as the magnetic field is reduced near the transition to 
rotating states. This intriguing phenomenon should be observable in experiments similar to 
those performed in the field of dynamical critical phenomena (\cite{halperinhohenberg}).
The exponents \( - \frac{1}{2} \) calculated here is a mean field exponent; renormalisation group 
calculations are required to determine how these exponents change when fluctuations are
incorporated.

For a permanent dipole, there is no transition between rotating and stationary states for an
oblique magnetic field ( \( \wH \neq 0 \)). The elements of the diffusion tensor, and their 
asymptotic behaviour 
for \( \Sigma \ll 1 \) and \( \Sigma \gg 1 \) are shown in table \ref{tab:table2}. In all
cases, one of the principal directions \( \beperp \), which is orthogonal to the vorticity
and the magnetic field, and the eigenvalues of the diffusion matrix in this direction are
all negative, indicating that there is strong amplification of concentration fluctuations
in this direction when a magnetic field is applied. For particles with permanent dipoles in the limit
\( \Sigma \gg 1 \), one eigenvalue of the diffusion tensor is negative and its magnitude 
increases proportional 
to \( \Sigma \) in one principal direction in the \( \barbomega-\barbH \) plane, but the
eigenvalue perpendicular to the \( \barbomega-\barbH \) plane is positive when \( \Ma \)
exceeds a threshold and it decreases
proportional to \( \Sigma^{-1} \). For \( \Sigma \ll 1 \), the eigenvalues in the 
\( \barbomega-\barbH \) plane increase proportional to \( \Sigma \). One of the eigenvalues
is negative, and the second could be positive or negative depending on the value of \( \Ma \).

For a suspension of particles with induced dipoles, there is a dynamical transition at the red line in figure 
\ref{fig:transobli}, which is different from the blue line where there is a transition between
static and rotating states in the single-particle dynamics. The dynamical transition is due to
inter-particle interactions. Of course, the linearisation approximation is not applicable close
to the transition where the disturbance to the orientation vector diverges, and more analysis
is required to examine how the divergence in the diffusion coefficients is cut-off due to 
non-linear effects. 

The eigenvalues of the diffusion matrix for particles with induced dipoles are qualitatively
similar to those with permanent dipoles. For \( \Sigi \gg 1 \), there is strong amplification 
of fluctuations perpendicular to the \( \barbomega-\barbH \) plane, and
in one principal direction in the \( \barbomega-\barbH \) plane, and the magnitude of
the eigenvalue increases proportional to \( \Sigi \) in this limit. There is amplification or 
damping in the third direction depending on the value of \( \chia \), and the magnitude of 
the eigenvalue is proportional to \( \Sigi^{-1} \). For \( \Sigi \ll 1 \), the magnitudes
of the eigenvalues decrease proportional to \( \Sigi \). The 
eigenvalue in the direction perpendicular to the \( \barbomega-\barbH \) plane is negative,
and concentration fluctuations are amplified in this direction. The eigenvalue in one 
principal directions in the \( \barbomega-\barbH \)
plane is also negative, and the second is positive or negative depending on the value
of \( \chia \) and the angle \( \wH \) between the magnetic field and the vorticity direction.

The effect of viscosity variations due to variations in the particle concentration
has also been analysed, considering a linear model for the dependence of the 
viscosity on the concentration. This results in a positive contribution to the 
isotropic part of the diffusion tensor, which dampens concentration fluctuations.
For particles with permanent dipoles, this contribution decreases proportional to 
\( \Sigma \) for \( \Sigma \ll 1 \), and proportional to \( \Sigma^{-1} \)
for \( \Sigma \gg 1 \). For particles with induced dipoles, the diffusion
coefficient decreases proportional to \( \Sigi^{-1} \) for \( \Sigi \gg 1 \),
and it decreases proportional to \( \Sigi \ll 1 \) for values of \( \wH \) 
where there is no dynamical transition.

The estimates for the hydrodynamic and magnetic contributions to the diffusion
tensor are as follows. The hydrodynamic contribution operates perpendicular to 
the magnetic field for a parallel magnetic field (equations \ref{eq:eq226},
\ref{eq:eq319}), and this scales as 
\begin{eqnarray}
 D_h \sim \phi d^2 |\barbomega|.
 \end{eqnarray}
 The characteristic diffusion time, the time for a particle to 
 diffuse a distance comparable to its diameter, is \( \tau_h = (\phi |\barbomega|)^{-1} \), 
 is independent of particle diameter. The magnetic contribution is proportional to 
\( D_m \sim \phi d^2 |\barbomega| \Ma \Sigma \sim (\phi \mu_0 M^2 / \pi \eta
d^4) \) (equation \ref{eq:eq224}).  This depends only on the magnetic moment 
of the particles, and not on the magnetic field or the particle angular velocity.
The magnetic moment per particle is the product of the magnetic moment 
per unit volume \( M_v \) and the particle volume, \( M = M_v (\pi d^3/6) \mbox{A 
m}^2 \). Based on this, the estimate for the diffusion coefficient is
\begin{eqnarray}
D_m \sim (\phi d^2 \mu_0 M_v^2 / \eta).
\end{eqnarray}
The characteristic diffusion time, which is the time taken to diffuse a
distance comparable to the particle diameter, is \( \tau_m = (\eta / \phi
\mu_0 M_v^2) \).
Thus, both \( D_h \) and \( D_m \) are proportional to \( d^2 \phi \), and the 
characteristic diffusion times are independent of diameter, if
\( D_m \) is expressed in terms of magnetic moment per unit volume. The vorticity
in magnetorheological applications, which is of the same magnitude as the strain
rate, could vary between \( 1-10^4 \mbox{s}^{-1} \). Therefore, the minimum 
hydrodynamic diffusion time is \( \tau_h \sim (10^{-4}/\phi) \mbox{ s} \). 
For particles with a permanent magnetic dipole, the dipole moment is usually 
expressed as the magnetic moment per unit mass, emu/gm, and the dipole moments are in 
the range 1-100 emu/gm. The magnetic moment per unit volume \( M_v \) is the product 
of the magnetic moment per unit mass and the mass density, which is in the range 
\( M_v \sim 1-10^3 \) emu/cm$^{3}$ \(\sim 10^3-10^6 \mbox{A/m} \), if we assume 
the material has a mass density of 
1-10 gm/cm$^3$. Therefore, the minimum magnetic diffusion time is \( (10^{-6}/\phi)
\mbox{s} \). Thus, the time required for a particle to diffuse across a distance
comparable to its diameter is in the \( \mbox{ms}-\mu \mbox{s} \) range if the strain rate
is sufficiently high. This provides a plausible mechanism for the formation of 
sample-spanning clusters when a magnetic field is applied.

The analysis of the diffusion due to interactions has been comprehensive, covering
particle suspensions with permanent and induced dipoles, and different relative
orientations of the flow and gradient directions and the applied magnetic field.
The diffusion tensors determined here can be incorporated in continuum equations
for magnetorheological fluids in order to capture the effect of interactions
on the particle dynamics and orientation. They provide an opportunity for a more 
granular design of magnetorheological devices, where the dimensionless parameters and 
the relative orientation of the flow and magnetic field could be designed for 
dispersion/clustering of desired magnitudes along specific axes. From a 
fundamental perspective, the present analysis reveals a rich dynamical 
landscape inviting detailed inspection of several interesting phenomena,
such as the divergence of the diffusivities at the transition between steady
and rotating orientation states for a suspension of particles with a permanent
dipole, and the dynamical transition where the diffusivities diverge due to 
collective effects for particles with an induced dipole. 
There is also scope for incorporating features such as spheroidal
particles, where the shape factor could lead to additional interesting 
phenomena not present for spherical particles.

The author would like to thank the Department of Science and Technology, Government of 
India for financial support.

The author reports no conflict of interest.

\bibliographystyle{jfm}
\bibliography{ref}

\end{document}